\documentstyle[12pt]{report}
\begin{document}

\centerline{\Large\bf 
Large fluctuations in multi-attractor systems}

\centerline{\Large\bf 
and the generalized 
Kramers problem}

\vskip 0.5truecm

\centerline {S M Soskin}

\noindent
Institute of 
Semiconductor Physics,
National Ukrainian Academy of Sciences,  Kiev, Ukraine.

\vskip 0.2truecm
\noindent
{\bf Abstract}

The main subject of the paper is an escape from a multi-well metastable potential on a time-scale of a 
formation of the quasi-equilibrium  between the wells. The main attention is devoted to such ranges of 
friction in which an external saddle does not belong to a basin of attraction of an initial attractor. A 
complete rigorous analysis of the problem for the {\it most probable escape path} is presented and a 
corresponding escape rate is calculated with a logarithmic accuracy. Unlike a conventional rate for a 
quasi-stationary flux, the rate on shorter time-scales strongly depends on friction, moreover, it may 
undergo oscillations in the underdamped range and a cutoff in the overdamped range.

A generalization of the results for inter-attractor transitions in stable potentials with more than two wells is also 
presented as well as a splitting procedure for a phenomenological description of inter-attractor transitions is suggested.

Applications to such problems as a dynamics of escape on time-scales shorter than an optimal fluctuation duration, 
prehistory problem, optimal control of 
fluctuations, fluctuational transport in ratchets, escapes at a periodic
driving and transitions in biased Josephson junctions and ionic channels are briefly discussed.

\vskip 0.2truecm


\chapter {Introduction}

The problem of rare fluctuational transitions in a classical system 
driven by a weak noise attracts an attention of theorists for more than half a
century (for a historical review see e.g. \cite{landauer}). Its treatment in 
a contemporary form should be counted probably from the celebrated work by 
Kramers  \cite{kramers} \footnote{In fact, there was an earlier work, by 
Pontryagin, Andronov and Vitt 
\cite{pav}, which forestalled in many respects the Kramers paper, but it 
remained practically 
unknown for western scientists and, historically, has not influenced the 
development of science.} 
in which, in particular, the noise-induced escape from a metastable potential well was 
considered. The principal result by Kramers is that, after a short initial period during which a 
quasi-stationary distribution formes within the well, the probability flux from it is an 
exponentially decaying function:

\begin {equation}
J = \alpha e^{-\alpha t}, 
\end{equation}

\noindent where the escape rate $\alpha$ contains, appart from the Arrhenius factor \cite 
{arrhenius} $\exp (-\Delta U/T)$ ($\Delta U$ is a height of the potential barrier and $T$ is a 
temperature) , a preexponential factor which relatively weakly depends on 
$T$, friction $\Gamma$ and some details of a potential $U(q)$

\begin {equation}
\alpha = A(T, \Gamma, [U]) \exp (-\frac {\Delta U} {T}), \ \
T \ll \Delta U, 
\end{equation}

\noindent and Kramers derived explicit asymptotic formulas for $A$ in the ultra-underdamped and moderate-to-overdamped limits.

I do not have a room here to review all developments and generalizations of the
Kramers problem (surveys of the state of art, at least by the end of 80th, are given in major
reviews \cite {hanggi}, \cite {meln}). Rather I shall mention just two activities which are quite
relevant to the subject of my paper.

One of them concerned the problem of filling the \lq\lq gap" between the ultra-underdamped and
moderate-to-overdamped limits for the expression of the preexponential factor $A$. This activity
was crowned by the work by Melnikov \cite {meln84} (see also the review \cite {meln} and
references therein) who developed a very beautiful method based on the reduction of the Fokker-Plank equation, in the underdamped regime, to some more simple integral equation which, in its
turn, was solved by the Wiener-Hopf method. We draw attention of a reader to that fact that, in
all friction ranges, a dependence of the escape rate on
friction is much weaker than exponential \cite {meln}.

However all these works considered only a {\it quasi-stationary} flux. A natural question is: how does
an escape flux evolves from zero at the initial instant to its quasi-stationary value at time-scales greatly exceeding a time 
of a formation of the quasi-equilibrium within the metastable part of the 
potential? There were few works on this problem  \cite {privman}, \cite {schneidman}, \cite 
{soskinetal} but they all concerned only a single-well case: the quasi-equilibrium  is established
in this case quickly (for a time of the order of a characteristic relaxation time in the well).

In contrast, the formation quasi-equilibrium in a {\it multi-well} metastable potential takes an 
exponentially longer time. Indeed, the formation has two essentially different stages in this case: the 
first stage, during which
quasi-equilibrium forms within the initial well, is short (similar to that one in a single-well case) while
the second stage, when an equilibrium between different wells forms, is exponentially longer.
Escapes for these two stages occur quite differently. {\it The present paper considers an escape}\footnote{ A closely related 
problem of inter-well transitions in a multi-well {\it stable} potential is also 
considered.} {\it flux just in the second stage}. It should be emphasized that, at small temperatures, just this stage is 
most relavant to real situations since the first stage ends
very quickly (its duration depends on temperature logarithmically while the second stage duration
increases exponentially sharply as temperature decreases) while larger time-scales, related to the quasi-stationary stage, may exceed significantly an observation/experiment time.

A duration of an escape/transition is typically much smaller than a time during which a system waits 
for the escape/transition \cite {freidlin84} so that the transition process may be considered at the relevant time-scales as 
an instantaneous one while a noise-driven multi-stable system on the whole may be described as a Markov chain \cite 
{freidlin84}  i.e. within an 
approximation of master equations for populations of attractors, with constant transition rates $\alpha_{ij}$ between 
states $i$ and $j$ (attractors or an unstable region). 

Fig.1 illustrates schematically an escape from a double-well metastable potential. If the system stays initially 
in 1 then, on time-scales preceding the formation quasi-equilibrium between wells 1 and 2, 
escapes occur most probable {\it directly} i.e. without relaxing into the bottom of the well 2, so that 
the flux at such time-scales is equal to $\alpha_{13}$ which may drastically differ from the 
quasi-stationary flux. 

A rate of a transition/escape  flux over more than one barrier, e.g. such as $\alpha_{13}$,  cannot be 
generally described by the Melnikov method, as it was recognized yet by Melnikov himself \cite{meln}. 
Instead, we use for their description the concept of {\it optimal large fluctuation}. Not only it provides 
a calculation of escape/transition rates but also allows to find a trajectory along which 
escapes/transitions occur with an overwhelming probability. The concept of optimal large fluctuation dates back to 
50th-60th \cite{onsmuch}-\cite{wenfr70} but the 
outburst of the interest to it falls onto the last decade which, apart from the logic of its own scientific development, is 
probably due to numerous recently 
appeared subjects related to large fluctuations and suggesting interesting 
applications in physics, 
biology, engineering, etc. These are first of all:  stochastic resonance (see for 
recent reviews 
\cite{sr95}, \cite{phystoda} and references therein), noise-induced transport 
in ratchets (e.g. 
\cite{magnasco}-\cite{drsv}), optimal control of fluctuations (e.g. \cite{sd}, 
\cite{vr}). A few hundreds papers on these subjects have been published for the last decade. As 
concers the 
problem of large fluctuations itself, it was treated by various methods: direct probabilistic methods 
(see \cite{wenfr70},  \cite{freidlin84} and references therein),  eikonal 
approximation of the 
Fokker-Plank equation (e.g. \cite{matkschu}-\cite{maierstein96}) and the path-integral method (e.g. 
\cite{dk}-\cite{prehistory}, 
\cite{drsv}-\cite {vr}). 
Usually the primary aim of all these methods 
is to derive an exponent in the most strong (activation-like) factor in the dependence of a 
transition rate on temperature (or any other quantity characterizing noise intensity):

\begin {equation}
\alpha_{\rm tr} \propto \exp (-\frac {S} {T}), \quad \quad
T \ll S, 
\end{equation}

\noindent where $S$ is an {\it action} of the Onsager-Muchlup type \cite{onsmuch} 
taken along an 
{\it optimal path} of a fluctuational transition.
For the escape from a single potential
well, the most probable escape path is time-reverse to the relaxational
trajectory from the saddle to the bottom of the well (see e.g. \cite
{gt}) which obviously provides that same Arrhenius factor in (1.2). A
corresponding noise realization is often called an {\it optimal
fluctuation} \cite{prehistory}.

In a {\it multi-well} case, a relaxational trajectory from an external saddle may relax into a well 
different from an initial one. Then an optimal path qualitatively differs from that one in a 
single-well case. Some important results on optimal paths in this case and on action along them  were 
obtained in \cite {gt}. However, apart from these results were obtained in a different context and by a 
method different from ours, they provided only a general type of a solution of the problem  for the 
path while the further analysis was far not complete and based mostly on intuitive ideas 
rather than on a rigorous treatment: just an absence of such treatment gave rise to the encountered by 
authors of \cite {gt} difficulties at numerical calculations.

In contrast, a {\it complete rigorous analysis} of optimal paths and action along them  is provided in the present work
which has allowed in particular to find characteristic non-trivial features of an evolution of direct 
escape rate (such as $\alpha_{13}$) as friction $\Gamma$ varies: unlike a conventional quasi-stationary 
flux, it may depend on $\Gamma$ exponentially sharply, moreover, it may undergo 
oscillations in the range of small $\Gamma$ and a cutoff at $\Gamma$ exceeding certain value from 
the moderate-to-high friction range.
The results for an escape flux are generalized for inter-attractor transitions in multi-well stable 
potentials, as well as applications to various other problems are discussed.

For a convenience of readers, quite detailed description 
of a contents of each chapter of this quite long paper is provide below.

A short {\it Chapter 2} has mostly an introductory aim, presenting a general description of a transition/escape flux in a multi-attractor system by means of master equations. The only novelty in this chapter is an introduction of a {\it splitting procedure} allowing to resolve transitions by numbers of returns from a final state.

The main part of the paper is concentrated in {\it Chapter 3} which concerns escape/transition rates 
and related optimal paths in potential systems driven by a weak white noise and by a linear friction of an arbitrary magnitude. The {\it sub-section 3.1} provides some very general estimations and 
conclusions based on the property of detailed balance. The {\it sub-section 3.2} presents in a general 
form a solution of the variational problem for the {\it most probable direct transition path} ($MPDTP$) 
for an arbitrary potential $U(q)$ and an arbitrary friction parameter $\Gamma$. Main results of this 
sub-section are equivalent to the results of \cite{gt} (obtained by a different method and in a different 
context). The central and the largest part of the paper is the {\it sub-section 3.3}: it presents the most 
interesting and non-trivial results of the paper. It consists of three parts. In the first {\it part, 3.3.1}, I 
formulate and prove 4 theorems which cover all possible types of $MPDTP$s relevant to the 
generalized Kramers problem (c.f. Fig.1). The second {\it part, 3.3.2}, concerns action $S$ (related to 
$\alpha_{13}$) and illustrates its main features at some typical example, presenting an evolution of 
action as $\Gamma$ varies. Finally, in the third {\it part, 3.3.3}, explicit asymptotic expressions for 
action and $MPDTP$ are derived for the underdamped range (they describe in particular oscillations of 
$S(\Gamma)$).

{\it Chapter 4} discusses briefly applications to various 
problems: inter-well transition rates in a 3-well stable potential, dynamics of an escape/transition flux on time-scales less or of the order of the time of 
a formation quasi-equilibrium within an initial well, 
prehistory problem, optimal 
control of fluctuations, fluctuational transport in ratchets, escapes at a 
periodic driving and transitions in biased Josephson junctions and ionic channels.

A summary and acknowledgements are presented in {\it Chapters 5 and 6} respectively. 

{\it Appendix A} illustrates the splitting procedure by an explicit description of higher-order partial 
probability fluxes in multi-stable systems. {\it Appendix B} describes the reduction of the 
Euler-Poisson equation for the $MPDTP$ to some much more simple one. {\it Appendix C } analyses a possibility to sew together different extremals 
(solutions of the variational problem).
{\it Appendix D} deals with the analysis of singularities in the solutions of the variational 
problem.

\chapter {Phenomenological description of transitions
in multi-stable systems}

Let us consider a dynamical system possessing more than one attractor. If a 
weak noise is added then one may formulate a problem of a fluctuational 
transition from a vicinity of a given initial attractor to a given final state which, 
generally speaking, may not belong to a basin of attraction of the initial 
attractor. It was shown in \cite {freidlin84} in a general form that, on a large time-scale, such 
system may be considered as a finite Markov chain in which an initially populated state 
corresponds to a given initial attractor of the dynamical system, a final state corresponds to a 
given final state of the transition while other states correspond to other attractors of the 
dynamical system.
The possibility for this 
is provided by that fact that a duration of an optimal fluctuation is 
exponentially smaller than a waiting time of such fluctuation. This allows 
to describe the dynamics in terms of transition rates and populations of 
attractors satisfying certain differential master equations.

For the sake of simplicity and clarity, we shall consider further, unless otherwise specified, the 
case when 
only 3 states are involved: an initial, final and one intermediate state - i.e. we shall consider a 
transition either from one of attractors of a 
bistable system to a non-attractor, or between attractors in a 
system possessing 3 attractors. The generalization to a larger number of 
involved attractors is not difficult though the resulting expressions are more 
cumbersome.

It will be assumed further in this Section that rates of direct transitions between the states, 
$\alpha_{ij}$, are known (for potential systems subject to white noise and linear friction, such 
rates will be calculated with a logarithmic accuracy in Sec.3).

\vskip 0.5truecm
\noindent
{\bf 2.1. SPLITTING PROCEDURE AND FIRST-PASSAGE FLUX}

\vskip 0.5truecm

Let us introduce a probability of that the 3-state system being intially at the state $1$ arrives at 
the state $3$ during an infinitesimal interval $[t,t+dt]$:
$dP(1 \rightarrow 3, t)$. Equivalently, one may consider a probability flux

\begin {equation}
J(1 \rightarrow 3, t)=\frac{dP(1 \rightarrow 3, t)}{dt}.
\end{equation}

The flux can be easily expressed via the populations $W_i$ of the states and corresponding direct 
transition rates:

\begin {equation}
J(1 \rightarrow 3, t)= \alpha_{13} W_1 + \alpha_{23} W_2 ,
\end{equation}

\noindent
while the populations obey conventional master equations \cite {vankampen}:

\begin{eqnarray}
\frac{dW_i}{dt}  & = &  - W_i \sum_{j \neq  i}\alpha_{ij} +  \sum_{j \neq  i}W_j\alpha_{ji},
\\
i, j  & = & 1, 2, 3,
\nonumber \\
W_1(0) & = & 1, \quad W_2(0)=0, \quad W_3(0)=0  \nonumber
\end{eqnarray}

\noindent
which can be easily solved.

In the most of applications, a final state of a transition is some attractor and its vicinity rather 
than just some point in a phase space of a dynamical system. In this case, rates of transitions {\it 
from} the final state may be of the same order or larger than rates of transitions {\it into} the 
final state. Hence, the flux (2.1) accounts both for those realizations in which the system visits 
the final state at a given instant $t$ for the first time and for those ones in which it already visited this state before $t$.
In many cases, one does need to resolve such transitions: for example, in the {\it mean first 
passage time} problem, one needs to account only for first-time transitions into $3$
while, in a prehistory problem, one needs to know the prehistory of the transition, in particular 
how many times the system visited the state $3$ before to arrive at it at the instant $t$. Then, one needs to split the probability flux into the corresponding partial fluxes

\begin {equation}
J(1 \rightarrow 3, t)= \sum_{n=1}^{\infty} J^{(n)}(1 \rightarrow 3, t),
\end{equation}

\noindent
where $ J^{(n)}(1 \rightarrow 3, t)$ corresponds to such transition at which the system visited 
the state $3$ $(n-1)$ times before to arrive at it for the $n$th time at the instant $t$. 

In order to be able to calculate $ J^{(n)}(1 \rightarrow 3, t)$ one needs to introduce {\it partial 
populations} of states $1$, $2$ and $3$, 
$W_1^{(n)}$, $W_2^{(n)}$ and $W_3^{(n)}$ respectively. Unlike conventional populations, 
these 
quantities satisfy the following condition: they account only for such 
realizations of noise at which the system entered the final state (i.e. the state $3$) before 
a current instant $t$ $n-1$ times. Obviously, $W_i^{(n)}$ satisfy the sum relation 

\begin {equation}
W_i= \sum_{n=1}^{\infty} W_i^{(n)}.
\end{equation}

I shall consider further in this Section only the most important partial flux, 
$J^{(1)}$. Note that, in the generalized Kramers problem, escape flux is equal to $J^{(1)}$ 
identically.  Thus the consideration below is equally relevant to stable and to metastable systems 
(in the latter case, a term \lq\lq state $3$" means a state far beyond a metastable part of the 
system). The higher-order fluxes in a stable system
are considered in the Appendix A\footnote {The splitting procedure
 is valid in particular for the simplest (and 
most frequently exploited) inter-attractor transition problem - in a bistable system. For the 
first-passage problem, the procedure does not lead in this case to new results since the first-passage 
problem in a bistable system is equivalent to an escape problem, which was solved before. However the results for higher-order-passage problems are non-trivial which is also demonstrated in 
the Appendix A.}.

The dynamics of the first-order partial 
populations is governed with the following master equations \footnote{The partial population 
$W_3^{(1)}(t) \equiv 0$, as it follows from its definition.}:

\begin{eqnarray}
\frac{dW_1^{(1)}}{dt} & = & - (\alpha_{12} + \alpha_{13})W_1^{(1)} + \alpha_{21}W_2^{(1)}, 
\nonumber \\
\frac{dW_2^{(1)}}{dt} & = & \alpha_{12}W_1^{(1)} - (\alpha_{21} + \alpha_{23})W_2^{(1)} 
, \\
& & W_1^{(1)} (0)=1, \quad W_2^{(1)} (0)=0 \nonumber
\end{eqnarray}

The system (2.6) is solved explicitly:

\begin{eqnarray} 
\lefteqn{
\stackrel{\rightarrow}{W}^{(1)} \equiv (^{W_1^{(1)}}_{W_2^{(1)}})=(^{\alpha_1}_{ 
\alpha_2}) e^{-\frac {t} {t_l}} +  (^{1-\alpha_1}_{ -\alpha_2}) e^{-\frac {t} {t_s}}, 
}
\\
& &
\alpha_1=(d-\alpha_{12}-\alpha_{13}+\alpha_{21}+\alpha_{23})/(2d), \quad
\alpha_2=\alpha_{12}/d,
\nonumber \\
& &
t_l=\frac {2}{\alpha_{12}+\alpha_{13}+\alpha_{21}+\alpha_{23}-d} \quad, \quad
t_s=\frac {2}{\alpha_{12}+\alpha_{13}+\alpha_{21}+\alpha_{23}+d} \quad,
\nonumber \\
& &
d \equiv \sqrt{(\alpha_{12}+\alpha_{13}-\alpha_{21}-\alpha_{23})^2 
+4\alpha_{12}\alpha_{21}} \quad. \nonumber
\end{eqnarray}

There are two time-scales in $\stackrel{\rightarrow}{W}^{(1)}(t)$ which are very different, in 
many cases: the short time ($t_s$) corresponds typically (though not always) to an establishment 
of a quasi-stationary 
distribution between states $1$ and $2$ while the long time ($t_l$)  corresponds typically to an 
escape 
from the system of states $1$ and $2$ on the whole.

Knowing $\stackrel{\rightarrow}{W}^{(1)}$, one can obtain all physical quantities 
which could be of interest in the first-passage problem.

First of all, it is a flux of first passages, $J^{(1)}$ 
\footnote {It will be meant by {\it flux} everywhere 
further in the text just $J^{(1)}$, unless it is specified otherwise. Note that, for any system, $J^{(1)}$ and a
conventional flux coincide at an initial instant while, in the Kramers problem 
i.e. for an escape from a metastable potential, they coincide at any time.}:

\begin{equation}
J^{(1)} \equiv \alpha_{13} W_1^{(1)}+ \alpha_{23} W_2^{(1)}= \alpha_{13} e^{-\frac {t} {t_s}}+(\alpha_{13}\alpha_1+\alpha_{23}\alpha_2)( e^{-\frac {t} {t_l}} - e^{-
\frac {t} {t_s}}) .
\end{equation}

The term $ \alpha_{13} \exp (-\frac {t} {t_s})$ dominates at the initial stage 
while only the term $\propto \exp (-\frac {t} {t_l})$ remains in 
the long-time scale.

It is worth to point out that, at the condition

\begin {equation}
\beta \equiv \frac {\alpha_{13}} {\alpha_{13}\alpha_1 + \alpha_{23}\alpha_2} <1,
\end{equation}

\noindent the flux is a non-monotonic function of time: it increases from $\alpha_{13}$ at $t=0$ to

\begin {equation}
J_m =(\alpha_{13}\alpha_1+\alpha_{23}\alpha_2)[\frac {t_l} {t_s} \beta]^{ \frac {t_s} {t_l - t_s} } (1-(1-\beta)[ \frac {t_l} {t_s} (1-\beta)]^{- \frac {t_l+t_s} {t_l-t_s} })
\end{equation}

\noindent at

\begin {equation}
t \equiv t_m= \frac {t_s} {1-t_s/t_l} \ln[\frac {t_l} {t_s} (1-\beta)]
\end{equation}

\noindent and then decreases to zero as $t$ becomes much larger $t_l$.

If

\begin {equation}
t_s \ll t_l,
\end{equation}

\noindent which is true in a majority of cases (a quasi-stationary 
distribution is being established much quicker than a transition to a final 
state occurs), then (2.10), (2.11) are simplified: 

\begin{eqnarray} 
J_m =  \alpha_{13}\alpha_1+\alpha_{23}\alpha_2, 
\quad\quad
t_m  =  t_s \ln[\frac {t_l} {t_s} (1-\beta)], \\
t_s  \ll  t_l. \nonumber
\end{eqnarray}

In the context of prehistory experiments \cite{prehistory} and optimal control of fluctuational 
transitions \cite{sd}, \cite{vr}, it is important to know from which of 
attractors most probable the system arrives at a final attractor for the first time. The ratio of the 
corresponding integral probabilities

\begin{eqnarray}
R \equiv \frac {\int^\infty_0 dt \alpha_{13} W_1^{(1)}} {\int^\infty_0 dt \alpha_{23} 
W_2^{(1)}}
=\frac {\alpha_{13} (\alpha_{1}t_l+(1-\alpha_{1})t_s)} {\alpha_{23} \alpha_{2} 
(t_l-t_s)} 
\nonumber \quad \quad \quad \quad \quad \quad \\
\\
=\frac {\alpha_{13} (\alpha_{21}+\alpha_{23})  \sqrt{ 
(\alpha_{12}+\alpha_{13}-\alpha_{21}-\alpha_{23})^2 
+4\alpha_{12}\alpha_{21}}} {\alpha_{23} \alpha_{12} 
(\alpha_{12}+\alpha_{13}+\alpha_{21}+\alpha_{23})}.
\nonumber
\end{eqnarray}

One more important characteristic of the transition is a mean first 
passage time\footnote{The 
formula (2.15) can be also considered as a partial case of a more general formula given in \cite 
{freidlin84}.}:

\begin{eqnarray}
MFPT \equiv \int^\infty_0 dt t (\alpha_{13} W_1^{(1)}+ \alpha_{23} W_2^{(1)}) 
=\alpha_{13} (\alpha_{1}t_l^2+(1-\alpha_{1})t_s^2)+
\alpha_{23} \alpha_{2} 
(t_l^2-t_s^2)
\nonumber \\
\\
=\frac {\alpha_{12} +\alpha_{21}+\alpha_{23}}
{\alpha_{21}\alpha_{13} + \alpha_{12}\alpha_{23} + \alpha_{13}\alpha_{23}}.
\quad \quad \nonumber
\end{eqnarray}

\newpage

\noindent
{\bf 2. 2. LIMIT CASES}
\vskip 0.5truecm

It is useful to analyze three quite typical limit cases

\vskip 0.5truecm
\noindent
1. 

\begin {equation}
\alpha_{13} \ll \alpha_{12}, \quad \quad \alpha_{23} \ll \alpha_{21}.
\end{equation}

As a simple illustration, one can bear in mind a potential system shown in 
Fig.2 (c.f. also Fig.4). 

In the case when (2.16) is satisfied, 

\begin {equation}
\alpha_1 \approx \frac {\alpha_{21}} {\alpha_{12}+ \alpha_{21}}, \quad
\alpha_2 \approx \frac {\alpha_{12}} {\alpha_{12}+ \alpha_{21}}, \quad 
t_s \approx \frac {1}{\alpha_{12}+ \alpha_{21}}, \quad 
t_l \approx \frac {\alpha_{12}+ \alpha_{21}} {\alpha_{21}\alpha_{13}+\alpha_{12}\alpha_{23}}.
\end{equation}

Correspondingly,

\begin{equation}
J \approx \alpha_{13} e^{-t (\alpha_{12}+ \alpha_{21})}+\frac {\alpha_{21}\alpha_{13}+\alpha_{12}\alpha_{23}
 }{\alpha_{12}+ \alpha_{21}}(e^{-
t \frac {\alpha_{21}\alpha_{13}+ \alpha_{12}\alpha_{23}} {\alpha_{12}+ 
\alpha_{21}}} - e^{-t(\alpha_{12}+ \alpha_{21})}) ,
\end{equation}

The condition (2.12) is obviously satisfied if (2.16) holds true. Correspondingly, 
if the inequality (2.9), which is very simplified in this limit,

\begin{equation}
\alpha_{13} < \alpha_{23},
\end{equation}

\noindent
is satisfied then  $J^{(1)}(t)$ is non-monotonic: it has a maximum at

\begin {eqnarray}
t_m  & \approx & \frac {1}{\alpha_{12}+ \alpha_{21}} \ln[\frac 
{\alpha_{13}(\alpha_{12}+ \alpha_{21})^3} {(\alpha_{21}\alpha_{13}+ 
\alpha_{12}\alpha_{23})^2}],
\nonumber \\
&&\\
J_m &  \approx & \frac {\alpha_{21}\alpha_{13}+ \alpha_{12}\alpha_{23}} 
{\alpha_{12}+ \alpha_{21}}.
\nonumber
\end{eqnarray}

\noindent
The condition (2.19), providing an increase of $J^{(1)}(t)$ at an initial stage, is 
quite clear: if $\alpha_{23} > \alpha_{13}$ then the flux 
should increase as the population of the state $2$ grows.

The expression for $R$, indicating a state from which most probable the 
system arrives at a final point, also becomes in the limit (2.16) very simple: 

\begin{equation}
R=\frac {\alpha_{21} \alpha_{13}} {\alpha_{12} \alpha_{23}}.
\end{equation}

\noindent It is just an established at $t \sim t_s$ ratio of populations in $1$ and 
$2$ multiplied by the ratio of the corresponding rates of transitions to $3$.

At last,

\begin{equation}
MFPT= \frac {\alpha_{12}+ \alpha_{21}} {\alpha_{21}\alpha_{13}+\alpha_{12}\alpha_{23}} \approx t_l.
\end{equation}

\vskip 0.5truecm
\noindent
2. 

\begin {equation}
\alpha_{13} \ll 
\alpha_{12}, \quad \quad 
\alpha_{21} \ll \alpha_{23} \ll \alpha_{12}.
\end{equation}

An illustration of this limit could be a potential system shown in the Fig.3(a).

In the case when (2.23) is satisfied,

\begin {equation}
\alpha_1 \approx \frac {\alpha_{21}} {\alpha_{12}}, \quad \quad \alpha_ 2
\approx 1, \quad \quad t_s \approx \frac {1} {\alpha_{12}}, \quad \quad t_l 
\approx \frac {1} {\alpha_{23}}.
\end{equation}

Correspondingly,

\begin{equation}
J^{(1)} \approx \alpha_{13} e^{-\alpha_{12}t}+\alpha_{23}
 ( e^{-\alpha_{23}t} - e^{-
\alpha_{12}t}) .
\end{equation}

The condition for a non-monotonicity $J^{(1)}$ coincides with (2.19).

The expression for $R$ and $MFPT$ are respectively

\begin {equation}
R \approx \frac {\alpha_{13}} {\alpha_{12}} \ll 1, \quad \quad \quad \quad 
MFPT \approx \alpha_{23}^{-1} \approx t_l.
\end{equation}

\vskip 0.5truecm
\noindent
3. 

\begin {equation}
\alpha_{21} \ll \alpha_{23} \ll \alpha_{12} \ll \alpha_{13}.
\end{equation}

For an example, see Fig.3(b).

In this case,

\begin {equation}
\alpha_1 \approx \frac {\alpha_{12}\alpha_{21}} {\alpha_{13}^2}, \quad 
\quad \alpha_ 2
\approx  \frac {\alpha_{12}} {\alpha_{13}}, \quad \quad t_s \approx \frac {1} 
{\alpha_{13}}, \quad \quad t_l 
\approx \frac {1} {\alpha_{23}}.
\end{equation}

Correspondingly,

\begin{equation}
J^{(1)} \approx  \alpha_{13} e^{-\alpha_{13}t}+\frac {\alpha_{12}\alpha_{23}} {\alpha_{13}}
 ( e^{-\alpha_{23}t} - e^{-
\alpha_{13}t}) .
\end{equation}

The flux (2.29) monotonically decreases: first, for a short time $\sim t_s$, from 
a large value $\alpha_{13}$ to a small value 
$\alpha_{12}\alpha_{23}/\alpha_{13}$, and then, for a long time  $\sim t_l$, to 
zero. There is, however, such paradox. From one side, the system transits to $3$ 
most probable, obviously, via the \lq\lq direct" route (which is characterized by 
the short time scale $t_s$):

\begin{equation}
R \approx \frac {\alpha_{13}}{\alpha_{12}} \gg 1,
\end{equation}

\noindent so that one could expect that $MFPT$ is equal to $t_s$. However it is not so:

\begin{equation}
MFPT \approx \frac {\alpha_{12}}{\alpha_{13}\alpha_{23}} \gg t_s \approx \frac {1} {\alpha_{13}}.
\end{equation}

The physical reason is that, although the probability for a system to come to
 $2$ before to get into $3$ is small (namely $\alpha_{12}/\alpha_{13}$), the time which 
it spends in $2$ is very large ($\sim \alpha_{23}^{-1}$), so that the contribution into 
the MFPT is larger than that one from realizations corresponding to the direct 
transition, notwithstanding an overwhelming probability of the latter. Note also 
that $MFPT$ is not equal to $t_l$ either.

\chapter {Transition rates. Potential systems}

As it follows from Sec.2, a transition problem in a multi-stable system driven by weak noise can 
be described in terms of direct transition rates and, therefore, the most fundamental task for a 
theory in the context of a transition problem is to calculate transition rates. We shall consider in 
this Section such rates in potential systems. But, first, let us review briefly the concept of {\it large 
fluctuation} in general. With this aim, we write in
a path-integral representation \cite {feinman} a transition probability density in a space of dynamical variables:

\begin{equation}
\alpha_{\rm tr}(x_{\rm f},t_{\rm f};x_{\rm i},t_{\rm i})= \int_{x(t_{\rm
 i})=x_{\rm i}} Df(t) {\rm P}[f(t)] \delta (x(t_{\rm f})-x_{\rm f})
\end{equation}

\noindent where $x_{\rm f}$  and $x_{\rm i}$ are respectively final (at $t=t_{\rm f}$) and 
initial (at $t=t_{\rm i}$) values of dynamical variables (many-dimensional, 
generally speaking) while ${\rm P}[f(t)]$ is a functional characterizing a 
probability density of a given noise realization $f(t)$. The dependence of 
${\rm P}$ on noise intensity $D_{\rm noise}$ is usually of the activation-like type \cite 
{feinman}:

\begin{equation}
{\rm P}[f(t)] = \frac {1} {Z} {\rm e}^{-\frac {\tilde{S}[f]} { D_{\rm noise}}},
\end{equation}

\noindent where $Z$ is a normalization factor. 

In particular, for a {\it white} noise \cite 
{feinman},

\begin{equation} 
\tilde S [f] = \frac {1} {2} \int_0^t d\tau f^2 (\tau).
\end{equation}

Transforming from noise variables to dynamical variables (using Langevin 
equations: c.f. for example eq.(3.6)), we derive 

\begin{equation}
\alpha_{\rm tr}= \int_{x(t_{\rm
 i})=x_{\rm i}}^{x(t_{\rm
 f})=x_{\rm f}}  Dx(t) J(x) \frac {1} {Z} {\rm e}^{-\frac {\tilde {\tilde S}[x]} { D_{\rm noise}}},
\end{equation}

\noindent where $J$ is a Jacobian of the transformation $\{f\rightarrow x\}$ while $\tilde {\tilde S} [x(t)] \equiv \tilde{S}[f(t)]$.

If we consider only a {\it direct} transition i.e. a transition which does not follow an intermediate 
attractor\footnote{If a path which provides a minimum of $\tilde {\tilde S} [x(t)] $ does follow 
an intermediate attractor then, unlike  $\alpha_{\rm pre}$ in (3.5), a \lq\lq preexponential" 
factor depends on $D_{\rm noise}$ also activation-like because a system stays in an intermediate 
attractor during a period whose characteristic duration depends on  $D_{\rm noise}$ activation-
like so that a portion of trajectories contributing into  $\alpha_{\rm tr}$ also depends on  
$D_{\rm noise}$ activation-like. At the same time, as it has been mentioned in Secs.1,2, a 
transition which follows an intermediate attractor should be described as a succession of transitions 
like in a corresponding Markov chain \cite{freidlin84} so that such paths are not relevant to 
{\it direct transition rates}.} then it follows from (3.4) that

\begin{equation}
\alpha_{\rm tr}^{(direct)}= \alpha_{\rm pre} {\rm e}^{-\frac { S^{(direct)}_{\rm min}} { D_{\rm noise}}},
\end{equation}

\noindent where $ S^{(direct)}_{\rm min}\equiv \tilde {\tilde S} [x_{\rm opt}(t)] $ is a
minimum of the functional $\tilde {\tilde S} [x]$ among all trajectories providing a {\it direct} 
transition while $\alpha_{\rm pre}$ is a preexponential factor which depends on noise intensity 
relatively weakly.

A {\it direct transition rate} (1.3) and a direct transition probability density (3.5) are closely related so that activation-like factors in them obviously coincide.
It is usually most important for a theory of a direct transition rate to determine just the most strong,
exponential, factor\footnote{The problem of a 
pre-factor is usually yet more complicated than the problem of an exponential factor. However, 
allowing for a comparatively weak dependence on noise intensity,  a pre-factor may be put a 
phenomenological constant in a wide range of transition rates.} in $\alpha_{\rm tr}$ i.e. to find $ S_{\rm min}^{(direct)}$. The problem for a minimum of 
a functional is a well defined mathematical problem. However it is very difficult (in a majority 
cases impossible) to obtain its solution in an explicit form while a purely numerical solution does 
not allow to come to general conclusions on characteristic features of solutions and, besides, it 
consumes a lot of computer time. That is why each explicit solution is very valuable especially if 
it reveals non-trivial features. 
An important class of systems for which an explicit (or at least partly explicit) solution of the 
transition problem is possible are potential systems subject to linear friction and white noise:

\begin{eqnarray}
& & \ddot {q} +\Gamma \dot {q} + dU/dq =f(t)  , \\
& & \langle f(t) \rangle = 0, \quad \langle
f(t)f(t^\prime)\rangle = D_{\rm noise}\delta (t - t^\prime) , \quad D_{\rm noise} \equiv 
2 \Gamma T, \nonumber 
\end{eqnarray}

\noindent where $T$ has a meaning of temperature.

Apart from that such model has numerous applications in physics, chemistry, 
engineering, etc. (see e.g. \cite {hanggi}, \cite {meln}, \cite {fpe}) it is distinguished from 
others by the property of {\it detailed balance} \cite {fpe}. Some general consequences of the 
detailed balance in potential (generally, multi-well) systems are analysed below in Sec.3.1. A 
reduction of a general variational problem for a minimal action to some 
much more simple problem is presented in Sec.3.2 while a major part of the paper is concentrated in Sec.3.3 which is devoted to an application of general results to transitions from a stationary attractor and particularly to a {\it generalized 
Kramers problem}.

\vskip 0.5truecm

\noindent
{\bf 3.1. DETAILED BALANCE}

\vskip 0.5truecm

Basing on the property of the detailed balance, let us show that transition 
rates between states which are not connected by a relaxational trajectory contain a small multiplier, in addition to the conventional factor $\exp (-\Delta E/T)$ where $\Delta E$ is a 
difference between final and initial energies, if the former is larger than the latter, 
or zero otherwise.

The property of the detailed balance reads for the system (3.6) \cite {fpe}

\begin{equation}
W_{\rm st}(1) \alpha_{\rm tr}(1\rightarrow 2) =W_{\rm st}(2) \alpha_{\rm tr}( 
2^{*}\rightarrow 1^{*})
\end{equation}

\noindent where $W_{\rm st}$ is a stationary probability density, which is 
Gibbsian \cite {fpe}

\begin{equation}
W_{\rm st}\propto \exp (-\frac {E} {T}), \quad, \quad E= \dot {q}^2/2+U(q),
\end{equation}

\noindent  $\alpha_{\rm tr}(i\rightarrow j)$ is a rate for a transition from a state $i$ to a state 
$j$, and the star $^*$ means that a conjugate state $i^{*}$ has the same coordinate as $i$ but an opposite velocity.

In the case when there is a relaxational trajectory from $2^{*}$ to $1^{*}$, $\alpha_{\rm 
tr}(2^{*}\rightarrow 1^{*})$ is of the order of 1, up to a logarithmic accuracy, because the noise is not necessary for such transition and the action (3.3) is equal to zero. Then it follows from 
(3.7), (3.8) that

\begin{equation}
\alpha_{\rm tr}(1\rightarrow 2) \sim {\rm e}^{-\frac {E(2)-E(1)} {T}}, \quad\quad\quad
2^{*}\stackrel {rel}{\rightarrow} 1^{*}. 
\end{equation}

It can be shown that a conventional trajectory 
time-reverse to a relaxational trajectory provides the equality of the action functional $\tilde {\tilde S}/(2 \Gamma)$ just to the difference of energies, thus indicating that it is just the most 
probable direct transition path (c.f. e.g. \cite{gt}).

The eq. (3.9) is relevant e.g. to a conventional (single-well) Kramers problem: states $1$ and $2$ correspond then to the bottom of the well and to the top of the barrier respectively.

If, on the contrary, $2^{*}$ and $1^{*}$ are not connected by a relaxational trajectory then 
$\alpha_{\rm tr}(2^{*}\rightarrow 1^{*})$  is exponentially small (if
temperature is small enough) because some finite noise is 
necessary in order to get from $2^{*}$ to $1^{*}$ and hence the action (3.3) is non-zero:

\begin{eqnarray}
& & \alpha_{\rm tr}(2^{*}\rightarrow 1^{*}) \sim {\rm e}^{-\frac {\Delta S} {T}}, \quad \quad
T \ll \Delta S, \\
& & 2^{*}\stackrel {rel}{\not \rightarrow} 1^{*}. \nonumber
\end{eqnarray}

Then it follows from (3.7), (3.8), (3.10) that

\begin{eqnarray}
& & \alpha_{\rm tr}(1\rightarrow 2) \sim  {\rm e}^{-\frac {E(2)-E(1)+\Delta S} 
{T}}, \quad \quad
T \ll \Delta S, \\
& & 2^{*}\not \rightarrow 1^{*}. \nonumber
\end{eqnarray}

Even if we do not know a concrete value $\Delta S$, we can carry out a 
qualitative (and, partly, even quantitative) analysis based on eqs. (3.10), (3.11) and 
results of Sec.2.

Let us consider as an example the problem of a 
first passage from $1$ to $3$ in a 
potential shown in Fig.2. Within a logarithmic accuracy, it is equivalent to 
the problem of a first passage from $1$ to $S_2$.

For the case shown on Fig.2(a), a noise-free trajectory from $S_2$  goes into $2$. Hence,

\begin{equation}
\alpha_{23} \sim  {\rm e}^{-\frac {U_{S_2}-U_2}{T}}, \quad \quad \alpha_{13} \sim  
{\rm e}^{-\frac 
{U_{S_2}-U_1+\Delta S}{T}}.
\end{equation}

Allowing also for 

\begin{equation}
\alpha_{12} \sim  {\rm e}^{-\frac {U_{S_1}-U_1}{T}}, \quad \quad \alpha_{21} \sim  
{\rm e}^{-\frac 
{U_{S_1}-U_2}{T}} 
\end{equation}

\noindent 
and for the hierarchy of energies

\begin{equation}
U_2<U_1<U_{S_1}<U_{S_2},
\end{equation}

\noindent
this case corresponds to the limit case (1) considered in Sec.2 and eqs. (2.16)-
(2.22) hold true. In particular, 

\begin{eqnarray}
& & R= \frac {\alpha_{21} \alpha_{13}}{\alpha_{12} \alpha_{23}} \sim {\rm e}^{-\frac 
{\Delta S} {T}} \ll 1, \quad \quad 
MFPT= \frac {\alpha_{12}+ \alpha_{21}}{\alpha_{12}\alpha_{23}+ \alpha_{21}\alpha_{13}} 
\sim {\rm e}^{-\frac 
{U_{S_2}-U_2} {T}},  
\\
& & S_2 \stackrel{rel}{\rightarrow} 2. \nonumber
\end{eqnarray}

Similarly, for the case shown on Fig.2(b) which differs from that one on Fig.2(a) 
only by friction so that the relaxational trajectory from $S_2$ goes into the well 
$1$,

\begin{eqnarray}
& & R \sim {\rm e}^{\frac {\Delta S} {T}} \gg 1, \quad \quad 
MFPT \sim {\rm e}^{-\frac {U_{S_2}-U_2} {T}},  \\
& & S_2 \stackrel{rel}{\rightarrow} 1. \nonumber
\end{eqnarray}

Thus, the system chooses in any case the \lq\lq easiest" way - the \lq\lq direct" 
route if the relaxational trajectory $S_2 \stackrel{rel}{\rightarrow} 1$ exists or the \lq\lq 
successive" route if it does not - so that, within a logarithmic accuracy, $MFPT$ does not depend 
on friction and the \lq \lq activation energy" in the expression for $MFPT$ is equal in both 
cases to the energy difference between a top of the highest barrier and a bottom of the deepest (among $1$ and $2$) well. 

The cases shown on Fig.3, (a) and (b), correspond respectively to the limit 
cases (2) and (3) in Sec.2. The analysis similar to that one for Fig.2  shows that, unlike the case (a), $MFPT$ in the case (b) essentially depends on $\Delta S$:

\begin{eqnarray}
& & MFPT \sim {\rm e}^{\frac {U_{S_2}-U_2-\Delta S} {T}},\quad \quad
\Delta S < U_{S_2}-U_2, \\
& &  S_1\stackrel{rel}{\rightarrow} 3. \nonumber
\end{eqnarray}

And obviously, in all cases, $\Delta S$ determines a flux at an \lq\lq initial" (but still exponentially long) stage. For many other problems e.g.  an optimal 
control of fluctuations and a directed diffusion in periodically driven ratchets (see 
Sec.4), 
one needs to know, apart from action, the most probable direct transition path $MPDTP$. The rest of Sec.3 is devoted to a derivation of the $MPDTP$ and 
calculation action along it.

\vskip 0.5truecm
\noindent
{\bf 3.2. GENERAL SOLUTION OF THE VARIATIONAL PROBLEM}

\vskip 0.5truecm

In order to find $ S^{(direct)}_{\rm min}$ for a direct transition between some given states, $i \rightarrow j$, one needs to express $f(t)$ via dynamical variables from (3.6), to substitute it into the action functional $\tilde S [f] $ (3.3) and then to find such
trajectory $q(t)$ which provides a minimum of $S[q] \equiv \tilde S [f] / (2 \Gamma)$ among all direct trajectories:

\begin{equation} 
S_{\rm min}= {\rm min} (S), \quad \quad S=\int_0^{t_{tr}} dt L(q, \dot {q}, \ddot {q}), \quad \quad
L = \frac {1}{4\Gamma} (\ddot {q} +\Gamma \dot {q} + dU/dq)^2, 
\end{equation}

\noindent where it is assumed that $q(t)$ does not follow intermediate attractors while

\begin {equation}
(^{q(0)}_{\dot {q} (0)}) = i, \quad  (^{q(t_{tr})}_{\dot {q} (t_{tr})}) = j, 
\end{equation}

\noindent and a duration of the transition $ t_{tr}$ should be 
varied too, in order to minimize $S$. 

The necessary condition for the extremum of a functional is an equality of its variation to 
zero. In the case of the functional (3.18), it is reduced to the Euler-Poisson equation 
\cite {elsgolc}, for $q(t)$ possessing finite derivatives up to the 4th order:

\begin {equation}
\frac{\partial L}{\partial q} - \frac {d}{dt} (\frac {\partial L}{\partial \dot 
{q}}) + \frac {d^2}{dt^2} (\frac {\partial L}{\partial \ddot {q}}) = 0,
\end{equation}

\noindent with the  boundary conditions (3.19).

In order to minimize $S$ over $ t_{tr}$ one needs to equal the derivative to zero:

\begin {equation}
\frac{\partial S}{\partial t_{tr}}=0.
\end{equation}

\noindent 
Substituting $S$ (3.18) into (3.21), carrying out an integration by parts twice
 and using (3.20), one can derive the 
equivalent condition \cite {elsgolc}:

\begin{equation} 
\tilde E  =  0, \quad\quad\quad
\tilde E \equiv  -L+(\frac{\partial L}{\partial \dot {q}} - \frac {d}{dt} (\frac {\partial 
L}{\partial \ddot {q}})) \dot {q} + \frac{\partial L}{\partial \ddot {q}} \ddot 
{q}.
\end{equation}  

\noindent Here, $\tilde E$ is analogous to mechanical energy \cite {landau} and is conserved at a solution of (3.20). 

Let us seek the 
solution $q_{opt}(t)$ of the Euler-Poisson equation (3.20) with boundary conditions (3.19) as a function time-reverse to a solution of such equation

\begin{eqnarray} 
\ddot {q} +\Gamma^{\prime}(t) \dot {q} + dU/dq =0, \\
(^{q(0)}_{\dot {q} (0)}) = j^*, \quad  (^{q(t_{tr})}_{\dot {q} (t_{tr})}) = i^*, 
\nonumber \\
q_{opt}(t)=q(t_{tr}-t), \quad \quad \quad \nonumber 
\end{eqnarray}

\noindent where $\Gamma^{\prime}(t)$ is so far unknown, a 
constant $ t_{tr}$ is so far arbitrary and  will be determined later from the condition (3.22), and a star $^*$ has the same meaning as in (3.7).

Note that there is no danger to miss a true solution of the variational problem when we use the representation (3.23). Indeed, assume that we know a true solution 
$q_{true}(t)$ (with a transition time $t_{true}$). Then,  a function $q(t)=q_{true}( 
t_{true}-t)$ satisfies (3.23) if

\begin{equation} 
\Gamma^{\prime}(t)= -\frac {\ddot {q}_{true}(t_{true}-t) + \frac {dU(q_{true}(t_{true}-t))}{d 
q_{true}(t_{true}-t)}}{ \dot {q}_{true}(t_{true}-t) }, \quad \quad \quad t_{tr}=t_{true},
\end{equation}

\noindent so that $q_{opt}(t)=q_{true}(t)$ i.e. $ q_{true}(t)$ is necessarily among solutions of the type (3.23).

Of course, one could seek a solution of the Euler-Poisson equation using a different 
representation but it is just the representation (3.23) which allows to reduce the complicated 
4th-order differential equation (3.20) to a much more simple equation. Indeed, 
putting $q_{opt}(t)$ (3.23) into the Euler-Poisson equation (3.20) with the Lagrange function L 
(3.18) one can derive, after some transformations (see the Appendix B),

\begin{equation}
\phi \frac {d^2 q}{dt^2} + \frac {1}{2} \frac {d \phi}{dt} \frac {dq}{dt}=0, \quad 
\quad \quad 
\phi \equiv \frac {(\Gamma^{\prime})^2 - \Gamma^2}{2} -\frac {d \Gamma^{\prime}}{dt}.
\end{equation}

\noindent 
where $q \equiv q(t)$ is determined by (3.23).
The equation (3.25) has solutions of three types:

\begin{eqnarray} 
& (1) &  \quad \quad \phi =0, \nonumber \\
& (2) &  \quad \quad \frac {dq}{dt} =0, \\
& (3) &  \quad \quad \phi (\frac {dq}{dt})^2= C, \quad \quad C \not = 0. \nonumber
\end{eqnarray} 

The second type, obviously, does not suit us because it cannot 
satisfy the conditions (3.19). The third type is not 
suitable either: it does not satisfy the condition for a minimization over a 
transition time (3.22). Indeed, if one substitutes into $\tilde E$ (3.22) the Lagrange 
function $L$ (3.18) with $q_{opt}(t)$ (3.23) in which 
$\Gamma^{\prime}(t)$ satisfies the equation (3) in (3.26), one obtains (c.f. the derivation 
eq.(3.25) in Appendix B)

\begin {equation}
\tilde E=-\frac {C}{2 \Gamma}.
\end{equation}

Thus, in order to satisfy both the Euler-Poisson equation (3.20) (together with the 
boundary conditions (3.19)) and the condition of zero quasi-energy $\tilde E$ (3.22),  one 
needs to choose in (3.26) the equation of the type (1). This equation can be solved 
explicitly:

\begin {equation}
\Gamma^{\prime}(t)= \Gamma \frac {1+Ae^{ \Gamma t}}{1-Ae^{ \Gamma t}}
\end{equation}

\noindent where a constant of integration $A$ should be chosen so that the 
proper relaxation $j^* \stackrel{relA}{\rightarrow} i^*$ (3.23) takes place 
\footnote {After the initial version of this paper had been prepared my 
attention was drawn to papers \cite {grt}, \cite {gt} in 
which equations equivalent to (3.23), (3.28) were obtained from the Hamilton equations corresponding to the Lagrangian $L$ (3.18), in a problem of a {\it nonequilibrium potential} which determined a quasi-stationary distribution (note that, in these papers, the auxiliary friction 
is written for a direct rather than time-reverse path so that their $A$ corresponds to my $-A{\rm exp}(\Gamma t_{tr})$). Apart from I obtain (3.23), (3.28) by a different method and in a different context, I advance in many important respects much more far than authors of \cite {grt}, \cite {gt}: 1) I prove that the type of a 
solution (3.23), (3.28) is the {\it only} type which can provide an extremum of 
action, as well as I prove that the $MPDTP$ cannot be sewed from trajectories of the type (3.23), 
(3.28) 
with {\it different} $A$ unless they are sewed in sadles (see Appendix C and the rest of the sub-section), 2) I provide a detailed 
analysis of an influence of the singularity in (3.28) at the $MPDTP$ (see Appendix D), 
3) unlike \cite {grt}, \cite {gt}, my analysis provides also a description of the $MPDTP$ with a given (rather than just optimal) duration of the transition 
(see eqs. (3.26.3), (3.27) and the item 3 in Sec.4), 4) I apply the general solution 
to the generalized Kramers problem and provide a detailed {\it rigorous}
analysis for the corresponding $MPDTP$ and action (see Sec. 3.3), 5) I derive {\it explicit} expressions for action and the $MPDTP$ in overdamped and underdamped regimes (see Sec. 3.3.2 and 3.3.3  respectively), 6) I suggest various other applications of the general results  (see Sec.4).}.

If $A<0$ then the function $\Gamma^{\prime}(t)$ (3.28) has a zero (in which 
$\Gamma^{\prime}$  changes its sign) at

\begin {equation}
t=t_0 \equiv \frac {1}{\Gamma} \ln (\frac {1}{\mid A \mid})
\end{equation}

If $A>0$ then the function $\Gamma^{\prime}(t)$ (3.28) has a singularity (a pole of the first 
order) at

\begin {equation}
t=t_p \equiv \frac {1}{ \Gamma} \ln (\frac {1}{A}).
\end{equation}

\noindent Due to this, a velocity of the auxiliary system (3.23), $\dot {q}$, drops to zero at 
$t=t_p$ while  $\Gamma^{\prime}$ changes its sign.

At any sign of $A$, $\Gamma^{\prime}$ turns into $-\Gamma$ as $t$ grows to infinity, 

\begin {equation}
\Gamma^{\prime} (+\infty)=-\Gamma.
\end{equation}

The next important question is whether an {\it extremal} can be \lq\lq sewed" from trajectories 
of the type (3.23), (3.28) with {\it different} $A$. Such sewing could seem natural since the 
Euler-Poisson equation would be satisfied everywhere except possibly the very sewing point. 
However it 
is proved in Appendix C that such sewing necessarily breaks the equality of the variation of the functional 
$S$ (3.18) to zero (which is just a definition of an extremal) unless the sewing point is either a 
stationary point of the potential system,

\begin {equation}
dU/dq=0, \quad\quad\quad \dot {q} =0
\end{equation}

\noindent
or a turning point ($\dot {q} =0$) whose coordinate is a coordinate of a discontinuity $dU/dq$.

Taken that we are interested only by direct transitions, extremals which include attractors as 
sewing points are not relevant. For the sake of brevity, let us call the remaining extremals i.e. 
extremals which  {\it do not follow intermediate attractors} as {\it direct 
extremals}.
A next step is to find that direct extremal (if it is)
along which action is smaller than along any other direct extremal i.e. to find the {\it most probable direct transition path} ($MPDTP$) if it is. 
The further analysis differs essentially for cases when 1) neither an initial nor final state of 
transition is a non-stationary (periodic) attractor while at least one of them is a stationary 
point, 2) one of the states is a periodic attractor, 3) neither of the states is neither a periodic 
attractor nor a stationary point. The case 3) 
is more of a formal rather than practical interest and will be discussed briefly in the end of this 
sub-section. As concerns a periodic attractor which may exist 
in a tilted periodic potential \cite{fpe}, a nonequilibrium potential 
(which is closely related to action) in the presence of such attractor was briefly analysed in \cite 
{gt}. However the analysis of \cite {gt} concentrated on a quasi-stationary distribution rather 
than on the transition problem and even that analysis
was far from being complete. Some partial case was analysed numerically and by 
Monte-Carlo 
simulations in \cite{kautz}. A consistent rigorous analysis of the associated variational problem 
is planned to be done by me elsewhere.

In the present paper, I shall consider mostly the case when $i$ is a stationary attractor while $j$ is a saddle: just this case is relevant 
to an escape from a multi-well metastable potential i.e. to the {\it generalized 
Kramers problem} (c.f. Figs.1, 4), which is the main subject of the present paper, as well as to inter-attractor transitions in a stable multi-well potential (it does not matter for an exponential factor whether  a final point is an attractor or a saddle from which a system may
relax to the final attractor noise-free).

\vskip 0.5truecm
\noindent
{\bf 3.3. 
GENERALIZED KRAMERS PROBLEM
}

\vskip 0.5truecm

For the sake of clarity and brevity, I restrict the analysis to the case when only two smooth 
adjacent potential wells are involved (see Fig.4(a) as an illustration), namely let the following conditions are satisfied:

\begin{itemize} 
\item[1)] 
a potential function $U(q)$ possesses at least two adjacent parabolic local minima $1$ and $2$
and, apart from a local maximum $S_1$
between 1 and 2, there is at least one more adjacent to it local maximum  $S_2$;

\item[2)] 
in the energy-coordinate plane $E-q$, noise-free trajectories emanating from $S_1$ go either into 
$1$ or into $2$ and do not pass above any local maximum of the curve $E=U(q)$; 

\item[3)] (a) if $U_{S_2}> U_{S_1}$ then the bit $S_2 O$, where $O$ is the nearest to $S_2$ in the direction 
$q(S_2)\rightarrow q(S_1)$ intersection of the horisontal line $E=U_{S_2}$ with the curve 
$E=U(q)$, lies above only one local maximum of the curve $E=U(q)$, namely $S_1$;

(b) if $U_{S_2}<U_{S_1}$ then the bit $S_1 O$, where $O$ is the nearest to $S_1$ in the direction 
$q(S_2)\rightarrow q(S_1)$ intersection of the horisontal line $E=U_{S_1}$ with the curve 
$E=U(q)$, does not lie above any 
local maximum; 

\item[4)] the initial state of the transition is $1$ while the final one is
$S_2$.

\end{itemize} 

It should be emphasized that Fig.4(a) is just an illustrating example while the consideration below is valid 
also in the case when $S_2$ is adjacent to $1$ rather than to $2$ as well as when 
$U_{S_2}<U_{S_1}$ (but the condition 2 above is still valid).

A generalization for a larger number of involved attractors is 
straightforward. In particular, a case of inter-attractor transitions in a 3-well potential (see Sec.4) is immediately reduced to the case considered in the present sub-section.

\vskip 0.5truecm

\noindent
{\bf 3.3.1. Most probable direct transition path}

\vskip 0.5truecm

First of all let us prove that if $S_2$
belongs to a basin of attraction of $1$ then a conventional time-reverse 
relaxational trajectory does provide an absolute minimum of action. 
Substituting $q_{opt}(t)$ (3.23) into the action functional (3.18), we obtain

\begin {eqnarray}
S_{\rm min} \equiv S_{\rm min}(1 \rightarrow S_2) & = & 
\frac {1}{4 \Gamma}\int_0^{t_{tr}} dt (\Gamma + 
\Gamma^{\prime}(t))^2 \dot {q}_{opt}^2  (t_{tr}-t)
\nonumber
\\
& = & U_{S_2}-U_1+ \frac {1}{4 \Gamma}\int_0^{t_{tr}} dt (\Gamma - 
\Gamma^{\prime}(t))^2 \dot {q}_{opt}^2  (t_{tr}-t),  
\end{eqnarray}

\noindent where, at the derivation of the second equality, it has been allowed for

\begin {equation}
\Gamma^{\prime}(t) \dot {q}_{opt}^2  (t_{tr}-t)= - \frac {dE(q_{opt} (t_{tr}-t))}{dt},
\end{equation}

\noindent where $E \equiv \dot {q}^2/2 + U(q)$ is an energy along the auxiliary
relaxational trajectory (3.23).

If we put $A=0$ into $\Gamma^{\prime}$ (3.28) we provide both a relaxation from $S_2$ just to $1$ \footnote {Note that $\dot {q} =0$ both in $S_2$ and in $1$ so that they coincide with $S_2^*$ and $1^*$ respectively.}, $S_2 \stackrel {relA}{\rightarrow}  1$, and an equality $\Gamma^{\prime}$ to $\Gamma$  which obviously provides the minimal possible action equal just to a difference of energies. 

Now, we pass to the most interesting case when a final state $S_2$ does not belong to a basin of 
attraction of an initial attractor 1. Let us show first that the $MPDTP$ $1\rightarrow S_2$ goes necessarily 
through the saddle\footnote{For the sake of brevity, we use here and thereafter a term {\it saddle} in relation to $S_1$ 
(and analogously, to $S_2$) both in a case of a smooth maximum when it is a true saddle and in a 
more formal case of a cusp-like maximum  when $S_1$ is even not a stationary point (though, like 
a true saddle, it possesses two incoming and two outcoming manifolds).
}
 $S_1$. Let us assume that a path which is 
time-reverse to the $MPDTP$ includes some point $B$ of the boundary of basin of attraction of $1$ 
which differs from $S_1$. Then, taken into account that this point belongs to the 
basin of  attraction of 1 and, therefore, the trajectory time-reverse to the $MPDTP$ $1\rightarrow B^*$ is just a relaxational trajectory $B\stackrel{rel}{\rightarrow}1$ which, in this case, necessarily follows 
the saddle $S_1$, we come to the conclusion 
that the saddle $S_1$ is definitely followed by the $MPDTP$ $1\rightarrow S_2$\footnote {In fact, 
as it follows from the consideration below, $S_1$ is the only point of the
boundary of the basin of attraction of $1$ which belongs to the time-reverse to the $MPDTP$ 
trajectory.} (c.f. also \cite {pl94}).

Thus, if $S_2$ does not belong to a basin of 
attraction of $1$ the $MPDTP$ from $1$ to $S_2$ consists of two bits:

\noindent 1) from $1$ to $S_1$, following the conventional 
time-reverse relaxational path $1 \stackrel{A=0}{\rightarrow}S_1$ ;

\noindent 2) from $S_1$ to $S_2$.

Taken that, for a smooth potential $U(q)$, direct extremals may be sewed only in saddles while there are no saddles other than 
$S_1$ and $S_2$ in our problem, the $MPDTP$ $S_1 \rightarrow S_2$ is definitely a path with 
a {\it single} $A$, $S_1 \stackrel{A}{\rightarrow} S_2$.
The question about a proper choice of $A$ is one of central questions of the 
present paper. We need to choose from all {\it direct} extremals $S_1 \rightarrow S_2$ 
(they have been described in Sec.3.2  in a general form and their number may be infinite) that one which provides a minimum for action. An algorithm for the choice depends
on a satisfaction such {\it 
2 conditions} \footnote {As $\Gamma$ varies, a satisfaction of these 
conditions changes at some critical values which will be discussed in the next sub-section.}:

\begin{itemize} 

\item[{\bf Condition 1.}]
{\it A noise-free trajectory $S_2 \stackrel {rel}{\rightarrow} 2$ possesses points both with $q<q(S_1)$ 
and with $q>q(S_1)$} (it is equivalent to that the trajectory passes in the 
energy-coordinate plane above the saddle $S_1$ at least once: c.f. the thin dashed line in 
Fig.4(a)).

\item[{\bf Condition 2.}]
{\it A noise-free trajectory $S_2\stackrel {rel}{\rightarrow} 2$ and a trajectory $2 \stackrel 
{A=0}{\rightarrow} S_1 $ (which is time-reverse to $S_1 \stackrel 
{rel}{\rightarrow} 2$) intersect at least once, apart from the obvious common point $2$} (c.f. 
intersections of the thin dashed and solid lines in Fig.4(a); note that, in the energy-coordinate 
plane, any trajectory overlaps a time-reverse to it trajectory, so that $2 \stackrel 
{A=0}{\rightarrow} S_1 $  and $ S_1 \stackrel {rel}{\rightarrow} 2$ are presented in Fig.4(a) by one 
and the same line).

\end{itemize} 

Correspondingly, any case is described by one of  4 theorems presented below. For their 
formulation, it is convenient to introduce the following {\it definitions}.

\vskip 0.3truecm

\noindent
{\bf Definition 1.} Let us define a bit of a trajectory as  a {\it passage} if a velocity is equal to zero in the beginning and in the end of the bit while it does not 
change its sign in between them (e.g. the trajectory shown in Fig.4 by the dotted line consists of 2 passages).

\vskip 0.3truecm

\noindent
{\bf Definition 2.} Let us define a point $I$ as the highest in energy 
intersection of the trajectories $S_2\stackrel {rel}{\rightarrow} 2$ and $2 \stackrel 
{A=0}{\rightarrow} S_1 $ (c.f. Fig.4(a)).

\vskip 0.3truecm

\noindent
{\bf Definition 3.} Let us define  $A_-$ as such negative value 
that the number of passages in $ S_1 \stackrel {A_-}{\rightarrow} S_2 $, $n_-$, is equal to a number 
of passages in the bit of the noise-free trajectory $ S_2 \stackrel {rel}{\rightarrow} I $, $n_{rel}$.

\vskip 0.3truecm

\noindent
{\bf Definition 4.} Let us define $A_+$ as such positive value that the number of passages in $ S_1 \stackrel 
{A_+}{\rightarrow} S_2 $, $n_+$, is smaller than $n_{rel}$ by one: $n_+=n_{rel}-1$.

\vskip 0.5truecm

\noindent
{\bf Theorem 1.}
{\it Let both conditions 1 and 2 be satisfied.
Then, the $MPDTP$  $S_1\rightarrow S_2$ is either $ S_1 \stackrel {A_-}{\rightarrow} S_2 $ or
$ S_1 \stackrel {A_+}{\rightarrow} S_2 $ while 
action along the $MPDTP$ is less 
than $U_{S_2}-U_2$.}

\vskip 0.2truecm

\noindent
{\it Remark 1.} Theorem 1 is relevant typically to small $\Gamma$: an example is shown in Fig.4 ($ 
S_1 \stackrel {A_-}{\rightarrow} S_2 $ and $ S_1 \stackrel {A_+}{\rightarrow} S_2 $ are shown by 
the thick dashed and dotted lines respectively).
Intuitively, the specified in the theorem choice of extremals is rather clear: as it follows from (3.33), $\Gamma^{\prime}(t)$ should 
differ from $\Gamma$ as little as possible - and the above described paths provide such 
$\Gamma^{\prime}$ which, on a major part of a path, is only slightly either smaller (for $A_-$) or larger (for $A_+$) than $\Gamma$. But the rigorous proof is quite tricky and uses mainly geometrical arguments. 

\vskip 0.2truecm

{\it Proof}.

Let us first prove that a minimal action is smaller than $U_{S_2}-U_2$. The conventional  successive path $S_1 \stackrel {rel}{\rightarrow}2 
\stackrel {A=0}{\rightarrow} S_2 $ provides action equal just to $U_{S_2}-U_2$. Let us construct the path $S_1 \stackrel 
{rel}{\rightarrow}I^*\stackrel {A=0}{\rightarrow} S_2$. It is a part of $S_1 \stackrel 
{rel}{\rightarrow}2 \stackrel {A=0}
{\rightarrow} S_2 $ and, obviously, action along the latter path exceeds action along the former one. 
Thus, a minimal action should be definitely less than  $U_{S_2}-U_2$. 

As concerns the $MPDTP$, let us consider separately negative and positive $A$.

\vskip 0.2truecm

\noindent (a) $A<0$.

First of all, let us prove an existence of  the described above $A_-$. 

Consider first a case when the noise-free trajectory  $ S_2  \stackrel {rel}{\rightarrow} 2$ 
possesses at least one turning point\footnote{We use a term {\it turning point} in a conventional for a physical literature meaning of a point on a trajectory where a velocity changes its sign.} between  $S_1$ and $2$. If we decrease $A$ 
continuously from zero a dissipation of energy along the relaxational trajectory $S_2 \stackrel 
{relA}{\rightarrow}$  decreases continuously so that the highest in energy turning point continuously  moves up in energy and necessarily reaches 
$S_1$ at some finite $|A|$, when the whole path lies above the trajectory $S_1 \stackrel {rel}{\rightarrow} 2$ (except the very $S_1$, obviously). And just this $A$ is $A_-$ as it follows from the definition $A_-$.

In that case when the trajectory $ S_2 \stackrel {rel}{\rightarrow} 2$ does not possess any turning point 
between  $S_1$ and $2$, the trajectory  $S_1 \stackrel {rel}{\rightarrow} 2$ necessarily 
possesses at least one turning point to the right from $2$ (otherwise the {\it condition 2} would 
not hold true). Then if we decrease $A$ continuously from zero a gain of energy along the path 
$S_1 \stackrel 
{A}{\rightarrow}$ increases continuously so that the highest in energy turning point continuously  moves up in energy until 
it meets $ S_2$ which will correspond just to $A=A_-$, by the definition $A_-$ ($n_-=1$ in this case).

Note that if the {\it condition 2} did not hold true the above proof of an existence $A_-$ would not be valid (see the {\it theorems 3, 4} below).

Let us show that if $S_1$ is a smooth maximum then there are no $A<A_-$ at which $S_1 \stackrel 
{A}{\rightarrow}S_2$ could exist. Indeed, in this case, $S_1$ is an unstable stationary point and
the approach of the relaxational trajectory $ S_2 \stackrel {rel A_-}{\rightarrow}S_1 $
towards the 
saddle $S_1$ occurs infinitely slowly: energy first decreases to a
certain 
minimal value $E_{min}(A_-)$ (which is slightly less than $U_{S_1}$) at
an 
instant $t_0$ (3.29) and then starts to increase to $U_{S_1}$. 
If $|A|>|A_-|$ then the change of the sign $\Gamma$ occurs too early and energy along the trajectory $ S_2 \stackrel {rel A}{\rightarrow} $
starts to increase before it reaches the level $E_{min}(A_-)$ so that the
trajectory passes above $S_1$.

In the case when $S_1$ is a cusp-like maximum, paths $S_1 \stackrel 
{A}{\rightarrow}S_2$ with $A<A_-$ may exist. Let us prove that neither of them  can be the $MPDTP$. Indeed, such path contains less number of passages than $n_-$ which means that the time-reverse to it path inevitably intersects the boundary of basin of attraction of $1$ in some
point $P_1$ which is not  $S_1$  (c.f. the Fig.4(b)). But the $MPDTP$ $1\rightarrow P_1^*$ is time-reverse to $P_1\stackrel{rel}{\rightarrow}1$ while the latter just follows the boundary of attraction 1 (until it meets the saddle $S_1$) rather than just intersects the boundary.
This proves that $ 1 \stackrel 
{A=0}{\rightarrow}S_1 \stackrel 
{A<A_-}{\rightarrow}S_2$ cannot be the $MPDTP$ $1\rightarrow S_2$.

Let us prove that a path $ S_1 \stackrel 
{A<0}{\rightarrow}S_2$ cannot provide the $MPDTP$ if $|A|<|A_-|$. Indeed,
if we decrease $|A|$ continuosly 
then $\Gamma^{\prime}(t)$ increases for all $t$ i.e. a dissipation of energy along the relaxation trajectory $S_2 \stackrel 
{rel A}{\rightarrow}$ increases too and, 
hence, the trajectory lowers and necessarily intersects in 
some point $P$ the trajectory $2 \stackrel {A=0}{\rightarrow} S_1$. Let $A$ is such that, after one or more oscillations in the well $2$, the trajectory does come into $S_1$.
Let us construct the path $S_1\stackrel 
{rel}{\rightarrow}P^*\stackrel {A}{\rightarrow} S_2$. Action along the bit $S_1\stackrel 
{rel}{\rightarrow}P^*$ is equal 
to zero while $ P^*\stackrel {A}{\rightarrow} S_2$ is only a part of $ S_1 \stackrel {A}{\rightarrow} S_2$ 
and action along the completing part, $ S_1 \stackrel {A}{\rightarrow}P^*$, is definitely non-zero. Thus, action along
$S_1\stackrel {A}{\rightarrow} S_2$ is certainly not minimal. 

Thus, {\it among all negative $A$, only $A_-$ may provide a minimum of action}.

\vskip 0.2truecm

\noindent (b) $A>0$.

First of all, let us show that $A_+$ exists. Due to the {\it condition 1}, 
$ S_2\stackrel {rel}{\rightarrow}2$ does pass in the energy-coordinate plane above $S_1$. If to 
increase $A$ continuously from zero then a dissipation of energy along the 
trajectory $ S_2 \stackrel {rel A}{\rightarrow}$ increases and the trajectory lowers so that it 
necessarily meets $S_1$ at some $A$ which is just $A_+$, by the definition $A_+$.

If $A<A_+$ then the relaxation trajectory first goes into the well $2$ and, at the 
corresponding instant $t_p$ (3.30), the velocity $\dot {q}$ drops to zero while energy drops 
to some value $E_{min}(A)=U(q(t_p))<U_{S_1}$.
As it is shown in Appendix D, the system goes then \lq\lq back in time" along the same 
trajectory by which it arrived at $q= q(t_p)$ unless $dU/dq=0$ at $q= q(t_p)$. In the former 
case, it cannot arrive at $S_1$ while, in the latter case i.e. if $q(t_p)=q(2)$, it is not a {\it direct} extremal (moreover, action along a path following the attractor $2$ is $\geq U_{S_2}-U_2$).

If $A>A_+$ the number of passages decreases which means that $ S_2\stackrel {rel A}{\rightarrow}S_1$ 
inevitably intersects the boundary of basin of attraction of $1$ in some
point $P_2$ which is not  $S_1$  (c.f. the Fig.4(b)) while a trajectory time-reverse to the $MPDTP$ $1\rightarrow P_2^*$ 
should follow the boundary (until it meets $S_1$) rather than just intersect the boundary.

Thus, {\it among all positive $A$, only $A_+$ may provide a minimum of action}.

Thus, the {\it theorem 1} has been proved.

\vskip 0.2truecm

\noindent
{\it Remark 2.} 
It is interesting to notice that if $U(q)$ is smooth in $S_1$ (which is a typical case) then
the relaxational trajectory  $ S_2 \stackrel {rel A_+}{\rightarrow} $ 
approaches $S_1$ at a finite instant of time (note the footnote 7 above), unlike the trajectory  $ S_2 \stackrel {rel A_-}{\rightarrow} $: the former trajectory reaches the saddle just at the instant $t_p^{(+)} \equiv t_p (A_+)$ (3.30). Indeed, if the
transition instant $t_{tr}$ was smaller than $t_p^{(+)}$ then
$\Gamma^{\prime}(t)$ would be positive and finite at any $t \leq t_{tr}$
- but, at any finite positive friction, the approach to the saddle
should take an infinite period which contradicts to the original
assumption that $ t_{tr}< t_p^{(+)}$. On the other hand,
if $ t_{tr}$ was larger than $t_p^{(+)}$ then we would also come to a
controversy. Indeed, at the instant $t= t_p^{(+)}$, the velocity $\dot
{q}$ must drop to zero. It is shown in the Appendix D that, unless
the coordinate $q(t_p^{(+)})$ coincides with a coordinate of the saddle $S_1$ or of the bottom 
of the well $2$, the trajectory returns along
the same trajectory to $S_2$ and therefore cannot provide the
transition $S_2 \rightarrow S_1$. But, due to the assumption that $ t_{tr}> t_p^{(+)}$, the 
saddle $S_1$ cannot be reached at the instant $ t_p^{(+)}$. The bottom of the well $2$ is not 
suitable either since a path following $2$  cannot be the $MPDTP$, by the definition of the $MPDTP$.
Thus, we have again come to a controvercy with the original assumption and, hence, the transition time is just $t_p^{(+)}$. 

\vskip 0.2truecm

\noindent
{\it Remark 3.}  Let us describe briefly how to find constants $A_-$, $A_+$. In 
an underdamped limit ($\Gamma$ is small in comparison with characteristic eigenfrequencies),
they can be found explicitly which will be done in Sec.3.3.3. In a general case, they have to be 
found numerically by a trial-and-error method in which a constant $A$ is being fitted until a 
relaxation trajectory (3.23), (3.28) $S_2 \stackrel {relA}{\rightarrow}$  arrives at 
$S_1$ while the 
resulting trajectory $S_2 \stackrel {relA}{\rightarrow}S_1$  consists from $n_{rel}$ or $n_{rel}-
1$ passages, for $A_-$ and $A_+$ respectively (note the footnote 7). This procedure is 
incomparably easier than a 
direct numerical solution of the variational problem (c.f. \cite {kautz}) and takes very little of 
computer time. A concrete algorithm may vary. E.g. one may use a standard 
method of successive approximation starting from the ranges $[-1, 0]$ and $[0, 1]$, for $A_-$ and $A_+$ 
respectively. A convergence is typically quick. 

\vskip 0.5truecm

\noindent
{\bf Theorem 2.}
{\it Let the condition 2 be satisfied while the condition 1 be not. Then the $MPDTP$ $S_1\rightarrow S_2$ is the path $S_1\stackrel {A_-}{\rightarrow}S_2$ while action along it is 
$<U_{S_2}-U_2$.}

{\it Proof.}

The proof is nearly identical to that one in the {\it theorem 1}. The only difference is that there are no direct paths 
corresponding to positive $A$.

\vskip 0.5truecm

Before to pass to two other theorems (whose proof is probably the most non-trivial part of the 
paper) let us present qualitative arguments in favor of that an extremal with a single negative $A$ does 
not exist at large enough friction: such arguments will facilitate an understanding the rigorous 
proof of theorems 3, 4. For the sake of simplicity, let us consider the most typical case, when $U(q)$ is smooth both in $S_1$ and $S_2$. For the overdamped case, a noise-free 
trajectory emanating from $S_1$ or $S_2$ follows
nearly the very slope (left or right, respectively) of the potential
well $2$. A friction  $\Gamma^{\prime} (t)$ (3.28) varies along the trajectory  $S_2 \stackrel 
{relA}{\rightarrow}$ from $ \Gamma$ at 
$t=-\infty$ to $ -\Gamma$ at $t=\infty$. A transition between these two regimes occurs for an 
interval $\sim \Gamma^{-1}$ and, hence, in order for the trajectory to manage to come to 
$S_1$ rather than to pass above it it should manage for this interval $\sim \Gamma^{-1}$ to 
pass from the right slope to the left one. However such passage (a large part of which takes place 
in the regime $\Gamma^{\prime} \ll \Gamma $) would require an interval $\sim 
\omega_{osc}^{-1}$ where $\omega_{osc}$ is a characteristic frequency of an eigenoscillation in 
the well 2. At large enough $\Gamma$, this time is much larger than $\Gamma^{-1}$ and 
therefore the passage between the different slopes does not manage to occur which means that an 
extremal $S_1 \stackrel{A} {\rightarrow} S_2$ cannot exist at large enough 
$\Gamma$.

Now, let us formulate the theorem 3 and prove it rigorously.

\vskip 0.5truecm

\noindent
{\bf Theorem 3.}
{\it Let neither of the conditions 1 and 2 be satisfied. 
Then there is no any direct extremal which would provide the transition i.e. the $MPDTP$ does not exist.}

{\it Proof.} 

An absence of direct extremals with a {\it positive} $A$ is obvious: a dissipation along  a trajectory $ S_2 \stackrel 
{relA>0}{\rightarrow}$ exceeds that one along a noise-free trajectory so that  the trajectory $ S_2 \stackrel 
{relA>0}{\rightarrow}$ cannot reach $S_1$.
	
Let us now prove an absence of a path $S_1\stackrel {A}{\rightarrow} S_2$  with a {\it negative} 
$A$. If such path with some hypothetic negative $A=A_h$ did exist then, for any negative $A$ 
with $|A|<|A_h|$, a dissipation of energy along the relaxational trajectory $ S_2 \stackrel 
{relA}{\rightarrow}$ should be larger than that one along $ S_2 \stackrel 
{relA_h}{\rightarrow}$, so that the trajectory in energy-coordinate plane would inevitably meet 
the slope of $U(q)$ (i.e. the line 
$E=U(q)$) below $S_1$, in other words, the trajectory would have a turning point somewhere 
between $q(S_1)$ and $q(2)$. In particular, this would hold true at $|A| \rightarrow 0$. At the same time, if $|A| \rightarrow 0$ then a point of minimal energy on the trajectory $ S_2 \stackrel {relA}{\rightarrow}$ (corresponding to $t_0$ (3.29)) 
approaches the bottom of the well $2$ because the trajectory $ S_2 \stackrel {relA \rightarrow 
0}{\rightarrow}$ approaches the noise-free trajectory $ S_2 \stackrel {rel}{\rightarrow}$ at 
any instant $t\ll \Gamma^{-1}{\ln}(1/|A|) \stackrel {A \rightarrow 
0}{\rightarrow}\infty$. A potential near $2$ may be approximated by a parabola:

\begin {equation}
U(q)=\frac{\Omega^2 (q-q(2))^2}{2},  \quad\quad\quad q \approx q(2),
\end{equation}

\noindent and an analysis of eqs.(3.23), (3.28) is simplified. We are to prove that the trajectory $ S_2 \stackrel {relA \rightarrow 
0}{\rightarrow}$ cannot possess a
turning point between $S_1$ and $2$ which will be equivalent to the proof 
of a non-existence the path  $S_1\stackrel {A_h}{\rightarrow} S_2$. 
This task is still non-trivial as it requires an {\it explicit} solution of eqs.(3.23), (3.28) 
while, even for such simple 
potential as (3.35), the equations are non-trivial at an arbitrary $\Gamma$. Instead of their solution, one 
may return to the original Euler-Poisson equation (3.20) which is reduced for the case of a 
parabolic $U(q)$ to a linear differential equation of the forth order with constant coefficients 
and, obviously, is easily solved. However, the immediate result of such solution is quite 
inconvenient for the required proof. That is why a different method is used below: I show that 
the proof can be reduced to certain partial case of eqs. (3.23), (3.28) at which the equations have a 
solution in a very convenient for the final proof form.

A noise-free equation of motion in a parabolic potential is merely a linear differential equation with 
constant coefficients and is easily solved \cite {landau}. If $\Gamma\ge \Gamma_{\rm min}$, where

\begin {equation}
\Gamma_{\rm min}=2 \Omega,
\end{equation}

\noindent
then the solution does not possess turning points in an infinitesimal vicinity of 
the bottom of the well.

In all cases to which the {\it theorem 3} relates, there is necessarily a small enough (but non-zero) vicinity of $2$ in which noise-free trajectories do not have any turning 
point (otherwise the {\it condition 2} would hold true which, in its turn, 
would contradict to the condition of the theorem). Hence, relevant values of $\Gamma$ are 
necessarily not less than $\Gamma_{\rm min}$ (3.36) \footnote{Typically, a minimal 
$\Gamma$ at which trajectories $S_2\stackrel {rel}{\rightarrow}2$ and $2 \stackrel 
{A=0}{\rightarrow}S_1$ do not intersect (apart from $2$) is equal just to $\Gamma_{\rm 
min}$ (3.36). However, it may be larger: in those cases when $U(q)$ has steep slopes 
intersections may still occur at large energies notwithstanding their absence in a vicinity of 2, at 
$\Gamma > \Gamma_{\rm min}$.}. If we prove an absence of turning points between $S_1$ and $2$ in 
$S_2 \stackrel {rel A \rightarrow 0}{\rightarrow}$ for $\Gamma =\Gamma_{\rm min}$ it will garantee the 
same for any larger $\Gamma$ because the larger $\Gamma$ the steeper the trajectory $S_2
\stackrel {rel A }{\rightarrow}$ in the energy-coordinate plane and, all the more so, there 
are no turning points between $S_1$ and $2$ (a rigorous proof of this will be given further).

The dynamic equations (3.23), (3.28) for the relaxational trajectory  $S_2
\stackrel {rel A \rightarrow -0}{\rightarrow}$ in a close vicinity of $2$, for $\Gamma=2 \Omega$, can be written as:

\begin{eqnarray} 
\frac{d^2 {\tilde q}}{d \tau ^2} - 2{\tanh} (\tau) \frac {d {\tilde q}}{d\tau} + {\tilde q} = 0,
\\
{\tilde q}= q-q(2), \quad \tau=\Omega(t-t_0),
\nonumber
\end{eqnarray}

\noindent
where $t_0$ is given in (3.29).

The differential equation (3.37) can be solved explicitly:

\begin {equation}
{\tilde q}(\tau)=C_1 {\sinh}( \tau) + C_2 (\tau {\sinh}( \tau) - {\cosh}( \tau) ),
\end{equation}

\noindent
where the integration constants $C_1, C_2$ can be determined from two additional conditions.

Let a point $r\equiv (q_r,\dot {q}_r)$ on the trajectory $S_2 \stackrel {rel}{\rightarrow}2$ which is reached at an 
instant $t_r$ is close to $2$ enough for a parabolic approximation to be valid and for the trajectory not to possess turning points at $t\geq t_r$. If 
$|A|\rightarrow 0$ the trajectory $S_2 \stackrel {relA}{\rightarrow}$ approaches $S_2 \stackrel 
{rel}{\rightarrow}$ so that a deviation of $r$ from a state on $S_2 \stackrel 
{relA}{\rightarrow}$ corresponding to the same moment $t_r$ becomes negligible. Thus, at

\begin {equation}
\tau=\tau_r \equiv t_r-t_0,
\end{equation}

\noindent
the trajectory (3.38) should pass a state very close to $r$ so that $C_1, C_2$ can be easily found:

\begin {eqnarray}
C_1=\frac{{\tilde q}_r\tau_r {\cosh}( \tau_r)-\dot {\tilde q}_r(\tau_r{\sinh}( \tau_r)- 
{\cosh}( \tau_r)) }{{\cosh}^2 ( \tau_r)}, \quad
C_2=\frac{-{\tilde q}_r {\cosh}( \tau_r)+ \dot {\tilde q}_r{\sinh}( \tau_r)}{{\cosh}^2 
( \tau_r)},
\nonumber
\\
{\tilde q}_r\equiv q_r - q(2), \quad  \dot {\tilde q}_r\equiv \dot {q}_r.
\end{eqnarray}

All close to $2$ points of $S_2 \stackrel {rel}{\rightarrow}2$ (including $r$) necessarily satisfy 
certain condition 
which will be particularly important for the further proof. In order to derive it let us write down 
the equation of motion along the noise-free trajectory $S_2 \stackrel {rel}{\rightarrow}2$ in a 
vicinity of $2$:

\begin {equation}
\frac{d^2 {\tilde q}}{d \tau ^2} + 2 \frac {d {\tilde q}}{d\tau} + {\tilde q} = 0,
\end{equation}

\noindent
in which the same notations as in (3.37) are used. Its solution is

\begin {equation}
{\tilde q}(\tau)=[{\tilde q}(\tau_r) + (\tau -\tau_r) (\dot{\tilde q}(\tau_r)+ {\tilde q}(\tau_r)) ]{\rm e}^{-(\tau -\tau_r)}.
\end{equation}

Differentiating (3.42), we obtain 

\begin {equation}
\dot {\tilde q}(\tau)=[\dot {\tilde q}(\tau_r) - (\tau -\tau_r) (\dot{\tilde q}(\tau_r)+ {\tilde 
q}(\tau_r)) ] {\rm e}^{-(\tau -\tau_r)},
\end{equation}

\noindent
from which it follows that in order for a velocity to keep its sign for all finite $\tau\ge\tau_r$
the following condition should be satisfied

\begin {equation}
\frac {{\tilde q}(\tau_r) }{\dot {\tilde q}(\tau_r)}<-1, \quad \quad \quad \dot {\tilde 
q}(\tau_r)\not=0
\end{equation}

\noindent
(the dashed lines in Fig.5(a) correspond to ${\tilde q}(\tau_r)/\dot {\tilde q}(\tau_r)=-1$).

Using (3.44), we show below that a path which possesses a turning point between $S_1$ and 
$2$ cannot be $S_2 \stackrel {rel A \rightarrow 0}{\rightarrow}$. It is convenient to consider 
separately a case when the turning point is approaching $2$ as $|A|$ is approaching zero and a 
case when the turning point remains at a finite distance from $q(2)$ at $|A|\rightarrow 0$.

\vskip 0.2 truecm
\noindent
1). Let a path (3.38) possesses at $\tau=\tau_1$ a turning point at ${\tilde q}<0$. Then (3.38)
can be written as 

\begin {equation}
{\tilde q}(\tau)=\frac {{\tilde q}(\tau_1) }{{\cosh}( \tau_1)}[ (\tau_1-\tau){\sinh}( \tau) +{\cosh}( \tau)]
\end{equation}

\noindent
(an example of such path is shown in Fig.5(a)).

A coordinate-to-velocity ratio along the path is

\begin {equation}
R\equiv \frac {{\tilde q}}{\dot{\tilde q}}={\tanh} (\tau) +\frac{1}{\tau_1-\tau }.
\end{equation}

The function $R(\tau)$ is monotonously increasing at any $\tau \not =\tau_1 $:

\begin {equation}
\frac{dR}{d \tau }= \frac{1}{{\cosh}^2 ( \tau)}+ \frac{1}{(\tau_1-\tau)^2 }>0,
\quad\quad
\tau\not = \tau_1.
\end{equation}

Allowing for 

\begin {equation}
R(-\infty)=-1,
\end{equation}

\noindent we may conclude that, at any finite $\tau<\tau_1$ (i.e. along that part of the path 
which precedes the turning point),

\begin {equation}
\frac {{\tilde q}}{\dot{\tilde q}}>-1.
\end{equation}

The incompatibility  of (3.49) with the condition (3.44) proves that the path 
(3.45) cannot coincide with the path (3.38), (3.40) and therefore 
$S_2 \stackrel {rel A \rightarrow 0}{\rightarrow}$ cannot possess a turning point at an infinitesimal distance to the left from 
$2$.

\vskip 0.2 truecm
\noindent
2). Let us show that an assumption that $S_2 \stackrel {rel A \rightarrow 0}{\rightarrow}$ 
possesses a turning point between $S_1$ and $2$ at some {\it finite} distance from 2 leads to a 
contradiction too. So, let us assume that $S_2 \stackrel {rel A \rightarrow 0}{\rightarrow}$ possesses 
the above described turning point $h$ at some $\tau=\tau_h$:

\begin {equation}
\dot{\tilde q}(\tau_h)=0, \quad \quad q(S_1)-q(2)< {\tilde q}(\tau_h)<0, \quad \quad 
{\tilde q}(\tau_h) \stackrel {A \rightarrow 0}{\not\rightarrow}0.
\end{equation}

A parabolic approximation may not be valid near $h$ but, near the bottom of the well $2$, the 
trajectory should still be approximated by the path (3.38), (3.40) which should not possess 
turning points in a vicinity of the bottom of the well and should be sewed at large positive 
$\tau\equiv \tau_l$ together with the trajectory which is time-reverse to the noise-free trajectory 
emanating from the hypothetic turning point $h\stackrel {rel}{\rightarrow}$, similar to the sewing together 
with $S_2\stackrel {rel}{\rightarrow}$ at large negative $\tau=\tau_r$. 
The condition of a sewing with a time-reverse trajectory can be immediately obtained on the basis of the condition (3.44) for a direct trajectory if to use that fact that a velocity in any point of a time-reverse trajectory is just opposite to a velocity in the same point of energy-coordinate plane for a direct trajectory. Thus,

\begin {equation}
\frac {{\tilde q}(\tau_l) }{\dot {\tilde q}(\tau_l)}>1, \quad \quad \quad \dot {\tilde 
q}(\tau_l)\not=0.
\end{equation}

At the same time, it follows from eq.(3.38)

\begin {equation}
R\equiv \frac {{\tilde q}}{\dot{\tilde q}}={\tanh} (\tau) -\frac{C_2}{C_1+C_2 \tau }.
\end{equation}

\noindent The function $R(\tau)$ increases monotonously everywhere except

\begin {equation}
\tau\equiv \tau_{turn}=-\frac{C_1}{C_2},
\end{equation}

\noindent which corresponds to a turning point of ${\tilde q}(\tau)$. Allowing for the absence of 
turning points in the vicinity of the bottom of the well (which has been proved above in 1)), there 
should be either $\tau_{turn}>\tau_l$ or $\tau_{turn}<\tau_r$. However, in any of these cases, 
at least one of the conditions (3.44) and (3.51) is not satisfied: in the former case, $R(\tau_r)>-1$ 
while, in the latter case, $R(\tau_l)<1$ (note that if $C_2=0$ then neither (3.44) nor (3.51) are 
satisfied since $-1<R(t)<1 \quad \forall t$).

This contradiction proves that the assumption (3.50) is wrong. Together with the proof of an impossibility for $S_2 \stackrel {rel A \rightarrow 0}{\rightarrow}$ to possess a 
turning point in the very vicinity of $2$ (see 1) above), this proves an absence of a transition 
path $S_1 \stackrel { A < 0}{\rightarrow}S_2$ at $\Gamma=\Gamma_{\rm min}$ (3.36).

Finally, we should prove that, for any larger $\Gamma$, this is all the more so. With this aim, 
let us write down eqs.(3.23), (3.28) for a case $A<0$ in the following form

\begin{eqnarray} 
\frac{d^2 q}{d {\tilde t}^2 } + \tilde \Gamma ({\tilde t})\frac{d q}{d {\tilde t} }  +
\frac{dU}{dq}=0,
\\
{\tilde t}=t-t_0, \quad\quad \tilde \Gamma ({\tilde t})=-\Gamma {\tanh}(\frac{\Gamma {\tilde 
t}}{2}).
\nonumber
\end{eqnarray}

An effective friction $ \tilde \Gamma ({\tilde t})$ is positive at ${\tilde t}<0$ and negative at 
${\tilde t}>0$ so that the instant ${\tilde t}=0$ corresponds to a minimal energy on a trajectory, 
for any $\Gamma$. At the same time, the larger $\Gamma$ the larger $ |\tilde \Gamma ({\tilde 
t})|$ and, therefore, the steeper a trajectory $E(q)$ in the energy-coordinate plane (see Fig.5(b)).

If, analogously to the analysis for a case $\Gamma/(2\Omega)=1$, to consider for 
$\Gamma/(2\Omega)>1$ a case $|A|\rightarrow 0$ which can be described in a vicinity of $2$ in 
a parabolic approximation (3.35) then the corresponding trajectory (3.54) should be sewed at 
$\Omega {\tilde t}= \tau_r \rightarrow - \infty$  with a noise-free trajectory and at $\Omega 
{\tilde t}= \tau_l \rightarrow + \infty$ with a trajectory time-reverse to a noise-free one. It is easy 
to derive for a general case $\Gamma \geq 2 \Omega$ the conditions similar to (3.44), (3.51):

\begin{equation} 
R(\tau_r)\equiv \frac{{\tilde q}(\tau_r)} {d{\tilde q}(\tau_r)/d\tau_r}< - 
\frac{\Gamma}{2\Omega} \leq -1, \quad \quad \quad
R(\tau_l)\equiv \frac{{\tilde q}(\tau_l)} {d{\tilde q}(\tau_l)/d\tau_l}> 
\frac{\Gamma}{2\Omega}
\geq 1.
\end{equation}

If $\Gamma > 2\Omega$ then the corresponding to (3.55)  lines in the $E-{\tilde q}$ plane lie necessarily 
lower than those ones for a case $\Gamma = 2\Omega$ (c.f. Fig.5(b)). At the same time, as it 
was shown above, lines corresponding to the trajectory (3.54) for $\Gamma > 2\Omega$ lie 
necessarily above those for $\Gamma= 2\Omega$ (c.f. Fig.5(b)). Taken that, even at $\Gamma 
= 2\Omega$, at least one of the sewing conditions (3.55) cannot be satisfied for the path (3.54), it 
is all the more so at $\Gamma > 2\Omega$.

Thus, the {\it theorem 3} has been proved.

\vskip 0.5truecm

\noindent
{\bf Theorem 4.}
{\it Let the condition 1 be satisfied while 2 be not. Then the $MPDTP$ $S_1\rightarrow S_2$ is the extremal $S_1 \stackrel {A_+} {\rightarrow}S_2$.}

\vskip 0.2truecm

\noindent
{\it Remark 4.} The case described by the {\it theorem 4} is not typical but it may occur if a bottom of the 
well $2$ is shallow while a slope of the well $2$ between $2$ and $S_2$ becomes rather steep at some distance from the bottom.

\vskip 0.2truecm

{\it Proof.}

The proof of a non-existence an extremal with a negative $A$ is identical to that one 
in the {\it theorem 3} while the proof concerning an extremal with $A=A_+$ is identical to that 
one in the {\it theorem 1}.

\vskip 0.5truecm

Finally, in this sub-section, I shall say very briefly about a problem of the $MPDTP$
$i 
\rightarrow j$ if both $i$ and $j$ are non-stationary points\footnote {This case is more of a formal rather than practical interest: if we consider a
transition 
from an infinitesimal vicinity of $i$ then, with an overwhelming
probability,  
the system will move first to an attractor and, only from there, will
transit to 
$j$, without ever return to the infinitesimal vicinity of $i$. However
if one seeks 
the most probable fluctuational transition from an attractor to $j$
which would obligatory follow $i$ then the 
variational problem for the $MPDTP$ $i \rightarrow j$ is important. This
can be 
necessary at an evaluation of a prehistory probability density \cite
{prehistory} 
(c.f. also \cite {nonstationary}).}. This case has certain differences from the case when
at least one of states is a stationary point. Thus, the $MPDTP$ may not obligatory follow
the 
saddle $S_1$. Rather it may be a path with a single $A$
(rather than with $A$ switching in the saddle $S_1$  from $A=0$ to non-zero $A$). Such
transition takes a finite time even if a maximum  $S_1$ is smooth. In the case $A>0$, this time is less than $t_p$
(3.30). In 
fact, most of the above-said in this paragraph concerns also a transition within
one and 
the same basin of attraction.

\vskip 0.5truecm

\noindent
{\bf 3.3.2. Action}

\vskip 0.5truecm

The final goal of the variational problem is action $S_{\rm min}$ (3.18) which, as it has been 
found in previous sub-sections, can be presented in any of two forms (3.33) in which 
$q_{opt}(t)$ is sewed from trajectories of the type (3.23) with $\Gamma^{\prime}(t)$ (3.28) in 
accordance with the algorithm described in the previous sub-section. I emphasize that the 
numerical procedure is incomparably easier than a direct numerical solution of the variational 
problem (c.f. \cite{kautz}) and takes typically very little computer time.

In some ranges of friction, action can be found explicitly. Thus, if $S_2$ belongs to a basin of attraction of $1$ then the $MPDTP$ $1\rightarrow S_2$ is just time-reverse to the relaxational trajectory $S_2\stackrel{rel}{\rightarrow} 1$ while action is just a difference of energies, $U_{S_2}-U_1$.
If $S_2$ does not belong to a basin of attraction of $1$ and friction is 
much less than characteristic eigenfrequencies, one can obtain explicit 
asymptotic formulas for action which will be done in Sec.3.3.3. 
If $\Gamma$ is slightly less than the upper limit for an existence of the $MPDTP$, 
$\Gamma_0$, then the $MPDTP$ is close to $1\stackrel{A=0}{\rightarrow}S_1\stackrel{rel}{\rightarrow}2\stackrel{A=0}{\rightarrow}S_2$ and action is:

\begin{equation}  
S_{\rm min}( 1 \rightarrow S_2) \approx U_{S_1}-U_1+ U_{S_2}-U_2,
\quad\quad
0<\Gamma_0-\Gamma \ll \Gamma_0.
\end{equation}

If $\Gamma\geq\Gamma_0$ then a direct transition rate $\alpha_{1S_2}$ (as well as 
$\alpha_{13}$ in phenomenological formulas of Sec.2) is equal to $0$:

\begin{equation}  
\alpha_{1S_2}=0,
\quad\quad\quad
\Gamma\geq\Gamma_0.
\end{equation}

Generally, $S_{\rm min}$ is to be calculated numerically. For the potential shown in Fig.4,

\begin{equation} 
U(q)=0.06 (q+1.5)^2 - \cos (q), 
\end{equation}

\noindent
we plot in Fig.6 an excess of action over the difference of energies \footnote{Note that action for the 
reverse transition is equal just to $\Delta S_{\rm min}$: $ S_{\rm min}( S_2 \rightarrow 1) =\Delta S_{\rm min}$.}:

\begin{equation} 
\Delta S_{\rm min} \equiv S_{\rm min}( 1 \rightarrow S_2) - (U_{S_2}-U_1)= \frac {1}{4 \Gamma}\int_0^{t_{tr}} dt (\Gamma - 
\Gamma^{\prime}(t))^2 \dot {q}_{opt}^2  (t_{tr}-t) .
\end{equation}

\noindent
A magnitude of $\Delta S_{\rm min}$ varies from\footnote{If $U_{S_2}$ was smaller than 
$U_{S_1}$ then the lower value of $\Delta S_{\rm min}$ would be $ U_{S_1}-U_{S_2}$. } $0$ 
to $ U_{S_1}-U_2$. The latter is approached as $\Gamma$ approaches the critical value 
$\Gamma_0$ at which a direct path disappears, as described in the {\it theorem 3}, and which is 
equal in this case (as well as in a majority of other cases) to a doubled frequency of 
eigenoscillation in a bottom of the well $2$:

\begin{equation} 
\Gamma_0=2\Omega.
\end{equation}

In the underdamped range, $\Gamma<\Gamma_1$, which is described by the {\it theorem 1},
the dependence $\Delta S_{\rm min} (\Gamma)$ undergoes characteristic oscillations (the inset shows them in an enlarged scale) which correspond to an alternation of ranges at which $S_2$ belongs to a 
basin of atraction of $1$ with ranges at which it does not. The ranges are separated by  critical 
values $\Gamma_n $ which correspond to saddle connections $S_2 
\stackrel{rel}{\rightarrow}S_1$ consisting of $n$ passages ($n=1, 2, ...$). Each oscillation has 
a cusp-like singularity in its maximum which corresponds to a jump-wise switch of the $MPDTP$ 
between paths corresponding to $A_+$ and $A_-$ (c.f. a discontinuity in a first derivative of 
nonequilibrium potential \cite{gt} and a fluctuational separatrix for optimal paths in a phase 
space \cite{jch94}, \cite{pl94}).

The range of moderate friction, $\Gamma_1<\Gamma<\Gamma_0$, is described by the {\it 
theorem 2}. A major variation of action occurs within just this range: the larger $\Gamma$ the 
deeper into the well $2$ the $MPDTP$ $1\stackrel{A=0}{\rightarrow}S_1 \stackrel{A_-
}{\rightarrow}S_2$ falls and, correspondingly, the larger $\Delta S_{\rm min}$.

\vskip 0.5truecm

\noindent
{\bf 3.3.3. Underdamped regime}

\vskip 0.5truecm

The goal of this sub-section is to derive explicit expressions for action and $MPDTP$ in the case 
when $\Gamma$ is small:

\begin {equation}
\beta  \equiv \frac {\Gamma}{{\rm min}(\omega_{osc},\omega_{osc}^{(1)},\omega_{osc}^{(2)})} 
\ll 1,
\end{equation}

\noindent where $\omega_{osc},\omega_{osc}^{(1)},\omega_{osc}^{(2)}$ are characteristic
 frequencies of eigenoscillation at energies between saddles and in wells $1$ and $2$ respectively.

\vskip 0.2truecm

\noindent
{\bf (a). Noise-free trajectory}

\vskip 0.2truecm

First, it is necessary to derive a formula for a critical value $\Gamma\equiv \Gamma_n$ which lies in the 
underdamped range (3.61) and provides a saddle connection $S_2\stackrel{rel}{\rightarrow}S_1$ with a given number of passages 
$n $. I shall derive also $n_{rel}$ as an explicit function of $\Gamma$. Integrating the 
equation for the energy along the trajectory, (3.34), in which $A=0$ (i.e. 
$\Gamma^{\prime}(t)\equiv\Gamma_n$) and allowing for a smallness of $\Gamma_n$ due to 
which energy may be considered as a constant at an integration of the left-hand side (3.34) 
along one passage, we obtain for a dissipation of energy along a given $m$th passage of a 
noise-free trajectory:

\begin{eqnarray} 
\Delta E_m\equiv E_m-E_{m+1}  =  \pi \Gamma_n I(E_m),
\\
m \leq  n-1,
\nonumber 
\end{eqnarray}

\noindent where $E_m$ is an energy in the beginning of the passage while $I$ is a mechanical action \cite {landau}

\begin {equation}
I(E) = \frac {1}{2 \pi}\oint dq \dot q, \quad \quad \dot q = \sqrt{2(E-U(q))}.
\end{equation}
 
For the last (i.e. $n$th) passage, we obtain similarly:

\begin{eqnarray} 
\Delta E_n \equiv E_n-U_{S_1}  =  \pi \Gamma_n I_k & , &
\\
& &
k  \equiv  \frac{1}{2} (3+s(-1)^n), 
\nonumber 
\\
& &
s  \equiv  {\rm sgn}(\frac{q(2)-q(1)}{q(2)-q(S_2)}), 
\nonumber 
\end{eqnarray}

\noindent where $I_1 \equiv I_1(U_{S_1})$ and $I_2 \equiv I_2(U_{S_1})$ 
are actions (3.63) in wells $1$ and $2$ respectively.

Dividing the eq.(3.62) by $\pi \tilde\Gamma_n I$, applying the resulting equation to the first $n-
1$ passages, summing up the results, exchanging the summation by the integration (the latter 
operation is justified by a smallness of $\Delta E_m$) and allowing for (3.64), we derive

\begin {equation}
\frac{1}{\pi}\int^{U_{S_2}}_{U_{S_1}+\pi \Gamma_n I_k} dE \frac {1}
{I(E) \Gamma_n }= n-1 .
\end{equation}

Allowing for 

\begin {equation}
\frac{1}{\pi}\int^{U_{S_1}+\pi \Gamma_n I_k } _{U_{S_1}}dE \frac {1}
{I(E) \Gamma_n }
\approx \frac{ I _k}{I_1+ I_2}=\frac{1}{1+(\frac{I_1}{I_2})^{s(-1)^n}},
\end{equation}

\noindent 
we derive

\begin {equation}
\frac{1}{\pi}\int^{U_{S_2}}_{U_{S_1}}dE \frac {1}
{ I(E) \Gamma_n }-\frac{1}{1+(\frac{I_1}{I_2})^{s(-1)^n}}= n-1 ,
\end{equation}

\noindent from which

\begin {eqnarray}
\Gamma_n = \frac{\Delta\omega_{osc}}{n-1+\frac{1}{1+(\frac{I_1}{I_2})^{s(-1)^n}}}, \quad\quad
\Delta\omega_{osc} \equiv  \frac{1}{\pi}\int_{U_{S_1}}^{U_{S_2}} dE \frac {1}{I(E)}, 
 \\
\Gamma_n  \ll  {\rm min} (\omega_{osc},\omega_{osc}^{(1)},\omega_{osc}^{(2)})) .
\nonumber 
\end{eqnarray}

If 

\begin {equation}
\Gamma \in ]\Gamma_{2l+(1+s)/2+1}, \Gamma_{2l+(1+s)/2}[, \quad\quad\quad
l\geq 0,
\end{equation}

\noindent
then  $S_2\stackrel{rel}{\rightarrow}$ goes just to $2$ rather than to $1$ and a number of passages in 
$S_2\stackrel{rel}{\rightarrow}I$ is 
 
\begin {equation}
n_{rel}= 2l+1+(1+s)/2.
\end{equation}

It follows from (3.68)-(3.70) that

\begin {eqnarray}
n_{rel}  = 
2 (n_{\Gamma}^{(-)}+1) +\frac{1}{2}(1-(-1)^{ n_{\Gamma}^{(+)} - n_{\Gamma}^{(-)} }), \quad 
n_{\Gamma}^{(\pm)}\equiv [\frac{1}{2}(\frac{\Delta \omega_{osc}}{\Gamma} 
\pm\frac{1}{1+ (I_2/ I_1)^s})], \\
s(-1)^{n_{rel}} = +1, 
\quad\quad\quad
\quad\quad\quad
\quad\quad\quad
\quad\quad\quad
\quad\quad\quad
\quad\quad\quad
\quad\quad\quad
\quad\quad\quad
\quad\quad\quad
\nonumber
\end{eqnarray}

\noindent
where the square brackets $[...]$ denote an integer part and the lower equality chooses those ranges of friction at which $S_2\stackrel{rel}{\rightarrow}$ goes just to $2$ rather than to $1$ (note that, in these ranges, $n_{rel}$ is odd if $q(2)$ is between $q(1)$ and $q(S_2)$ and even otherwise).

\vskip 0.2truecm

\noindent
{\bf (b). Action}

\vskip 0.2truecm

Let us find an explicit dependence of action on friction, in ranges (3.69).
The $MPDTP$ necessarily follows $S_1$. 
Action along the bit $1\rightarrow S_1$ is equal just to a difference of 
energies $U_{S_1}-U_1$ and, hence, a non-zero contribution into $\Delta 
S_{\rm min}$ 
is made only by the bit $S_1\rightarrow S_2$. Allowing for this and using the identity (3.34), we 
may write (3.59) as

\begin{equation}
\Delta S_{\rm min}  = 
\frac {1}{4 
\Gamma}\int_{ S_1}^{ S_2} dE \frac{(\Gamma -
\tilde \Gamma^{\prime}(E))^2}{\tilde \Gamma^{\prime}(E)} ,
\quad\quad\quad
S_{2}  \stackrel {rel}{\not \rightarrow}  1,
\end{equation}

\noindent where $\tilde \Gamma^{\prime}(E)$ is the auxiliary friction $\Gamma^{\prime}(t)$ 
(3.28) expressed as a function of energy $E(t)$ along the auxiliary relaxational trajectory $S_2  \stackrel {relA}{\rightarrow}S_1$ (3.23), (3.28).

In order to find $\tilde \Gamma^{\prime}(E)$ one needs to find $E(t)$.
In a general case of an arbitrary $\Gamma$, the function $E(t)$ cannot 
be found in an explicit form while, in an underdamped case (3.61), it can be found explicitly in the relevant range of 
energies. Indeed, the 
characteristic time-scale on which both $\Gamma^{\prime}(t)$ and $E(t)$ may change significantly is $\Gamma^{-
1}$. Correspondingly, in accordance with the conventional averaging method \cite {bog}, changes of energy on smaller 
time-scales are not essential for  $\tilde \Gamma^{\prime}(E)$.
In the underdamped case, $\Gamma^{-
1}$ is much larger than a characteristic duration of one passage\footnote {In case of smooth maxima $S_1$ and $S_2$,  
the present consideration does not cover the very beginning of a first passage and the very end of a last passage i.e. the 
very vicinities of $S_2$ and $S_1$ respectively: their contribution into action is negligible as it will be shown in the end 
of the sub-section.} which is equal approximately to half a period of eigenoscillation at an average energy on the 
passage. 
 Correspondingly, a change of energy along one passage is small  while the dynamic equation (3.34) (which is obeyed by 
energy along the trajectory) may be averaged over a passage or, 
equivalently, over a period of eigenoscillation. Thus, averaging the eq.(3.34) over a passage, allowing for \cite {kramers}

\begin {equation}
\overline {\dot{q}^2}= I \omega ,
\end{equation}

\noindent
where the overbar means an averaging over an oscillation while $I$ and $\omega$ are respectively a mechanical action 
(3.63) and frequency of 
eigenoscillation at a given energy $E$, and transforming from $E$ to $I$ (note that 
$dE/dI=\omega$ \cite{landau}), we obtain

\begin {equation}
\frac{dI}{dt}=-\Gamma^{\prime}I.
\end{equation}

\noindent After the substitution $\Gamma^{\prime}(t)$ (3.28), the eq.(3.74) can be integrated explicitly:

\begin {equation}
I=I_{t=0} {\rm e}^{-\int_0^t d\tau \Gamma^{\prime}(\tau)} = I_{t=0}  {\rm e}^{- \Gamma t}(\frac{1-A{\rm e}^{\Gamma t}}{1-A})^2 ,\quad\quad
\quad
I_{t=0} \equiv I(U_{S_2}).
\end{equation}

Expressing from (3.28) ${\rm e}^{\Gamma t}$ via $\Gamma^{\prime}$ and substituting it into 
(3.75), one obtains

\begin {equation}
I=I(U_{S_2}) \frac{4A }{((\frac{\Gamma^{\prime}}{\Gamma })^2-1)( 1-A)^2},
\end{equation}

\noindent from which it follows

\begin {equation}
\frac{\Gamma^{\prime}}{\Gamma } \equiv \frac{\tilde\Gamma^{\prime}(E)}{\Gamma }=
\sqrt {1+\frac{I(U_{S_2})}{I(E)}\frac{4A}{(1-A)^2} }.
\end{equation}

Thus, in order to find $\Gamma^{\prime}$ as a function of energy along the trajectory we need 
only to find $A$. It should be found from that condition 
that a trajectory $S_2 \stackrel{relA}{\rightarrow}$ goes just to $S_1$, moreover, a number of 
passages should be equal to either $n_{rel}$ (3.71) or $n_{rel}-1$, for $A_-$ and $A_+$ 
respectively. To find $A_{\pm}$ I use equations analogous to 
(3.62)-(3.67) which were used in order to find $\Gamma_n$ and $n_{rel}(\Gamma)$. The only 
difference is that I put in $\Gamma^{\prime}(t)$ $A= A_{\pm}$  instead of $A=0$ 
and, correspondingly, $\Gamma_n$ should be exchanged by $\tilde 
\Gamma^{\prime}(E)$ (3.77) while $n$ should be exchanged by $n_{rel}$ (3.70) or  $n_{rel}-1$, for 
$A_-$ and $A_+$ respectively. Thus, instead of (3.67) for the case $A=0$, I obtain for 
$A_{\pm}$ the following equation:

\begin {equation}
\frac{1}{\pi}\int^{U_{S_2}}_{U_{S_1}}dE \frac {1}
{ \Gamma \sqrt {I(I+4I(U_{S_2})\frac{A_{\pm}}{(1- A_{\pm})^2})} }-\frac{1}{1+(\frac{I_1}{I_2})^{\mp 1}}= n_{rel}-\frac{1}{2}(3\pm1).
\end{equation}

Generally, the integral in (3.78) cannot be found explicitly since $I(E)$ (3.63) is 
typically a complicated function which can be presented only in an integral form.
At the same time, in order to find $ A_{\pm} $ explicitly  we do need to express the 
integral as an explicit function of $ A_{\pm} $. Fortunately, in the 
underdamped case (3.61), one can split the whole variety of $U(q)$ into such two complementary 
classes of functions that an approximate value of $A_+$ can be found from (3.78) 
explicitly for both
classes. Obviously, the results match each other on the \lq\lq boundary" between the classes. 

All potentials $U(q)$ are splitted into the two classes by a very simple condition: whether at least 
one of wells is deep or not i.e. whether the parameter 

\begin {equation}
\mu  \equiv \frac { U_{S_2}-U_{S_1} }{U_{S_1}-{\rm min}(U_1,U_2)} 
\end{equation}

\noindent 
is small or not.

\vskip 0.2truecm

\noindent
1). Let us first consider the case when at least 
one of wells is deep\footnote {The case when the well $1$ is not deep while $2$ is 
seems to be the most interesting case in the generalized Kramers problem, both in the 
underdamped regime and in general: 1) though, after an escape from $1$, the system 
will most probable slide down into $2$, a period of stay there may be so long that it 
will exceed a realistic duration of an experiment; thus, the flux from a metastable part of potential will be formed on the 
time-scale of such experiment only by transitions which do not follow $2$ and just such transitions possess interesting 
features which form the main subject of this paper,
2)  the ratio between a maximal magnitude of oscillations of action (see below) and the Arrhenius factor (just the 
difference of energies $U_{S_2}-U_1$) is in this case the largest possible, as it will be shown below,
3)  the deeper the well $2$ the larger a range in which action varies as friction varies from small to large values.}:

\begin {equation}
\mu \ll 1 .
\end{equation}

In this case, a variation of $I$ within the region of integration (3.78), $[U_{S_1}, U_{S_2}]$, is 
small and therefore $\tilde\Gamma^{\prime}(E)$ may be considered approximately as a 
constant, from which it immediately follows that, at $\Gamma<\Gamma_1$, 
 
\begin {equation}
\tilde \Gamma^{\prime}(E)=\Gamma_{\pm}\equiv \Gamma_{n_{rel}-(1\pm 1)/2}
\end{equation}

\noindent 
(corresponding to  $A=A_{\pm}$ respectively) where $n_{rel}$ is given by (3.71) (equivalently, $\Gamma_{\pm}$ may be found from (3.68)-(3.70)). Substituting (3.81) into (3.72), we obtain

\begin {equation}
\Delta S_{\rm min}=\frac {U_{S_2}-U_{S_1}}{4}
\rm {min} (\frac{(\Gamma-\Gamma_+)^2}{\Gamma\Gamma_+}, \frac{(\Gamma-\Gamma_-)^2}{\Gamma\Gamma_-}), \quad\quad \Gamma<\Gamma_1, \quad\quad \mu \ll 1 .
\end{equation}

Similarly, at $\Gamma>\Gamma_1$ ($A_+$ does not exist in this range of $\Gamma$), 

\begin {equation}
\Delta S_{\rm min}=\frac {U_{S_2}-U_{S_1}}{4}
\frac{(\Gamma-\Gamma_-)^2}{\Gamma\Gamma_-},\quad\quad
\Gamma>\Gamma_1, \quad\quad \mu \ll 1 .
\end{equation}

If, apart from the constant (zero-order) term, we took into account in  the Taylor expansion of 
$\tilde\Gamma^{\prime}(E)$ a next term we would obtain corrections to $\Delta S_{\rm 
min}$ (3.82) and (3.83) of the order of $\mu^2$ and $\mu^3/\beta$ respectively. Taking into account that an accuracy of the averaging 
method is of the order of $\beta$ (3.61), the overal inaccuracy of (3.82), (3.83) is

\begin {equation}
r \sim {\rm max} (\beta, \mu^2).
\end{equation}

For the potential (3.58), an asymptote (3.82), (3.83) is shown in Fig.6 by the dotted line: it well approximates the exact action within the accuracy (3.84).

In addition, I present explicit expressions for constants $A_+$, $A_-$ which have an accuracy $\sim {\rm max} (\beta, \mu)$ and will be used 
further in an explicit formula for the $MPDTP$: 

\begin {equation}
A_{\stackrel{+}{-}}  \approx  \frac{\Gamma_{\stackrel{+}{-}}-\Gamma}
{\Gamma_{\stackrel{+}{-}}+\Gamma } \quad ,
\quad\quad
\quad
\mu  \ll  1.
\end {equation}

\vskip 0.2truecm

\noindent
2). In the case when neither of wells is deep i.e.

\begin {equation}
\mu \stackrel{>}{\sim} 1 ,
\end{equation}

\noindent
the condition of small friction (3.61) is equivalent to the condition of a large $n_{rel}$,

\begin {equation}
1 \ll n_{rel} \sim \frac {\omega_{osc}}{\Gamma} {\rm ln}\frac{U_{S_2}-{\rm min}(U_1,U_2)} {U_{S_1}-{\rm min}(U_1,U_2)}.
\end{equation}

Then, as it is obvious intuitively and will be confirmed by a result, a deviation of 
$\tilde\Gamma^{\prime}$ from $\Gamma$ in the relevant range of energies is small so that $|A_{\pm}|$ is small too,

\begin {equation}
|A_{\pm}| \ll 1 .
\end{equation}

Correspondingly, we may approximate $\tilde\Gamma^{\prime}/\Gamma$ (3.77) by the 
expression

\begin {equation}
\frac{\tilde\Gamma^{\prime}}{\Gamma }\approx
1+2\frac{I(U_{S_2})}{I}A_{\pm} , \quad\quad\quad
2\frac{I(U_{S_2})}{I}A_{\pm} \ll1 
\end{equation}

\noindent
(the latter inequality will be checked after the result for $A_{\pm}$ is obtained). 

Similarly, if to expand the integrand in (3.78) into a Taylor series over $A_{\pm}$ and omit all 
powers higher than the first one then we can easily express $A_{\pm}$  from the resulting 
approximation of (3.78) as

\begin {eqnarray}
A_{\pm} =  \frac{\Gamma}{2I(U_{S_2}) \Delta \omega_{osc}^{\prime}}\{
\frac{\Delta \omega_{osc}}{\Gamma}-(n_{rel}-1 \mp \frac{1}{1+\frac{I_1}{I_2}}
)\} & , & \\
\Delta \omega_{osc}^{\prime} & \equiv & \frac{1}{\pi} \int_{U_{S_1}}^{ U_{S_2}}dE \frac {1}{I^2}, \nonumber
\\
n_{rel} & \gg & 1.
\nonumber
\end{eqnarray}

\noindent The expression in braces does not exceed 1 and 
therefore, allowing for (3.86),

\begin {equation}
A_{\pm} \stackrel {<}{\sim} \frac{\Gamma}{\omega_{osc}} \frac{ I(U_{S_1})}{ I(U_{S_2})} \ll 1.
\end{equation}

\noindent Thus, both the condition (3.88) and the inequality in (3.89) are confirmed.

Substituting $A_{\pm}$ (3.90) into $\tilde\Gamma^{\prime}$ (3.89) and the resulting $\tilde\Gamma^{\prime}$
into $\Delta S_{\rm min} $ (3.72), keeping only the lower power of $n_{rel}^{-1}\propto \Gamma$, integrating 
the resulting integrand and choosing a minimum between the resulting actions for $A_-$ and $A_+$, we obtain

\begin {equation}
\Delta S_{\rm min}=\frac{\pi \Gamma^2}{4\Delta\omega_{osc}^{\prime}}
(|\frac{\Delta \omega_{osc}}{\Gamma}-n_{rel}+1|-\frac{1}{1+\frac{I_1}{I_2}})^2,
\quad\quad
n_{rel} \gg 1.
\end{equation}

\noindent An inaccuracy of (3.91)

\begin {equation}
r\sim {\rm max}(\beta, \frac{\beta}{\mu}).
\end{equation}

If both the condition of deep wells (3.80) and the condition of large number 
of passages (3.87) are satisfied i.e. if $\beta\ll\mu \ll1$ the expressions for $A_{\pm}$, (3.85) and 
(3.90),  and for action, (3.82) and (3.92), give identical results up to the leading terms (a relative 
difference due to higer-order terms is $\sim {\rm max}(n_{rel}^{-1}, \mu^2)$) and this provides a 
\lq\lq bridge" between the results for the cases of deep and non-deep wells. Moreover, accuracies of asymptotic expressions also can be matched on the boundary between the cases, namely at 

\begin {equation}
\beta\ll\mu^2 \ll1.
\end{equation}

\noindent
Thus, keeping at the treatment of the case (3.86) a next term in all relevant Taylor expansions 
on $\Gamma$, we derive a more accurate formula for $\Delta S_{\rm min}$:

\begin {eqnarray}
\Delta S_{\rm min}={\rm min}(\Delta S_{\rm min}^{(+)}, \Delta S_{\rm min}^{(-)}), 
\quad\quad
\Delta S_{\rm min}^{(\pm)}\equiv \frac{\pi \Gamma^2}{4\Delta\omega_{osc}^{\prime}}
\delta_{\pm}^2 (1+3\frac{ \Gamma\Delta\omega_{osc}^{\prime\prime}}{(\Delta\omega_{osc}^{\prime})^2}\delta_{\pm}), \\
\delta_{\pm}=\frac{\Delta \omega_{osc}}{\Gamma}-n_{rel}+1 \pm \frac{1}{1+\frac{I_1}{I_2}}, 
\quad\quad
 \Delta\omega_{osc}^{\prime\prime}\equiv \frac{1}{\pi} \int_{U_{S_1}}^{ U_{S_2}}dE \frac {1}{I^3}, 
\quad\quad
n_{rel} \gg 1.
\nonumber
\end{eqnarray}

\noindent An inaccuracy of (3.95)

\begin {equation}
r\sim {\rm max}(\beta, \frac{\beta^2}{\mu^2}) \sim \beta.
\end{equation}

In the case $\mu\ll 1$, in order to keep an inaccuracy on the minimal possible level (i.e. $\sim \beta$) at any $\Gamma$ 
from the underdamped range (3.61) one may use the formula (3.95) at $\beta\stackrel{<}{\sim}\mu^2$ while the formula 
(3.82) may be used at $\beta\stackrel{>}{\sim}\mu^2$.

\vskip 0.5truecm

Finally in this sub-section, let us show that corrections determined by close vicinities of smooth maxima $S_1$ and $S_2$ are exponentially small in the underdamped case.

Let us first consider a contribution into action from a bit of a trajectory 
$S_2\stackrel{relA_{\pm}}{\rightarrow}S_1$ close to the upper saddle, $S_2$. In principle, one could calculate this contribution 
explicitly. Indeed, in order to find a time dependence of energy on the trajectory, we can use the 
original (i.e. non-averaged) equation (3.34) for $dE/dt$ while $\dot q$ in the left-hand side of 
this equation can be calculated in the dissipationless approximation in which, besides, the 
potential $U(q)$ may be approximated by the inverted parabola,

\begin {equation}
U(q) \approx -\frac{1}{2}\omega_{S_2}^2 (q-q(S_2))^2 ,\quad\quad\quad
|q-q(S_2)| \ll q(S_2)-q(S_1) .
\end{equation}

However, in comparison with an inaccuracy of the averaging method ($\sim \beta$), the explicit account of the contribution from the 
vicinity of $S_2$ makes no sense because the contribution from this region 
into action is exponentially small.
Indeed, on the first 
passage, the approximation (3.73)-(3.77) (based on the averaging method) is valid starting from such point $P$ on the 
trajectory that a time of relaxation from this point 
to the lower end of the first passage is much less than $\Gamma^{-1}$. 
We may estimate roughly this time as $\Gamma^{-1}$. Then, one can easily show, using 
the dissipationless approximation, that a coordinate of this point is exponentially close to the 
saddle:

\begin {equation}
q(S_2)-q(P) \sim (q(S_2)-q(S_1)) {\rm e}^{-\sigma \frac{\omega_{S_2}}{\Gamma}},\quad\quad\quad
\sigma\sim 1.
\end{equation}

Correspondingly,

\begin {equation}
E(S_2)-E(P) \sim (U_{S_2}-U_2)\frac{\Gamma}{\omega_{S_2}} {\rm e}^{-2\sigma \frac{\omega_{ 
S_2 }}{\Gamma}},\quad\quad\quad
\sigma\sim 1.
\end{equation}

Taken that $\omega_{S_2}\sim \omega_{osc}$, 
a contribution of the vicinity of $S_2$ into action is 
exponentially small and may be omitted within accuracies of formulas (3.82), (3.92), (3.95).

As concerns the analysis of the contribution from the discussed vicinity of the lower saddle, $S_1$, it is 
more complicated and involves, in particular, an analysis of the singularity of the trajectory (c.f. 
the Appendix D) but the ultimate conclusion is the same: the contribution from the vicinity of 
$S_1$ is exponentially small and may be omitted.

\vskip 0.2truecm

\noindent
{\bf (c). Most probable direct transition path}

\vskip 0.2truecm

Let us derive an explicit expression for the $MPDTP$ in the underdamped regime.

If to make in the dynamical equation (3.23) the transformation of variables from the 
coordinate-velocity $q$-$\dot {q}$ to action-angle $I$-$\psi$ we shall derive such dynamical 
equations \cite{soskin89}

\begin {equation}
\dot {I}  =  -\frac {\Gamma^{\prime}}{\omega} \dot{q}^2, \quad\quad
\dot {\psi}  =  \omega + \Gamma^{\prime}\dot {q}\frac{\partial q}{\partial I},
\end{equation}

\noindent where

\begin {equation}
q \equiv q(I, \psi)
\end{equation}

\noindent is coordinate as a function of a mechanical action and of an angle: it is periodic on 
$\psi$ with a period $2 \pi$ and its concrete form depends on a concrete shape of a potential 
$U(q)$.

The averaging transforms the first of eqs.(3.100) into the eq.(3.74) while the another equation is 
transformed into 

\begin {equation}
\dot {\psi} =  \omega .
\end{equation}

\noindent The latter equation is readily integrated:

\begin {equation}
\psi \equiv \psi (t)  =  \psi_0 +\int_0^t d \tau \omega (I(\tau)), \\
\end{equation}

\noindent where $I(t)$ is given in (3.75) while $\psi_0 $, together with $I(U_{S_2})$, corresponds  nearly to $S_2$,

\begin {equation}
q(I(U_{S_2}), \psi_0) \approx q(S_2)
\end{equation}

\noindent (the expression \lq\lq nearly $S_2$" means here a state $P$ from (3.98), (3.99)).

Thus, it follows from (3.23), (3.75), (3.103), (3.104) that 

\begin {equation}
q_{opt}(t) \approx q(I(t-t_{rel}), \psi (t-t_{rel})), \quad\quad
0 \leq t \leq t_{rel} ,
\nonumber
\end{equation}

\noindent where functions $q$, $I$ and $\psi$ 
are given in the eqs. (3.101), (3.75), (3.103) and (3.104) respectively while $ t_{rel}$ corresponds to a relaxation (3.75) from $I(U_{S_2})$ to 
$I(U_{S_1})$:

\begin {equation}
t_{rel}= \frac{1}{\Gamma} {\rm ln}\frac{B-\sqrt {B^2-A^2}}{A^2} , \quad\quad
B \equiv A+ \frac{I(U_{S_1})(1-A)^2}{2 I(U_{S_2})},
\end{equation}

\noindent and $A$ is to be chosen between $A_{\pm}$ (see (3.85) and (3.90) for cases of 
deep and non-deep wells respectively) dependently on which of them provides a smaller action. 

The formula (3.105) describes correctly nearly the whole $MPDTP$ except the exponentially small 
vicinities of the saddles.

\chapter {Discussion and applications}

In this section, I discuss briefly a 
few connected with the present work items as well as applications whose detailed 
analysis is supposed to be done elsewhere.

{\bf 1.}

\noindent
Let us demonstrate how results of Sec.3.3 can be immediately applied to the problem of 
inter-attractor transitions in a 3-well stable potential (c.f. Figs.2,3).
First of all, we note that $MPDTP$ for a transition {\it attractor-attractor} necessarily follows a 
saddle from which a system can relax noise-free to a final attractor, so that the problem is 
reduced to the transition {\it attractor-saddle}. If the saddle belongs to a basin of attraction of 
the attractor then the problem is trivial: the $MPDTP$ is just a path time-reverse to a relaxational 
trajectory while action is just a difference of energies in the saddle and in the attractor. If the saddle (let us call it $S_2$) 
does not belong to a basin of attraction of the attractor 
(let us call it $1$) then the problem is closely related to that one considered in Sec.3.3 which, in its 
turn, is reduced to the transition $S_1 \rightarrow S_2$. However, in Sec.3.3, we considered only 
the case when neither of two noise-free trajectories emanating from $S_1$ reaches a coordinate 
$q(S_2)$. Generally speaking, it may be not so (if $U_{S_1}>U_{S_2}$ while friction is 
small). But, still, the problem is easily reduced to the case considered in Sec.3.3: due to the property 
of detailed balance, the $MPDTP$ $S_1 \rightarrow S_2$ is just time-reverse to the $MPDTP$ $S_2 
\rightarrow S_1$ while if the transition $S_1 \rightarrow S_2$ does not satisfy the above mentioned restriction then the 
transition $S_2 \rightarrow S_1$ necessarily satisfies an analogous 
restriction required for it to be described within the case considered in 
Sec.3.3 (results of Sec.3.3 for a 
saddle-saddle transition are applied to the transition $S_2 \rightarrow S_1$ if
to exchange notations: $S_1$ by $S_2$ and vice versa); corresponding actions differ merely  by a 
factor $U_{S_1}-U_{S_2}$.

{\bf 2.}

\noindent
Global bifurcations in a dynamical system ({\it saddle connections}) play a crucial role for
 various characteristics of fluctuational transitions.
Let us illustrate this for a system which possesses three attractors.
Assume that a basin of attraction of  an attractor $1$ possesses at a given friction only one saddle,
$S_1$. If to vary friction the connection to another saddle,
$S_1 \stackrel{rel}{\rightarrow} S_2$,
occurs at some friction which marks a change of a saddle via which the escape from an attractor $1$ to an attractor $3$
takes place (c.f. Fig.3) and a corresponding switch of the {\it most probable 
transition route} ($MPTR$) (c.f. $R$ in (2.26) and (2.30)). The
connection $S_1 \stackrel{rel}{\rightarrow} S_2$
leads also to sharp changes of the flux (c.f. (2.25) and
(2.29)) and of 
the {\it mean first passage time} ($MFPT$) (c.f. (2.26) and (2.31))
\footnote {Obviously, a crucial for fluctuational transitions role of saddle connections like those described above was found before in
related problems: see e.g. \cite{meln85} where, in particular, a strong
sensitivity of a probability of a multi-well jump in a tilted cosine
potential to a tilt when it is close to the threshold for an onset of
the running solution has been demonstrated.}.

In potential systems, reverse saddle connections, $S_2 \stackrel{rel}{\rightarrow}  S_1$, also 
play an important role for fluctuational transitions. 
They mark switches of the $MPTR$: the 
ultimate transition occurs most probable from that potential well to
which the trajectory emanating from $S_2$ relaxes (c.f. (3.15) and (3.16)). They also mark 
characteristic  changes of the initial transition flux ($\alpha_{13}$) which is well demonstrated 
by the inset in Fig.6: critical values $\Gamma_{n\geq 1}$ which correspond just to saddle 
connections $S_2\stackrel{rel}{\rightarrow}  S_1$ separate ranges $]\Gamma_{2m+2}, 
\Gamma_{2m+1}[$ in which $MPDTP$ $1 \rightarrow S_2$ is time-reverse to the noise-free 
trajectory $S_2\stackrel{rel}{\rightarrow} 1$ while action is equal just to the difference of 
energies $U_{S_2}-U_1$ from ranges $]\Gamma_{2m+3}, \Gamma_{2m+2}[$ in which $MPDTP$ necessarily follows $S_1$ while action exceeds  $U_{S_2}-U_1$. A maximum of each oscillation 
corresponds approximately to a largest distance from $\Gamma$ to the nearest to it critical 
value $\Gamma_n$. If to number oscillations from the right then, as it follows from (3.92), 
(3.69), (3.68),  a magnitude of an $m$th oscillation decreases at large $m$ by a quadratic law:

\begin {equation}
\Delta S_{\rm min}^{({\rm max})}(m)= \frac{1}{ m^2}\frac{\pi \Delta\omega_{osc} ^2}{16\Delta\omega_{osc}^{\prime} (1+\frac{I_1}{I_2})^2}, \quad\quad
m \gg 1. 
\end{equation}

The largest oscillation is the first one from the right i.e. in the range $]\Gamma_3, \Gamma_2[$ 
(c.f. Fig.6) for the case when $s=+1$ i.e. when $q(2)$ is between $q(1)$ and $q(S_2)$ (c.f. Fig.4) 
or in the range $]\Gamma_2, \Gamma_1[$ otherwise (i.e. when $s=-1$).
Its magnitude has the maximal value when a depth of the well $2$ is much larger both than a 
depth of the well $1$ and than a difference of energies in saddles since, just in this case, a 
relative deviation of friction $\Gamma$ from the nearest value providing a saddle connection $S_2 
\stackrel{rel}{\rightarrow}  S_1$ is the largest. It is easy to derive from (3.82), (3.81) that

\begin {eqnarray}
\Delta S_{\rm min}^{({\rm max})}(m=1)\equiv \Delta S_{\rm min}(\Gamma^{({\rm max})}) 
= (U_{S_2}-U_{S_1}) \frac{\sqrt{a}+\frac{1}{\sqrt{a}}-2}{4},  
\\
a\equiv \sqrt{\frac{\Gamma_{(1-s)/2+1}}{\Gamma_{(1-s)/2+2}}}, \quad\quad
\Gamma^{({\rm max})}\equiv  \sqrt{\Gamma_{(1-s)/2+1}
\Gamma_{(1-s)/2+2}
}, \quad\quad 
U_{S_2}-U_2  \gg  U_{S_2}-U_1. 
\nonumber
\end{eqnarray}

It is interesting to note that $\Delta S_{\rm min}^{({\rm max})}(m=1)$ differs drastically for 
the cases $s=-1$ and $s=+1$. It is demonstrated easier of all for the case

\begin {equation}
U_{S_2}-U_2 \quad \gg \quad U_{S_2}-U_1 \quad \gg  \quad
U_{S_2}-U_{S_1}\quad,  
\end{equation}

\noindent when $a$ in (4.2) can be found explicitly: $a=3$ if $s=-1$ and $a\approx 1+I_2/I_1 
\gg 1$ if $s=+1$. Correspondingly,

\begin {equation}
\Delta S_{\rm min}^{({\rm max})}(m=1) \approx (U_{S_2}- U_{S_1}) \{ ^{0.077\quad\quad\quad\quad {\rm at} \quad s=-1}_{0.25\sqrt{I_2/I_1}\stackrel{>}{\sim}1\quad {\rm at} \quad s=+1} \quad .
\end{equation}

At small enough temperatures, oscillations of action lead to exponentially strong oscillations of an
initial flux. Note however that, in the case $s=-1$, required for this temperatures should be very 
small, as obvious from (4.4), while, in the case $s=+1$, the first oscillation is much stronger 
and if $I_2/I_1$ is large enough then $\Delta S_{\rm min}^{({\rm max})}(m=1) $ may be much 
larger than the Arrhenius factor $U_{S_2}-U_1$ so that the oscillation of an initial flux is huge 
at any temperature from the relevant for the Kramers 
problem range $T\ll U_{S_2}-U_1$.

As concerns the case $s=-1$, a major variation of action occurs 
(monotonously) in the range $[\Gamma_1, \Gamma_0]$: $\Delta S_{\rm min}$ varies from $0$ 
to $U_{S_1}-U_2$  (see Fig.6) and, if (4.3) holds, the variation of action is much larger both 
than the Arrhenius factor and than the first oscillation  (4.4).

I emphasise that, since the Kramers paper \cite {kramers}, all works on
the escape from a metastable potential (or on transition rates in a stable potential) considered only power-like
dependences of the escape rate (flux) on friction\footnote {To the best
of my knowledge, there was only one work, \cite{meln85} (reproduced
also in the review \cite{meln}), in which some indirect and inexplicit
evidence of a strong dependence of transition rates in a multi-well
potential on friction was contained. However that work considered (by
the method very different from mine) only some very particular case: an
underdamped motion in a slightly tilted cosine potential close to the
threshold for an onset of the running solution. And even for that
particular case, the resulting expressions for rates of multi-well jumps
were not analysed in details, while the Kramers problem for an escape
from a multi-well metastable potential was not considered at all.},
while we have demonstrated that, in a multi-well metastable potential,
the dependence of the initial flux on friction is exponentially strong,
including in particular exponentially strong oscillations.

Unlike potential systems, in non-potential or periodically-driven
systems, 
switches of $MPTR$s and sharp changes in a flux and $MFPT$ should not, 
generally, be associated with saddle connections $S_2 \stackrel{rel}{\rightarrow} S_1$:
the 
detailed balance does not hold in such systems \cite {fpe} so that most probable fluctuational paths are no longer
associated 
with time-reverse relaxational paths and, hence, bifurcations of the
latter do not 
give rise to bifurcations of the fluctuational paths. Preliminary
results for 
periodically driven {\it zero-dispersion} \cite {soskin89},\cite {soskin94},\cite
{prl96} systems 
do confirm this \cite {global}: a trajectory time-reverse to $MPDTP$ in this multi-attractor system intersects a
boundary of 
a basin of attraction of an initial attractor in a point which is not $S_1$, so 
that saddle connections $S_2 \stackrel{rel}{\rightarrow} S_1$ may not lead to bifurcations of $MPDTPs$ (but reverse connections, $S_1 \stackrel{rel}{\rightarrow} S_2$, are still relevant).

{\bf 3.}

\noindent
One more item which I did not touch so far but which should be
discussed, at 
least briefly, concerns most probable transition paths with a given
(rather than 
optimal) time of the transition. Such paths may be necessary for example
for a 
calculation of a flux from any (both single- and multi-well) metastable
potential at the very initial stage (before the quasi-stationarity
within the initial well is formed). They are necessary also
for a calculation of tails of a {\it prehistory probability density}
\cite {prehistory}, 
\cite {nonstationary}. In \cite {nonstationary}, such paths\footnote
{They are 
called {\it nonstationary optimal paths} in \cite {nonstationary}.}
within one 
well of the overdamped double-well Duffing oscillator were considered 
numerically. I note that the method described in the Sec.3.3 can 
provide more explicit (rather than purely numerical) solutions of the
Euler-Poisson equations, in much more general case: for an arbitrary
potential and 
an arbitrary friction. With this aim, one should choose the type (3) of
the 
solution (3.26) of the Euler-Poisson equation, unlike most probable direct
transition 
paths considered in the present paper which correspond to the type (1) in
(3.26). 
Indeed, if a time of the transition is fixed by us the minimization of
action 
over a transition time (3.21) is not to be done and the condition
(3.22), for a 
zero quasienergy, is no longer valid. 
In order to provide a
given 
transition time it is necessary to fit an integration constant $C$ in the
solution (3) in 
(3.26).
Unlike the escape problem, the 
problem with a given transition time reduces to three rather than two
dynamic 
equations which, generally speaking, may display chaos, especially taking into account a
singularity in the dynamic equation for $\Gamma^{\prime}$, at turning
points of $q_{opt}(t)$. It would be very interesting to test this intriguing conjecture.

{\bf 4.}

\noindent
One more promising application of the results of the present paper is an
optimal 
control of fluctuations. As it was shown in \cite{sd}, a deterministic 
field which is to be applied in order for either to enhance or to suppress a
given 
fluctuational transition  optimally can be expressed explicitly via the most
probable 
transition path at the absence of an optimal field, at quite general
conditions. 
The papers \cite {sd}, \cite {vr} dealed with transitions within one and
the same 
basin of attraction (more exactly, from an attractor to a saddle). The
general 
approach of the present paper based on master equations describing
multiple 
returns between attractors in a multi-attractor system may provide a 
generalization of the methods of \cite {sd}, \cite {vr} for a case of 
multi-attractor systems. Besides, the exact solution of the variational
problem in a 
multi-well potential system may be directly used in order to find the
optimal 
field in such system.

{\bf 5.}

\noindent
The next item which I shall discuss in this section concerns a
noise-induced 
unidirectional motion in periodic potentials \cite {magnasco} - \cite
{drsv}. The 
effect was originally considered for potentials asymmetric within the
period 
(\lq\lq ratchets") \cite {magnasco} - \cite {doering1}. It may arise
also in 
symmetric potentials (e.g. \cite {chialvo}) and in periodically driven
systems 
which lack spatio-temporal symmetry \cite {drsv}. If a periodic
potential has 
more than one well within the period then the consideration should be
similar to 
that one which was developed in this paper, using master equations
governing a 
dynamics of first-order conditional populations of wells within one
period of the potential. 
This could be especially relevant to the resonant directed diffusion in 
non-adiabatically 
driven zero-dispersion systems. It was found recently \cite {drsv} 
that the directed diffusion in periodic potentials driven by a
non-adiabatic 
periodic force was enhanced significantly if the frequency of the force
was close 
to the frequency of eigenoscillation in the potential at such energy
which 
corresponds to the minimal absolute value of the derivative on energy 
$|d\omega (E)/dE|$. It obviously follows from results of \cite {drsv}
that the 
most strong enhancement should be expected for {\it zero-dispersion}
 systems: in such systems, $|d\omega (E)/dE|$
possesses 
a zero(s) at some energy(ies) \cite {soskin89},\cite {soskin94},\cite
{prl96}. Periodic potentials which possess the 
zero-dispersion 
property have typically two or more barriers of different heights 
within each period \cite 
{soskin89}, \cite {oleg} and, thus, the analysis of the directed
diffusion in 
such systems would need a use of the master equations.

{\bf 6.}

\noindent
Results of Sec.3 for the
most 
probable transition path may be important for the escape
problem in  
periodically driven multi-well potentials. Consider, for 
example, the periodic potential in Fig.7. At an absence of the driving, an escape
from each 
well to an adjacent period occurs most probable via the nearest high
barrier 
(since a relaxation trajectory from the top of a higher barrier
goes into 
just adjacent wells, in our particular example). Generalizing the
results \cite 
{drsv} (c.f. eqs.(6), (7) in \cite {drsv}), one may conclude that action
along $MPDTP$ to any of barriers 
decreases at a driving by a periodic force
$F \cos (\Omega t)$  ($F>0 $) by the factor 

\begin{equation}
-\Delta S  \equiv  F |\chi(\Omega)| >0, \quad\quad 
\chi (\Omega) =  \int_{-\infty}^{\infty} dt e^{i \Omega t} \frac
{\Gamma + \Gamma^{\prime}(t)}{2 \Gamma} \dot {q}(t),
\end{equation}

\noindent where $q(t)$ and $\Gamma^{\prime}(t)$ are described by (3.23),
(3.28) with $A$ corresponding to each particular $MPDTP$. Values
$\Delta S $ differ for different {\it MPDTP}s and, generally, the sign of
the difference between those corresponding to the transition to the
nearest high barrier and to the far one may turn out opposite to the
sign of the corresponding difference in the absence of the driving. For
example, in the case shown on Fig.7, it certainly  occurs if $\Omega \ll 1$:
$\Delta S$ is larger for that {\it MPDTP} which provides a transition to
the more far high barrier (c.f. \cite {drsv}) while, in the absence of
the driving, action along this {\it MPDTP} is larger than along the
another one, due to the absence of a relaxational trajectory.
Correspondingly, at some critical value of the amplitude of the driving force,
$F_c$, two effects exactly compensate each other and the optimal path
changes jump-wise: for $F>F_c$, the escape occurs via the more far high
barrier, unlike the undriven case, and in order to calculate its
probability one should use results of Sec.3 of the present
paper.

{\bf 7.}

\noindent
Finally, I would like to draw readers' attention to possible
applications of 
results of the paper to two important objects described as potential systems.
One of them is a biased Josephson junction \cite {barone}, which is
described by 
a \lq\lq washboard" potential \cite {fpe}. This system was investigated
by many 
authors. In particular, a stationary distribution (which accounts for
escape 
rates both from the running regime and from the locked one) was
investigated 
numerically in \cite {gt} using the formulas equivalent to our (3.23),
(3.28) (note 
also the explicit expressions for the escape rates in \lq\lq underdamped
and low-bias" case \cite 
{meln85}, \cite {bbms} and expressions in quadratures (though quite
complicated)  for transition rates in the underdamped case with the tilt
close to the threshold for an onset of the resistive state \cite 
{meln85}, as well as 
various numerical studies \cite {volris}, \cite {jr}, \cite {kautz},
\cite 
{fpe}). However the problem for the transition (rather than escape)
rates at arbitrary friction and bias was not considered and results of Sec.3 can
be used for this (see also the 
discussion above).
Note also that the developed here method (slightly modified for the
application to this problem) could provide an easy and reliable
numerical procedure for a calculation of transition rates from the
running solution (as well as the associated \lq\lq nonequilibrium
potential" \cite {gt}) while the authors of \cite {gt} reported an
instability and non-reliability of numerical results obtained by their
method.

The another important relevant application concerns ionic channels \cite 
{hille}, \cite {bob}. Motion of ions in channels may be
described in 
some cases \cite {bob} as an underdamped motion in a multi-well
potential so that results of the paper may be relevant to this system.

\chapter {Summary}

\hskip 7truecm {\bf 1.}

There has been introduced (see Sec.2) a {\it splitting procedure} for a phenomenological 
treatment of inter-attractor transitions in a multi-attractor system driven by a weak noise: an 
integral fluctuational transition flux is splitted into partial ones corresponding to different 
numbers of returns from a final attractor which may occur before an ultimate transition at a 
given instant. Such splitting allows to describe a dynamics of first-passage and prehistory 
problems by certain master equations whose solutions in terms of {\it direct} inter-attractor 
transition rates $\alpha_{ij}$ are found explicitly. Examples have been analysed and
a non-triviality of some of results has been demonstrated.

\hskip 7truecm {\bf 2.}

The classical Kramers problem for the escape from a metastable single 
potential well has been generalized for a case of a {\it multi-well}
metastable 
potential. If friction does not exceed certain limit a dynamics of the escape is described by means 
of master equations mentioned above and has more than one exponentially long time-scale, unlike the
conventional case \cite {kramers}. At the smallest of these time-scales, a system transits from an 
initial well $1$ {\it directly} (i.e. not following intermediate attractors) to that saddle $S_2$ 
from which it can leave a metastable part of the potential noise-free. Thus, the escape flux at this 
stage is equal to a transition rate $\alpha_{1S_2}$. 

If $S_2$ does not belong to a basin of attraction of $1$ (c.f. Fig.4) then such transition
cannot be described by a conventional time-reverse 
relaxational path because the corresponding relaxational path just does not 
exist: $S_2 \stackrel{rel}{\not \rightarrow}1$. In order to find the {\it
most probable direct transition path} ($MPDTP$) $ 1\rightarrow S_2 $ and 
{\it action} along it (the latter determines with a logarithmic accuracy a transition rate 
$\alpha_{1S_2}$) I have found {\it direct extremals} of the variational problem for an extremum of 
action i.e. extremals which do not follow intermediate attractors (see Sec.3.2). The solution is {\it valid 
for an arbitrary potential and an arbitrary friction parameter}. 
It may consist either of a single bit of certain type (see below) or of bits of this type 
sewed together in saddles while each bit is a path time-reverse to 
the auxiliary relaxational trajectory (3.23) corresponding to a time-dependent friction (3.28) in which 
a constant $A$ is to be chosen in such a way that the relaxation from an end of the bit to its 
beginning is provided.

The described above type of a single bit in the $MPDTP$ is equivalent to that one obtained in \cite {gt} by a different 
method and in a different context. But in order to find the $MPDTP$ it is particularly important to know how to 
choose among an infinite number of direct extremals just that one which provides a {\it 
minimal} among them action. Authors of \cite {gt} considered the latter problem only for one specific 
potential, namely a tilted cosine potential. Moreover their choice of a proper extremal was based 
mostly on intuitive arguments.

In contrast with \cite {gt}, I provide a complete {\it mathematicaly rigorous} analysis of 
the discussed above transition $1\rightarrow S_2$  when just two wells of arbitrary forms are 
involved (the number of wells is restricted by us to two just for the sake of clarity and brevity 
while a generalization to a larger number of wells is straightforward). First, the $MPDTP$ follows 
a conventional {\it escape} path from an initial attractor $1$ to a saddle of its basin of 
attraction, $S_1$, i.e. a path time-reverse to the noise-free trajectory 
$S_1 \stackrel{rel}{\rightarrow}1$. The main problem was to find the bit of the $MPDTP$ 
between saddles, $S_1 \rightarrow S_2$. All possible cases are covered by {\it theorems 1-4} in 
Sec.3.3.1. A typical $MPDTP$ is shown in Fig.4. The {\it theorem 3} states that if friction 
is not less than certain critical value $\Gamma_0$ (which is typically equal to a doubled frequency of eigenoscillation in a bottom of the well $2$)
then the $MPDTP$ $S_1 \rightarrow S_2$ 
does not exist at all i.e. $\alpha_{1S_2}=0$ in this case. 

A calculation of action $S_{\rm min}$ along the $MPDTP$ is provided by a simple numerical 
procedure which is incomparably easier than a purely numerical solution of the variational 
problem for a minimum of the action functional (c.f. \cite {kautz}) and, besides, our procedure 
provides a true solution for certain while the latter procedure as well as a purely numerical search 
of the constant $A$ in (3.28) (c.f. \cite{gt}) may miss an absolute minimum. 

Thus, {\it  I have found with a logarithmic accuracy a complete 
solution of the generalized Kramers problem in all ranges of friction}
(which have been distinctly separated). Unlike most of previous
works on the escape from a metastable potential, in which only a
power-like dependence on friction was found (note the footnote 2 in Sec.4), it has been 
demonstrated in the present paper that, in the case
of a {\it multi-well} metastable potential, the {\it dependence on friction can be
exponentially strong}, at small enough temperature.

\hskip 7truecm {\bf 3.}

In the underdamped range, both $MPDTP$ and $S_{\rm min}$ have been found {\it explicitly} for an arbitrary potential (see Sec.3.3.3, c.f. also an asymptote in Fig.6).

\hskip 7truecm {\bf 4.}

Generally, global bifurcations in dynamical
systems({\it saddle connections}) drastically
influence fluctuational transitions, at
a weak 
noise added, especially in potential systems. At small enough temperature, this gives rise, in
particular, to the 
characteristic {\it exponentially strong oscillations of an initial flux} from
a multi-well metastable potential as a 
friction 
parameter varies (see an example in the inset in Fig.6, see also the item 2 in Sec.4). Maxima of 
oscillations are cusp-like which corresponds to a jump-wise switch of the $MPDTP$ (c.f. a line of 
discontinuity for a non-equilibrium potential \cite{gt} and a fluctuational separatrix for  optimal 
paths \cite{jch94}, \cite{pl94}).

\hskip 7truecm {\bf 5.}

Results for the Kramers problem, listed in items 2-4 above, can be easily generalized for the 
problem of {\it inter-attractor} transition rates in a {\it stable} multi-well potential system (see the item 1 in Sec.4 for the case of 3-well potential).

\hskip 7truecm {\bf 6.}

I have sketched applications to various other problems: short-time dynamics of large 
fluctuations, prehistory probability density, optimal control of fluctuations,
noise-induced 
transport in ratchets, escapes in a multi-well potential at a periodic
driving
and inter-attractor transitions in biased Josephson junctions and ionic channels.

\chapter {Acknowledgements}

\vskip 1.0truecm

I appreciate discussions with
M.Array\'{a}s, M.I.Dykman, I.Kh.Kaufman, D.G.Luchinsky, R.Mannella,
P.V.E.McClintock, A.McKane, V.I.Sheka, V.N.Smelyanskiy and O.M.Yevtushenko as well as an assistance of D.G.Luchinsky in a design of a numerical code used at the drawing Fig.6
and a technical assistance of G.Pickett in drawing some figures.
I express a gratitude to Lancaster University for 
a hospitality during the stay in it when a large part of this work was
done. 
The work was supported in part by the EPSRC (UK) and by the INTAS. 

\appendix
\chapter {}

The master equations which govern partial populations of the $n$th order
for the transition $1 \rightarrow 3$ (i.e. such populations which
account only for those realizations of noise at which the system being
initially at $1$ visited $3$ before a current instant $n-1$ times) are
the following at $n \geq 2$, 

\begin{eqnarray}
\frac{dW_1^{(n)}}{dt} & = & - (\alpha_{12} + \alpha_{13})W_1^{(n)} +
\alpha_{21}W_2^{(n)} + \alpha_{31}W_3^{(n)}, 
\nonumber \\
\frac{dW_2^{(n)}}{dt} & = & \alpha_{12}W_1^{(n)} - (\alpha_{21} +
\alpha_{23})W_2^{(n)} + \alpha_{32}W_3^{(n)}, \\
\frac{dW_3^{(n)}}{dt} & = & - (\alpha_{31} + \alpha_{32})W_3^{(n)} +
J^{(n-1)}, 
\nonumber \\
W_i^{(n)}(0) & = & 0, \quad \quad i=1,2,3  \nonumber
\end{eqnarray}

\noindent
where the flux of the $(n-1)$st order

\begin{equation}
J^{(n-1)} \equiv \alpha_{13} W_1^{(n-1)}+ \alpha_{23} W_2^{(n-1)}
\end{equation}

\noindent
is assumed to be some known function of time (see below).

The equation for $ W_3^{(n)} $ is separated and easily solved:

\begin{equation}
W_3^{(n)}(t)=\int_0^t d \tau J^{(n-1)}(\tau) e^{-(\alpha_{31} +
\alpha_{32})(t-\tau)}.
\end{equation}

Substituting (A.3) into (A.1), we obtain for $W_1^{(n)}$ and $
W_2^{(n)}$ a closed system of two linear inhomogeneous $1$st-order
differential equations whose solution, with the account of the initial
conditions (A.1), is the following:

\begin{eqnarray}
\lefteqn{
\stackrel{\rightarrow}{W}^{(n)} \equiv (^{W_1^{(n)}}_{W_2^{(n)}})  = 
\stackrel{\rightarrow}{W}^{(n)}_l + \stackrel{\rightarrow}{W}^{(n)}_s, 
}
\\
& & 
\stackrel{\rightarrow}{W}^{(n)}_{l,s} = \frac { ( k^{(s,l)}\alpha_{31}
-\alpha_{32})
(^1_{ k^{(l,s)}})}{k^{(s,l)} - k^{(l,s)}}
\int_0^t d \tau e^{-\frac{t-\tau}{t_{l,s}}}
\int_0^{\tau} d \tau^{\prime} e^{-\frac{\tau- \tau^{\prime} }{t_3}}
J^{(n-1)}( \tau^{\prime}) , 
\nonumber \\
& & 
k^{(l,s)}= \frac{-t_{l,s}^{-1} + \alpha_{12}+\alpha_{13}}{\alpha_{21}},
\quad \quad 
t_3= (\alpha_{31} + \alpha_{32})^{-1}, 
\nonumber 
\end{eqnarray}

\noindent
where $ t_l, t_s$ are defined in (2.7).

Taken that $J^{(1)}$ is given by (2.8),
$\stackrel{\rightarrow}{W}^{(n)} $ and $J^{(n)}$  can be found explicitly
for any $n \geq 2$ by the successive application of formulas (A.4) and
(A.2), integrating explicitly at each successive step exponential terms
in the integrand. For example, 

\begin{eqnarray}
\lefteqn{
J^{(2)}  = J^{(2)}_l + J^{(2)}_s, 
}
\\
& & 
J^{(2)}_{l,s} = \frac{(\alpha_{13}+\alpha_{23}k^{(l,s)})
(\alpha_{31}k^{(s,l)} -\alpha_{32})}{ k^{(s,l)} - k^{(l,s)} } 
\{
c_{l,s} \frac{1}{t_3^{-1} - t_{l,s}^{-1}}
[t e^{-\frac{t}{t_{l,s}}} - 
\frac{1}{t_{l,s}^{-1} - t_3^{-1}}
(
e^{-\frac{t}{t_3}} -
e^{-\frac{t}{t_{l,s}}} 
)
]
+
\nonumber \\
& & 
\quad \quad
c_{s,l} \frac{1}{t_3^{-1} - t_{s,l}^{-1}}
[\frac{1}{t_{l,s}^{-1} - t_{s,l}^{-1}}
(
e^{-\frac{t}{t_{s,l}}} -
e^{-\frac{t}{t_{l,s}}} 
)
- 
\frac{1}{t_{l,s}^{-1} - t_3^{-1}}
(
e^{-\frac{t}{t_3}} -
e^{-\frac{t}{t_{l,s}}} 
)
]
\},
\nonumber \\
& & 
\quad \quad
\quad \quad
c_l =  \alpha_{13}\alpha_{1}+\alpha_{23}\alpha_{2}, \quad\quad
c_s=\alpha_{13}-c_l,
\nonumber 
\end{eqnarray}

\noindent
where $\alpha_{1,2}$ are defined in (2.7).

Two most essential differences of higher-order partial fluxes from
$J^{(1)}$ are the following:

\noindent
1) $J^{(n \geq 2)}(0)=0$ while $J^{(1)}(0)= \alpha_{13}$ ($J^{(n \geq
2)} \propto t^2$ at small $t$);

\noindent
2) an additional time-scale $t_3$ is present in the dynamics of $J^{(n
\geq 2)}(t)$, due to that an escape from the final state of the
transition, $3$, is involved, unlike the case of $J^{(1)}(t)$; for
example, if $t_3 \gg t_{l,s}$ then just higher-order fluxes prevail over
the first-order one, at $t \sim t_3$.

In order to demonstrate more clearly a non-triviality of the above
method
let us consider briefly the simplest multi-stable system - the system
with just $2$ states.
Introducing direct transition rates $\alpha_{12}, \alpha_{21}$,
conditional populations $W_{1,2}^{(n)}$ and partial fluxes for the transition $1 \rightarrow 2$

\begin{equation}
J^{(n)}( 1 \rightarrow 2, t) \equiv \alpha_{12} W_1^{(n)},
\end{equation}

\noindent
we can write for them the following master equations:

\begin{eqnarray}
\frac{dW_1^{(1)}}{dt}= - \alpha_{12}W_1^{(1)}, \\
W_1^{(1)}(0)=1,  \nonumber
\end{eqnarray}

\begin{eqnarray}
\frac{dW_1^{(n)}}{dt} & = & - \alpha_{12} W_1^{(n)} +
\alpha_{21}W_2^{(n)},
\nonumber \\
\frac{dW_2^{(n)}}{dt} & = & - \alpha_{21}W_2^{(n)} +
\alpha_{12}W_1^{(n-1)},
\\
& & 
W_1 ^{(n)}(0)= W_2 ^{(n)}(0)=0,
\nonumber \\
& & 
n \geq 2.
\nonumber 
\end{eqnarray}

The solution of (A.7) is a well-known result

\begin{equation}
W_1^{(1)}(t) = e^{-\alpha_{12} t}.
\end{equation}

It corresponds to the flux

\begin{equation}
J^{(1)}(t) = \alpha_{12} e^{-\alpha_{12} t}
\end{equation}

Obviously, the flux (A.10) coincides with the conventional result for
the flux in the escape problem \cite{kramers}. At the same time, the
case
$n>1$ was not considered before, to the best of my knowledge. The
solution of (A.8) is

\begin{equation}
W_1^{(n)}(t) = \alpha_{12} \alpha_{21} \int_0^t d \tau \int_0^{\tau} d
\tau^{\prime} e^{\alpha_{12} (\tau-t)+ \alpha_{21} (\tau^{\prime} -
\tau)}
W_1^{(n-1)}( \tau^{\prime}).
\end{equation}

Using the solution (A.9) for $ W_1^{(1)}$, substituting it into (A.11),
performing the integration, and then repeating this as many times as
necessary,  one can obtain an explicit expression for any order of $
W_1^{(n)}$ and therefore for the corresponding flux (A.6). For example,
at $\alpha_{12} \neq \alpha_{21} $,

\begin{equation}
J^{(2)}(t) = \alpha_{21} \left(\frac {\alpha_{12} }{\alpha_{12}-
\alpha_{21} } \right)^2 ( e^{-\alpha_{21} t} + e^{-\alpha_{12} t}
(t(\alpha_{21}-\alpha_{12})-1)).
\end{equation}

\noindent
E.g. if $\alpha_{21} \ll \alpha_{12} $, then $J^{(2)}$ significantly
exceeds $J^{(1)}$ already at $t>\alpha_{12}^{-1} \ln
(\alpha_{12}/\alpha_{21})$ which reflects that fact that the state $1$
is nearly depleted at this time-scale and the only possibility for a
system to be in $1$ may be for an account of rare returns from $2$.

\chapter {}

I shall derive in this Appendix the eq. (3.25).

First of all, let us write down explicitly the partial derivatives of
$L$ (3.18), with $ q_{opt}(t) $ as an argument:

\begin{eqnarray} 
\frac {\partial L}{\partial \ddot {q}} |_{ q_{opt}(t) } 
& = & \frac{1}{2 \Gamma} (\ddot {q}_{opt} +  \Gamma \dot {q}_{opt} +
\frac{dU(q_{opt})}{d q_{opt}}), 
\nonumber \\
\frac {\partial L}{\partial \dot {q}}|_{ q_{opt}(t) } 
& = & \Gamma \frac {\partial L}{\partial \ddot {q}}|_{ q_{opt}(t) }, 
\\
\frac {\partial L}{\partial q}|_{ q_{opt}(t) } 
& = & \frac{d^2 U(q_{opt})}{d q_{opt}^2} \frac {\partial L}{\partial
\ddot {q}}|_{ q_{opt}(t) }. 
\nonumber 
\end{eqnarray}

Let us put (B.1) into the Euler-Poisson equation (3.20), exchange the
\lq\lq blind" time variable $t$ for $t_{tr}-t$, use the definition of $
q_{opt}(t) $ (3.23) i.e. exchange $ q_{opt}(t_{tr}-t)$ for $q(t)$
(3.23), and take into account after this the symbolic relation

\begin{equation} 
\frac {d \quad \quad \quad}{d(t_{tr}-t)} \equiv -\frac {d}{dt} .
\end{equation} 

Then we shall derive

\begin{eqnarray} 
\frac{d^2 U(q)}{d q^2} \eta +\Gamma \frac {d \eta}{dt} + \frac {d^2
\eta}{dt^2} = 0,
\\
\eta \equiv \frac {d^2 q}{dt^2} - \Gamma \frac {dq}{dt} + 
\frac{d U(q)}{d q},
\nonumber 
\end{eqnarray}

\noindent
where $q \equiv q(t)$ satisfies the equation of motion (3.23).

Allowing for the eq.(3.23) for $q$, the identity in (B.3) can be written as

\begin{equation} 
\eta \equiv - (\Gamma + \Gamma^{\prime}(t)) \frac {dq}{dt}.
\end{equation} 

Substituting (B.4) into the equation for $\eta$ in (B.3), allowing for 

\begin{equation} 
\frac{d^2 U(q)}{d q^2} \frac{dq}{d t} = \frac{d}{d t} (\frac{d U(q)}{d
q}),
\end{equation} 

\noindent
and using (3.23) again, we obtain

\begin{equation} 
((\Gamma^{\prime}(t))^2 - \Gamma^2 - 2 \frac{d \Gamma^{\prime}(t)}{d t})
\frac{d^2q}{dt^2} + (\Gamma^{\prime}(t) \frac{d \Gamma^{\prime}(t)}{d t}
- \frac{d^2 \Gamma^{\prime}(t)}{d t^2}) \frac{d q}{d t} = 0.
\end{equation} 

Carrying out the differentiation of $\phi$ in (3.25) explicitly, one obtains the equation identical to (B.6).

\chapter {}

The goal of this Appendix is to study a possibility to sew together different \lq\lq single-$A$" 
extremals. 
Let us consider the functional (3.18) and assume that there is an
extremal (i.e. a trajectory providing an extremum of the functional)
$q_e(t)$ which is sewed in some intermediate point of the phase space $(q_{in}, \dot {q}_{in})$ 
from trajectories of the type (3.23), (3.28) with different A. By the definition, $q_e(t)$ must 
satisfy the condition of the equality to zero of the variation of the functional:

\begin{equation}
\delta S = 0, 
\quad \quad \quad
\delta S  \equiv  \int_0^{t_{tr}} dt[\frac{\partial L}{\partial q}
\delta q +
\frac {\partial L}{\partial \dot {q}} \delta \dot {q} +
\frac {\partial L}{\partial \ddot {q}} \delta \ddot {q}].
\end{equation}

Let us divide the whole interval of integration in the integral (C.1)
for three parts:

\begin {equation}
\delta S = \int_0^{t_{in}-\Delta t}... +\int_{t_{in}-\Delta t}^{t_{in}+\Delta
t}... +\int_{t_{in}+\Delta t}^{t_{tr}}... ,
\end{equation}

\noindent
where $t_{in}$ corresponds to the intermediate point ($q_e(t_{in})=q_{in}, 
\dot{q}_e(t_{in})=\dot {q}_{in}$) while
$\Delta t$ is some arbitrary small interval. If to repeat for the
first and third integrals the same procedure as is conventionally used
\cite {elsgolc} at the derivation of the Euler-Poisson equation (3.20)
(i.e. to carry out an integration by parts twice) and to take into
account that the Euler-Poisson equation is satisfied for (3.23), (3.28)
everywhere except possibly $t_{in}$ we shall obtain \footnote{We do not
vary $t_{in}$ and $t_{tr}$ in $S$ because solutions (3.23), (3.28) necessarily
satisfy the condition (3.22) and, thus, the corresponding variations are
equal to zero.}:

\begin {eqnarray}
& \delta S & = \\
         &   & [\frac {\partial L}{\partial \ddot {q}} \delta \dot {q} 
+(\frac{\partial L}{\partial \dot {q}} - \frac {d}{dt} (\frac {\partial 
L}{\partial \ddot {q}})) \delta q] |_{t_{in}-\Delta t}
+\int_{t_{in}-\Delta t}^{t_{in}+\Delta t}... -
[\frac {\partial L}{\partial \ddot {q}} \delta \dot {q} 
+(\frac{\partial L}{\partial \dot {q}} - \frac {d}{dt} (\frac {\partial 
L}{\partial \ddot {q}})) \delta q] |_{t_{in}+\Delta t}.
\nonumber
\end{eqnarray}

If to make $\Delta t$ infinitesimal the integral in (C.3) vanishes.
Allowing also for that $q_e(t), \dot {q_e(t)}$ should be continuous 
everywhere (otherwise a random force (3.6) would become infinite) i.e.

\begin{eqnarray} 
\delta q |_{t_{in}-\Delta t} & = & \delta q |_{t_{in}+\Delta t}   \equiv  
\delta q_{in}, 
\nonumber 
\\ 
\delta \dot {q} |_{t_{in}-\Delta t} & = & \delta \dot {q} |_{t_{in}+\Delta t} 
\equiv  
\delta \dot {q}_{in}, \\
\Delta t & \rightarrow & 0,
\nonumber 
\end{eqnarray}

\noindent
we derive

\begin {eqnarray}
\delta S & = & \\
         &   & \delta \dot {q}_{in} \left[\frac {\partial L}{\partial
\ddot {q}} |_{t_{in}-\Delta t} - \frac {\partial L}{\partial \ddot {q}}
|_{t_{in}+\Delta t} \right]+ \nonumber \\
         &   & \delta q_{in} \left[\frac {\partial L}{\partial \dot {q}}
|_{t_{in}-\Delta t} - \frac {\partial L}{\partial \dot {q}} |_{t_{in}+\Delta
t}+
\frac {d}{dt} (\frac {\partial 
L}{\partial \ddot {q}}) |_{t_{in}+\Delta t}-
\frac {d}{dt} (\frac {\partial 
L}{\partial \ddot {q}}) |_{t_{in}-\Delta t} \right].
\nonumber 
\end{eqnarray}

Coordinate and velocity are varied here independently. Hence the
variation $\delta S$ is identically equal to zero only if the
expressions in the square brackets are equal to zero. With the account
of that the following equalities are satisfied on the extremal,

\begin{equation} 
\frac {\partial L}{\partial \ddot {q}}  = \frac{1}{2 \Gamma} (\Gamma
+ \Gamma^{\prime} (t_{tr}-t)) \dot {q}_{e}(t), \quad\quad\quad
\frac {\partial L}{\partial \dot {q}}
 = \Gamma \frac {\partial L}{\partial \ddot {q}}, 
\end{equation}

\noindent
and allowing for the continuity of $\dot {q}_{e}(t)$, the condition for
the equality to zero of the expressions in square brackets in (C.5) can
be written as 

\begin{eqnarray} 
& & 
 [\Gamma^{\prime} (t_{tr}-t_{in}+\Delta t)- \Gamma^{\prime}
(t_{tr}-t_{in}-\Delta t)]
\dot {q}_{e}(t_{in})  =  0, \nonumber \\
& & 
\left[\dot {\Gamma}^{\prime} (t_{tr}-t_{in}+\Delta t)- \dot
{\Gamma}^{\prime} (t_{tr}-t_{in}-\Delta t) \right] \dot {q}_{e}(t_{in}) +
\\
& & 
\quad 
\quad 
(\Gamma + \Gamma^{\prime} (t_{tr}-t_{in}-\Delta t)) \ddot
{q}_{e}(t_{in}+\Delta t)-
(\Gamma + \Gamma^{\prime} (t_{tr}-t_{in}+\Delta t)) \ddot
{q}_{e}(t_{in}-\Delta t)
= 0. 
\nonumber
\end{eqnarray}

By the original assumption, the values of $A$ in $\Gamma^{\prime}$
(3.28) are different at $ t_{in}-\Delta t $ and at $ t_{in}+\Delta t$,
which means that

\begin {equation}
\Gamma^{\prime} (t_{tr}-t_{in}+\Delta t)  \not=  \Gamma^{\prime}
(t_{tr}-t_{in}-\Delta t).
\end{equation}

\noindent
Then, the first condition in (C.7) can be written as

\begin {equation}
\dot {q}_{e}(t_{in})=0.
\end{equation}

If $dU/dq$ is continious then it follows from (C.9), (3.23) and from the continuity $q_{e}(t)$ that
$\ddot {q}_{e}(t)$ is continuous in $ t_{in}$ as well:

\begin {equation}
\ddot {q}_{e}(t_{in}-\Delta t)  =  \ddot {q}_{e}(t_{in}+\Delta t)  \equiv  
\ddot {q}_{e}(t_{in}).
\end{equation}

If $\Gamma^{\prime}
(t_{tr}-t_{in}-\Delta t)$ and $\Gamma^{\prime} (t_{tr}-t_{in}+\Delta t)  $ are finite then, 
with the account of (C.8)-(C.10), the second of the conditions (C.7) is
equivalent to 

\begin {equation}
\ddot {q}_{e}(t_{in})=0.
\end{equation}

With the account of (3.23), conditions (C.9), (C.11) are equivalent to
(3.32).

For that case when, at least for one of sewed extremals, $\Gamma^{\prime} $ turns into infinity in the 
sewing point (c.f. Appendix D) then even a satisfaction (C.11) may not provide an equality of the 
variation $\delta S$ to zero. However, it is shown in Sec.3.3.1 that, for a transition from an attractor 
corresponding to a a bottom of one of wells to a state which does not belong to its basin of attraction, 
an extremal which provides an absolute minimum of action necessarily follows the saddle of the 
basin. Thus, in this case, the $MPDTP$  is {\it necessarily} sewed in the saddle from \lq\lq single-$A$" extremals with 
different $A$.

\chapter{}
The trajectory (3.23), (3.28) is analysed below in the context of an
existence the pole $t_p$ (3.30) (correspondingly, $A$ will be assumed
positive unless it is specified otherwise).

Let us recon a time from $t_p$:

\begin {equation}
\tau \equiv t-t_p
\end{equation}

\noindent
Then, (3.28) and (3.23) can be written respectively as

\begin {equation}
\tilde \Gamma (\tau) \equiv \Gamma^{\prime}(t)= \Gamma \frac {1+e^{
\Gamma \tau}}{1-e^{ \Gamma \tau}} ,
\end{equation}

\begin{eqnarray} 
\ddot {\tilde q} +\tilde \Gamma(\tau) \dot {\tilde q} + dU(\tilde q)/d
\tilde q =0, \\
\tilde q(\tau) \equiv q(t). \quad \quad \quad \nonumber 
\end{eqnarray}

It is easy to see that 

\begin {equation}
\tilde \Gamma (-\tau) = - \tilde \Gamma (\tau),
\end{equation}

If to denote

\begin {equation}
\tilde q_- (\tau) \equiv \tilde q(-\tau) 
\end{equation}

\noindent
we shall obtain for $\tilde q_- (\tau)$, with the account of (D.4), the
same equation as for $\tilde q(\tau)$ (i.e. (D.3)). Thus, if the initial
conditions coincide i.e. if

\begin {equation}
\tilde q_- (0) = \tilde q(0), \quad\quad  \dot {\tilde q}_- (0)=\dot
{\tilde q}(0),
\end{equation}

\noindent
and if the equation (D.3) with such initial conditions has a unique
solution then the trajectory for positive $\tau$ is just time-reverse to
that one for negative $\tau$ 

\begin {equation}
\tilde q(\tau) = \tilde q(-\tau).
\end{equation}

The condition (D.6) is satisfied only if 

\begin {equation}
\dot {\tilde q}(0)=0. 
\end{equation}

The condition (D.8) is obviously satisfied \footnote{Note that the
consideration similar to (D.1)-(D.6) is valid for negative $A$ too if to
change $t_p$ by $t_0$ (3.29). However, the condition (D.6) is not
satisfied for negative $A$ unless, incidentally, $t_0$ corresponds to a
turning point. Thus, typically, the trajectory
does not follow \lq\lq back in time", at negative $A$.} because,
otherwise, $\dot {\tilde q}$ and $\ddot {\tilde q}$ would diverge at
$\tau=0$, taking into account that $\tilde \Gamma (\tau)$ diverges at
$\tau=0$:

\begin {equation}
\tilde \Gamma (\tau) \approx -\frac{2}{\tau}, \quad\quad |\tau| \ll
\frac{1}{\Gamma}.
\end{equation}

As concerns the uniqueness of the solution, it depends on whether
$q(t_p)$ is a stationary point of the original dynamic equation (with
a true $\Gamma$) or not. Let us consider these cases separately.

\noindent
1).

\begin {equation}
\frac {dU}{dq} |_{\tilde q(0)} \not= 0.
\end{equation}

Let us expand the velocity into the Taylor series,

\begin {equation}
\dot {\tilde q}(\tau)  =  a_1 \tau + a_2 \tau^2 + ..., 
\end{equation}

\noindent
and substitute it into the eq.(D.3) \footnote{Strictly speaking, one
could try also an expansion $\dot {\tilde q}(\tau) = b(\tau) (a_1 \tau +
a_2 \tau^2 + ...)$ where $b(\tau)$ is a non-analytic function such that
$\tau b(\tau) \rightarrow 0$ at $\tau \rightarrow 0$ (e.g. $b(\tau)=\ln
(\tau)$). But it is easy to show that the eq.(D.3) cannot then be
satisfied.}. Then, keeping the leading order in $\tau$, we obtain

\begin {equation}
a_1=\frac {dU}{dq} |_{\tilde q(0)}.
\end{equation}

The higher-order coefficients in (D.11) can be easily (and uniquely) 
found using (D.3), (D.9)-(D.12) as well as higher-order terms in the
Taylor expansion of $\tau \tilde 
\Gamma (\tau)$.

Thus, the trajectory is uniquely defined if (D.10) holds and, therefore,
(D.7) holds true in this case too.

It is interesting to note also that the direction from which the system
(D.3) arrives at the turning point $(\tilde q(0), \quad  \dot {\tilde
q}(0)=0)$ depends only on a sign of $dU/dq|_{\tilde q(0)}$. Thus, if the
latter is negative while $q(t=0) > q(t_p)$ then the trajectory should
necessarily have at least one more turning point before (i.e. at
$t<t_p$) in which a velocity changes the negative sign for the positive
one.

\vskip 0.5truecm

\noindent
2).

\begin {equation}
\frac {dU}{dq} |_{\tilde q(0)} = 0.
\end{equation}

At small $\tau$, we may omit then the term $dU/dq$ in the eq.(D.3)
(which will be confirmed by the result) and obtain the closed equation
for $d \tilde q / d \tau$ which is easily integrated:

\begin {equation}
\frac {d \tilde q}{d \tau}  =  C \exp (-\int d \tau \tilde \Gamma
(\tau)).
\end{equation}

\noindent
where $C$ is an arbitrary constant.

With the account of (D.9),

\begin {equation}
\frac {d \tilde q}{d \tau}  \approx  C \tau^2.
\end{equation}

Correspondingly,

\begin {equation}
\tilde q (\tau)  \approx   \frac {1}{3} C \tau^3,
\end{equation}

\noindent
from which the validity of the omission of the term $dU/dq$ in (D.3) if
(D.13) holds follows.

Taken that $C$ is arbitrary, there is an infinite number of trajectories
which satisfy both the equation of motion (D.3) and the initial
conditions (D.8), (D.13). It follows from this, in particular, that (D.7) may not hold true.

It is interesting also that, at any non-zero $C$, the trajectory (D.16) approaches
$\tilde q (0)$ (or departs from it) for a finite time.

\newpage
\centerline {\underbar {Figure Captions}}

\begin{itemize} \item[1.] A double-well metastable potential and
schematically 
shown \lq\lq direct" transitions: $1\rightarrow3$ (dotted line) and $1\rightarrow2$, $2\rightarrow1$, $2\rightarrow3$ (dashed lines).

\item[2.] Energy-coordinate plane.  The solid line shows a 3-well potential $U(q)$ (numbers 
indicate attractors, dots indicate saddles) while the dashed line shows a relaxational 
(noise-free) trajectory  emanating from the saddle $S_2$ to the left.
Figures (a) and (b) demonstrate a consequence of the saddle connection $S_2\stackrel{rel}{\rightarrow} S_1$: a 
switching (as friction varies) of an attractor a direct transition rate from which 
is determined by the Arrhenius factor.

\item[3.] Energy-coordinate plane.  The solid line shows a 3-well potential $U(q)$ (numbers 
indicate attractors, dots indicate saddles) while the dashed line shows a relaxational 
(noise-free) trajectory  emanating from $S_1$ to the right.
Figures (a) and (b) demonstrate a consequence of the saddle connection $S_2\stackrel{rel}{\rightarrow}  S_1$: a 
switching (as friction varies) of a saddle through which a fluctuational transition into the attractor $3$ occurs 
if the initially occupied attractor is $1$.

\item[4.] Energy-coordinate (a) and phase (b) planes for the generalized
Kramers 
problem. Thick solid lines 
show: (a) a potential curve $E=U(q)\equiv 0.06(q+1.5)^2 - cos(q)$ (potential of such type
describes 
a r.f. SQUID \cite {barone} - \cite {pre98}), (b) boundaries of 
basins of attraction. Dots
indicate: (a) local maxima of the potential ($S_1$ and $S_2$), an intersection $O$ of the horisontal line 
$E=U_{S_2}$ (dash-dotted line) with the potential curve $E=U(q)$, and an intersection $I$ between trajectories $S_2 \stackrel{rel}{\rightarrow}2$ and $2 \stackrel{A=0}{\rightarrow} S_1$, (b)
saddles $S_1, S_2$. Thin solid/dashed lines show the 
relaxational (for $\Gamma=0.045$) trajectories from 
$S_1/S_2$: the 
trajectory from $S_2$ goes to the well $2$. Thick dotted/dashed line 
corresponds to the auxiliary relaxational trajectory (3.23), (3.28) with
$A=A_+$ 
/ $A=A_-$. The $MPDTP$ $1 \rightarrow S_2$ can be seen in the figure (a):
it 
follows first the 
thin solid line $1 \stackrel{A=0}{\rightarrow} S_1$ and then the dotted line $S_1
\stackrel{A_+}{\rightarrow} S_2$.

\item[5.] Energy-coordinate plane for a motion in the parabolic potential 
$U({\tilde q})= {\tilde q}^2/2$ (solid line): ${\tilde q}\equiv q-q(2), \quad E\equiv (d{\tilde 
q}/d \tau)^2/2+ U({\tilde q})$. (a). The dashed lines show asymptotic noise-free trajectories (as well 
as time-reverse to them trajectories) at 
$\Gamma=2\Omega$ and, at the same time, indicate lines ${\tilde q}/ \dot {\tilde q} =\pm1$. The dotted line shows an example of a path (3.45) 
which possesses a turning point to the left from the bottom of the well (only that part which 
corresponds to a motion preceding the turning point is shown since the motion following the 
turning point is not relevant).
(b). The lines indicating the sewing 
conditions for $\Gamma/(2\Omega)=1$  (${\tilde q}/ \dot {\tilde q} =\pm1$) and for 
$\Gamma/(2\Omega)=\alpha=1.5>1$ (${\tilde q}/ \dot {\tilde q} =\pm\alpha$) are 
shown by respectively thick and thin dashed lines. Examples of paths (3.54) with identical initial 
conditions but different $\Gamma$ are shown by dotted lines: by the thick line for 
$\Gamma/(2\Omega)=1$ and by the thin one for $\Gamma/(2\Omega)=\alpha$.

\item[6.] A dependence of an excess of action over the difference of energies, for an escape from 
$1$ to $S_2$ in the potential shown in Fig.4. The solid line is calculated numerically by (3.59) in which $q_{opt}$ and 
$\Gamma^{\prime}(t)$ are calculated by the algorithm described in Sec.3.3.1. 
The dotted line shows the explicit underdamped asymptote (3.82), (3.83). The horisontal and vertical  
dashed lines indicate respectively the upper limit for $\Delta S_{\rm min}$ and the lower limit for 
$\Gamma$ at which the $MPDTP$ $1 \rightarrow S_2$ does not exist. The 
inset shows an underdamped range in an enlarged scale.

\item[7.] An example of a symmetric periodic potential (solid line) with more than
one well 
within a period. Dashed lines show relaxational trajectories from higher 
barriers (indicated by dots). An escape to an adjacent period occurs
most probable, for this 
concrete example, via the {\it nearest} high barrier, following the
time-reverse 
relaxational trajectory. If to apply a periodic force the most probable
escape 
path may switch to the \lq\lq {\it more far}" high barrier, thus,
changing the direction of an initial noise-induced flux for the
opposite one.

\end{itemize} 

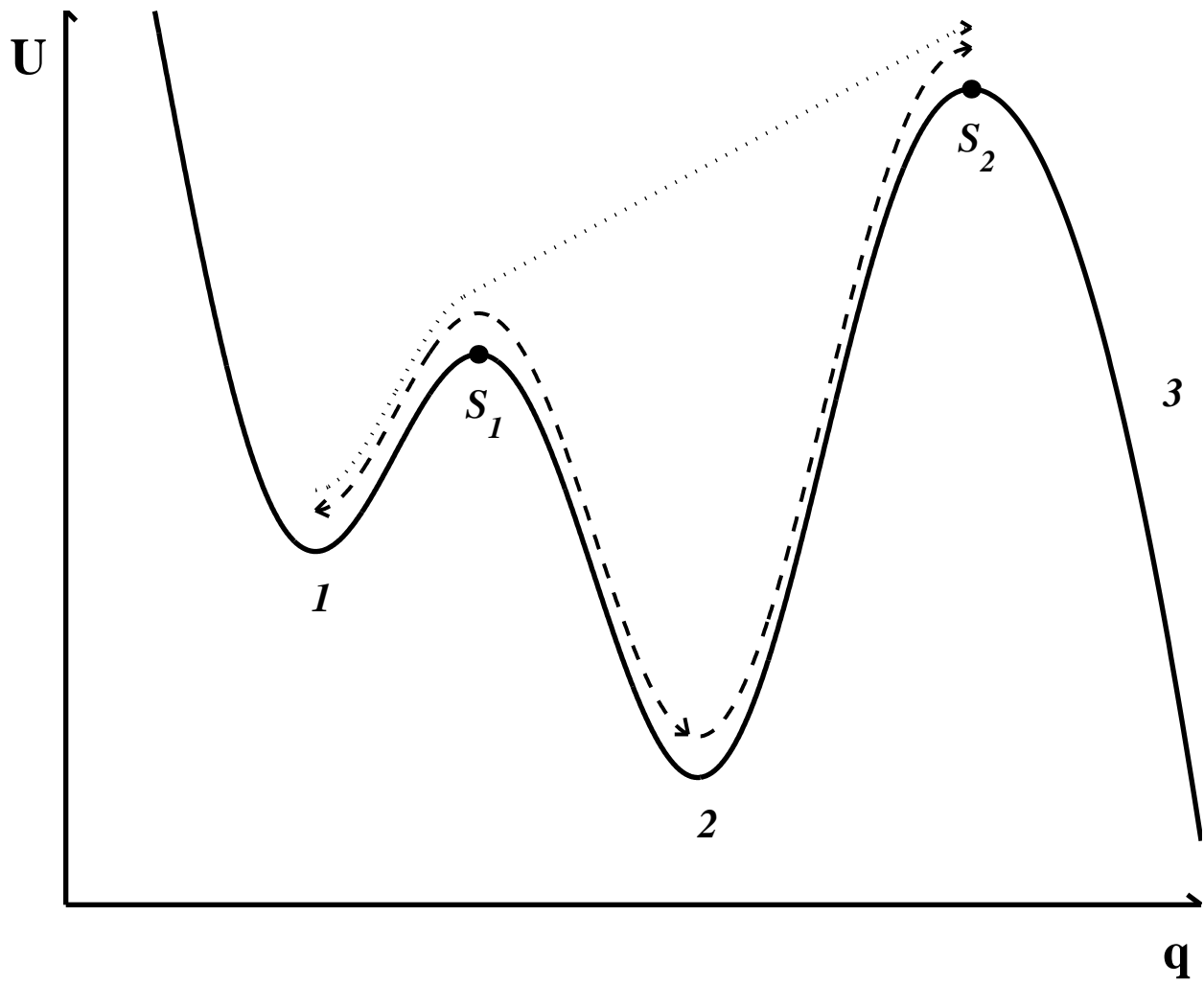
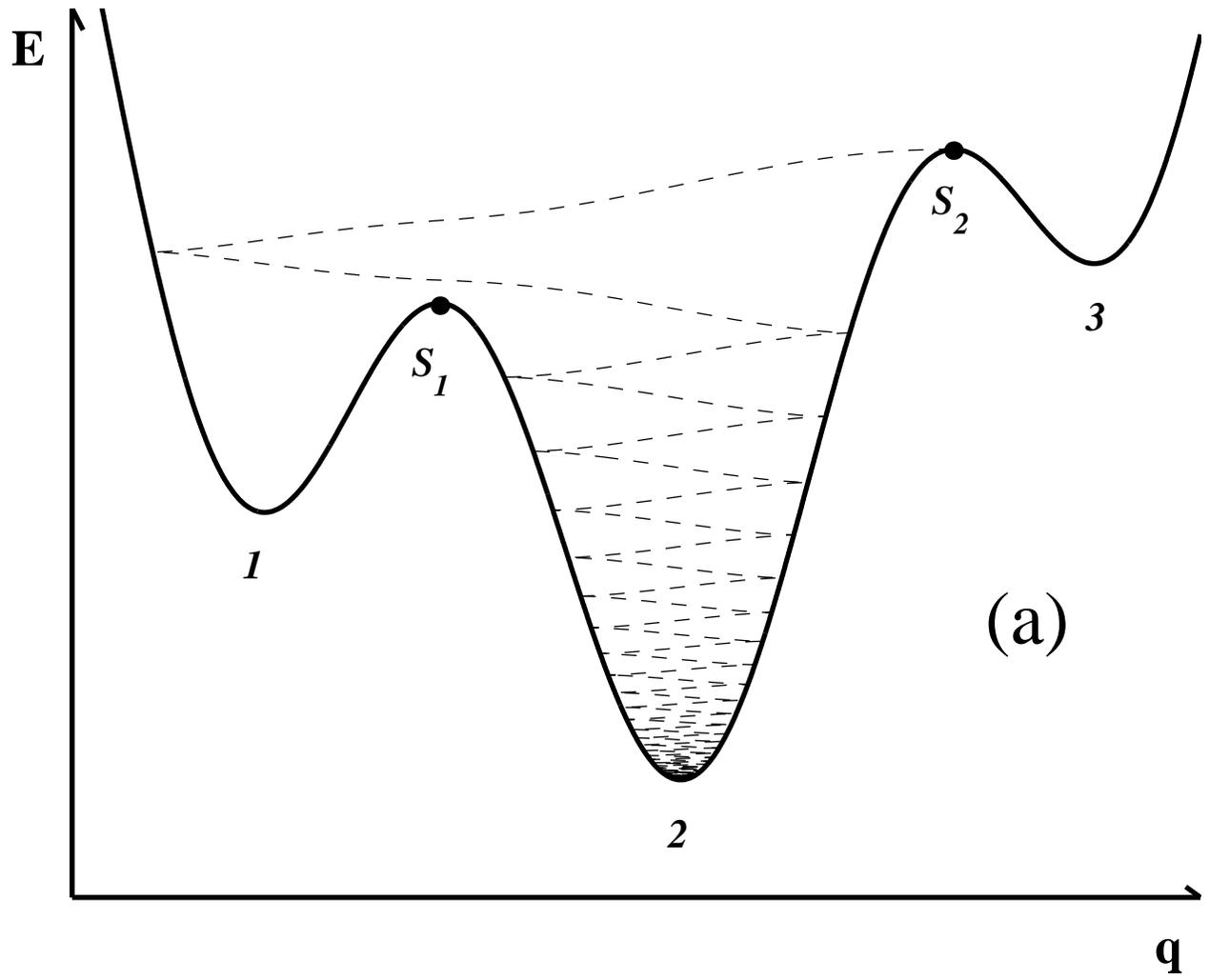
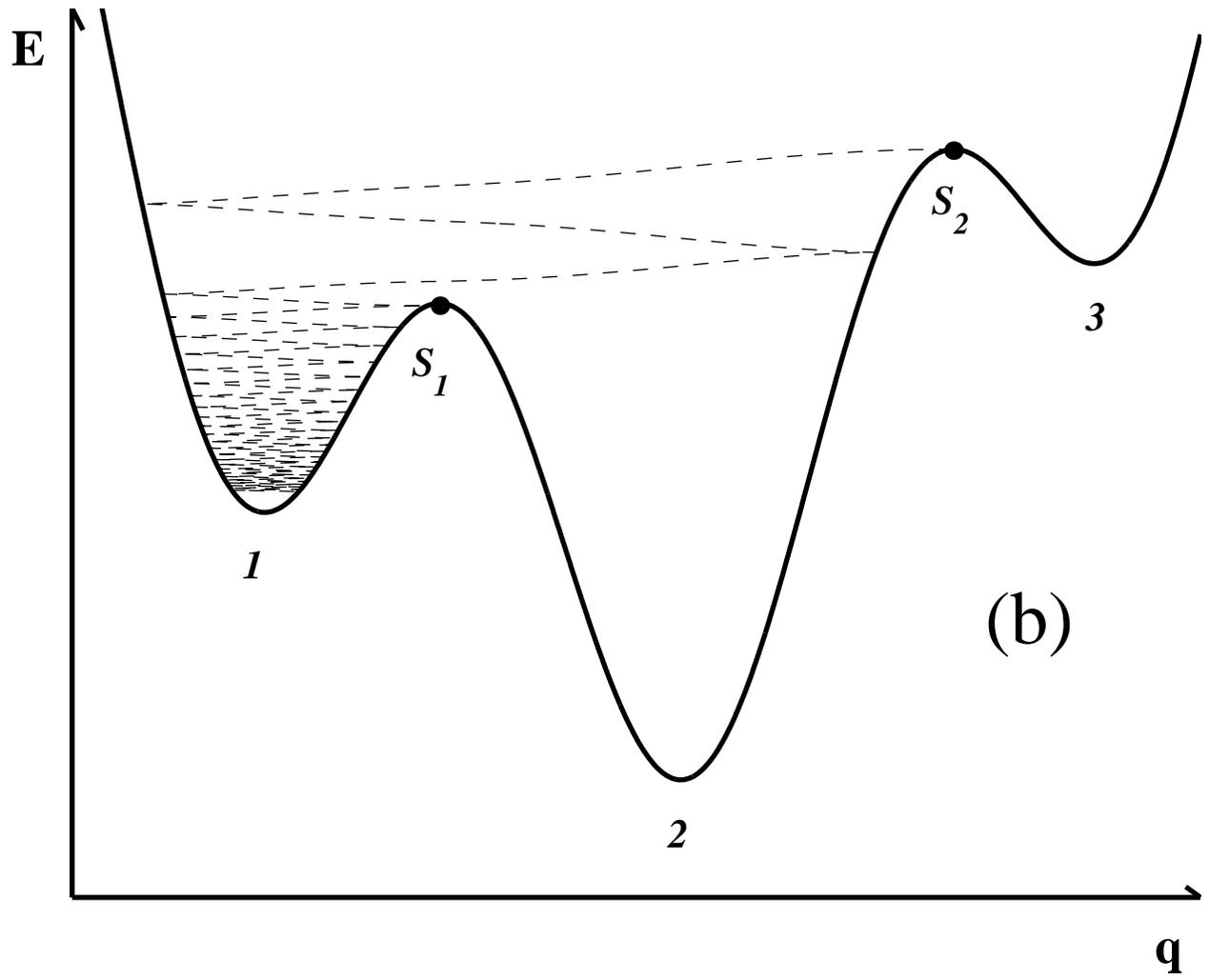
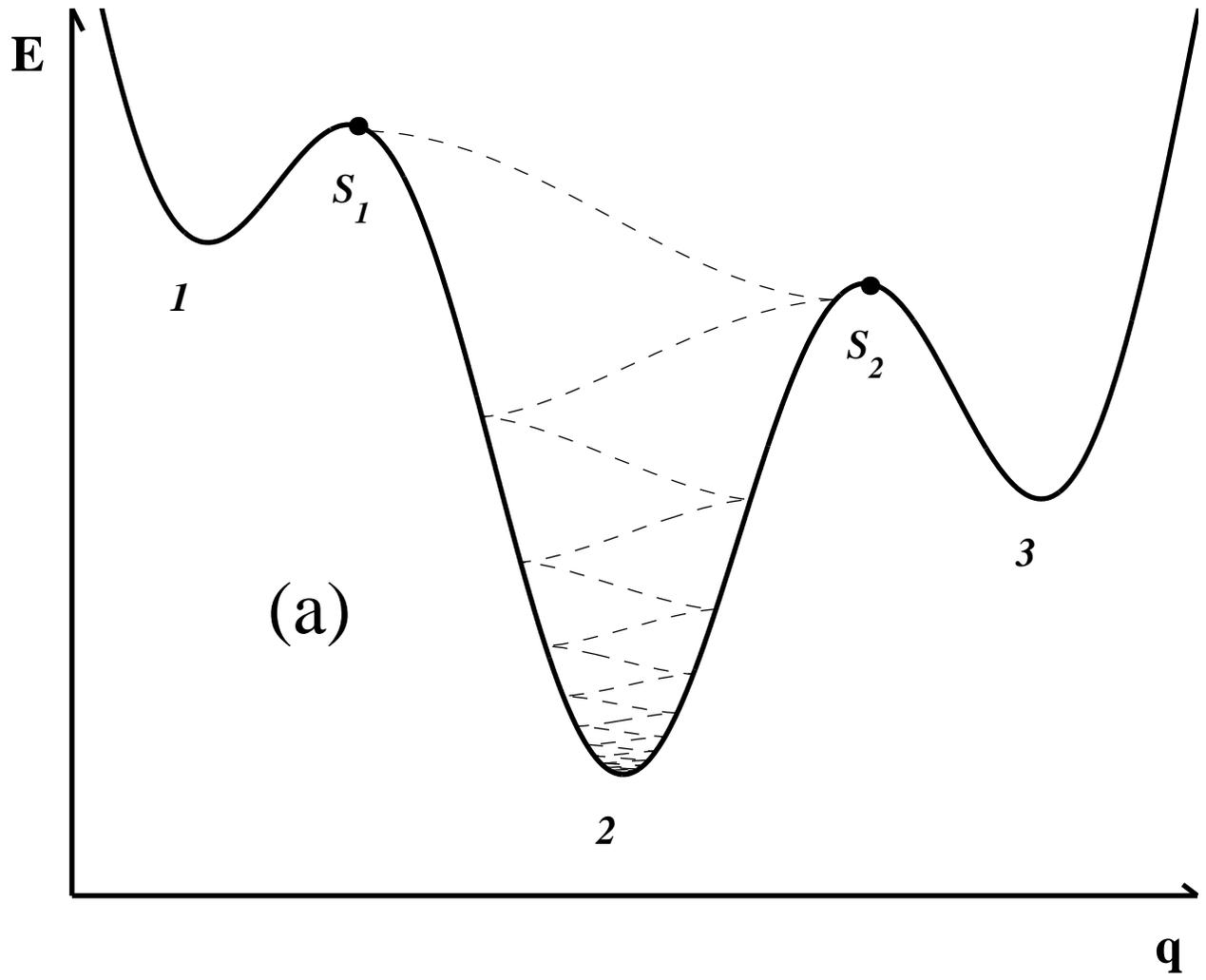
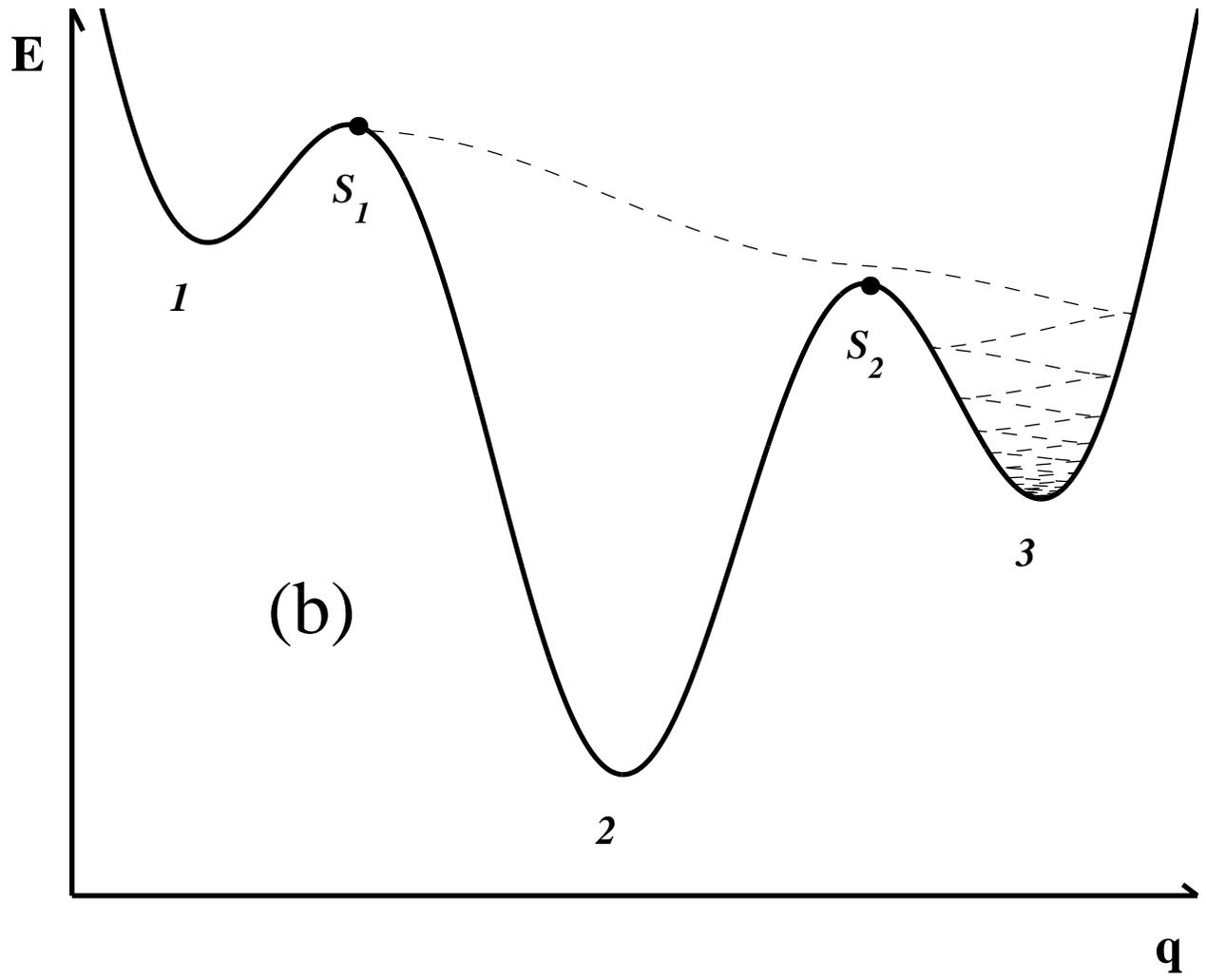
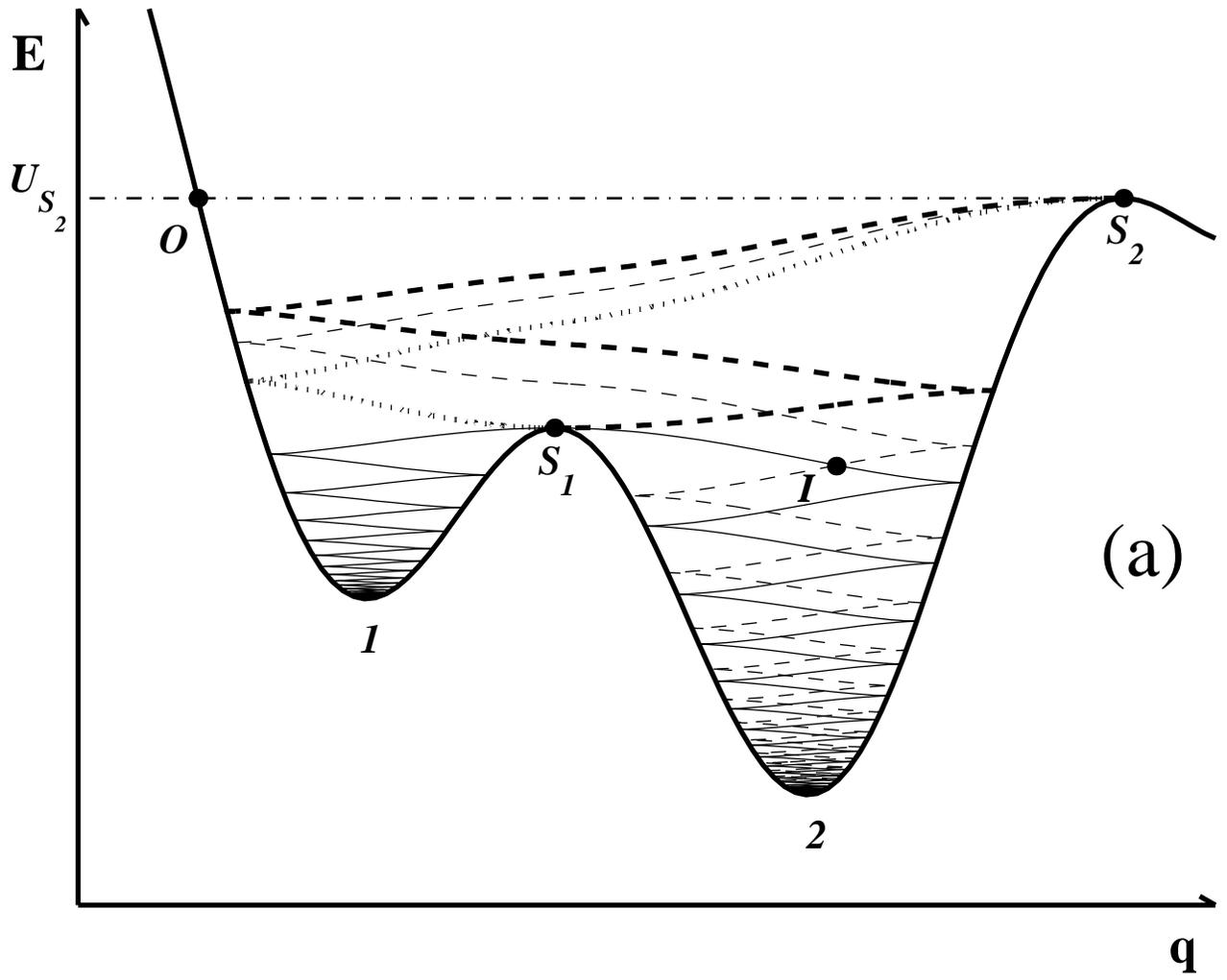
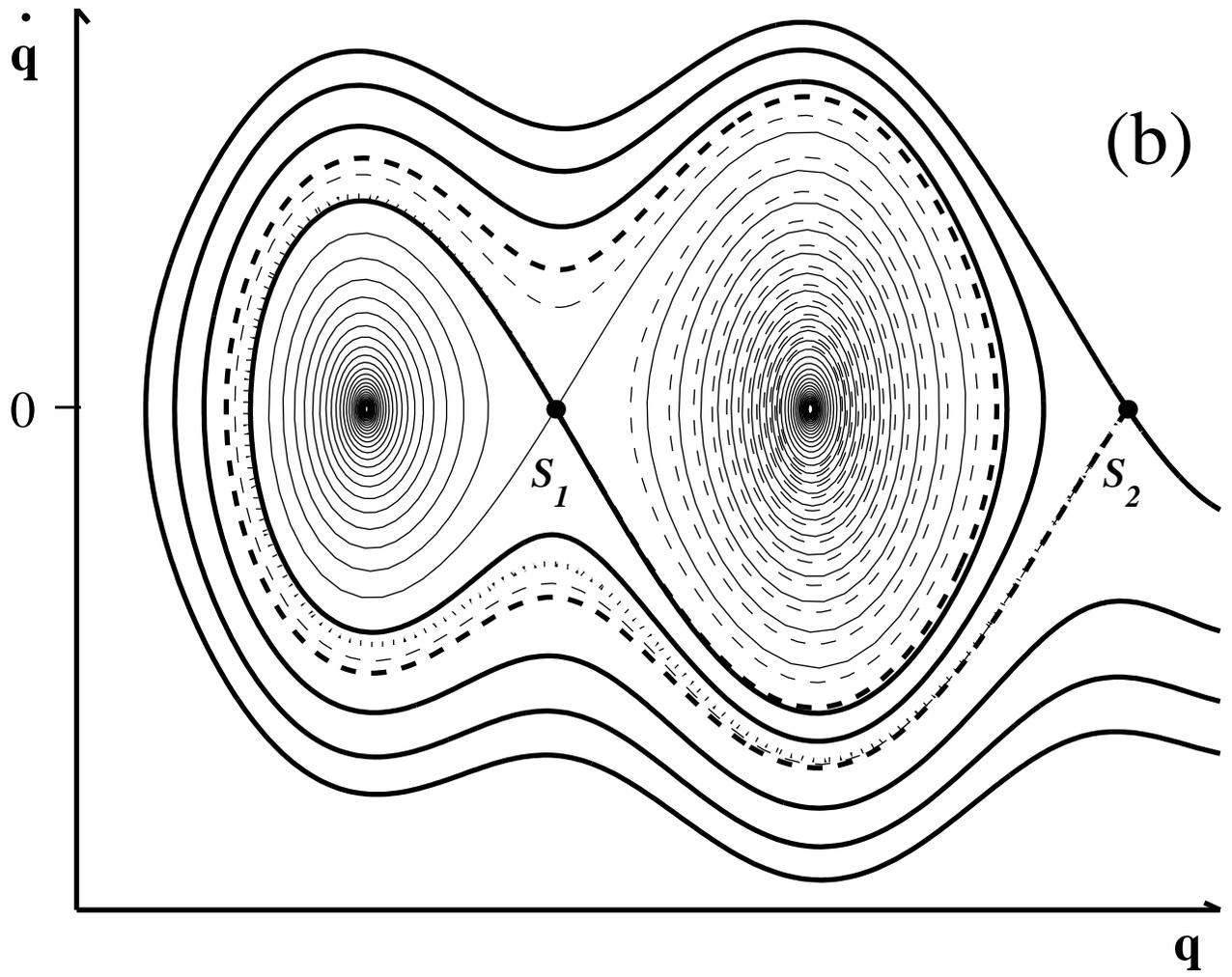
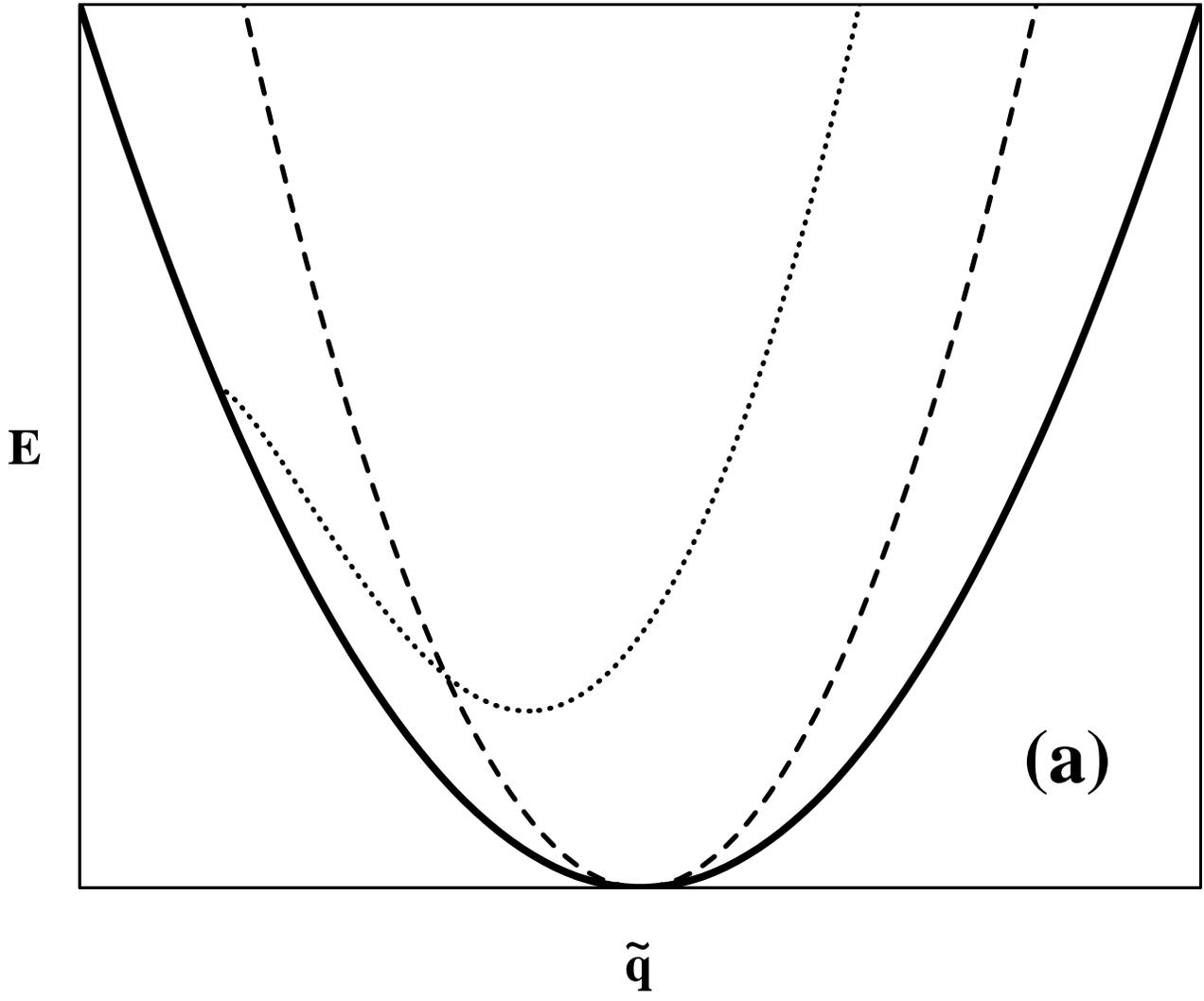
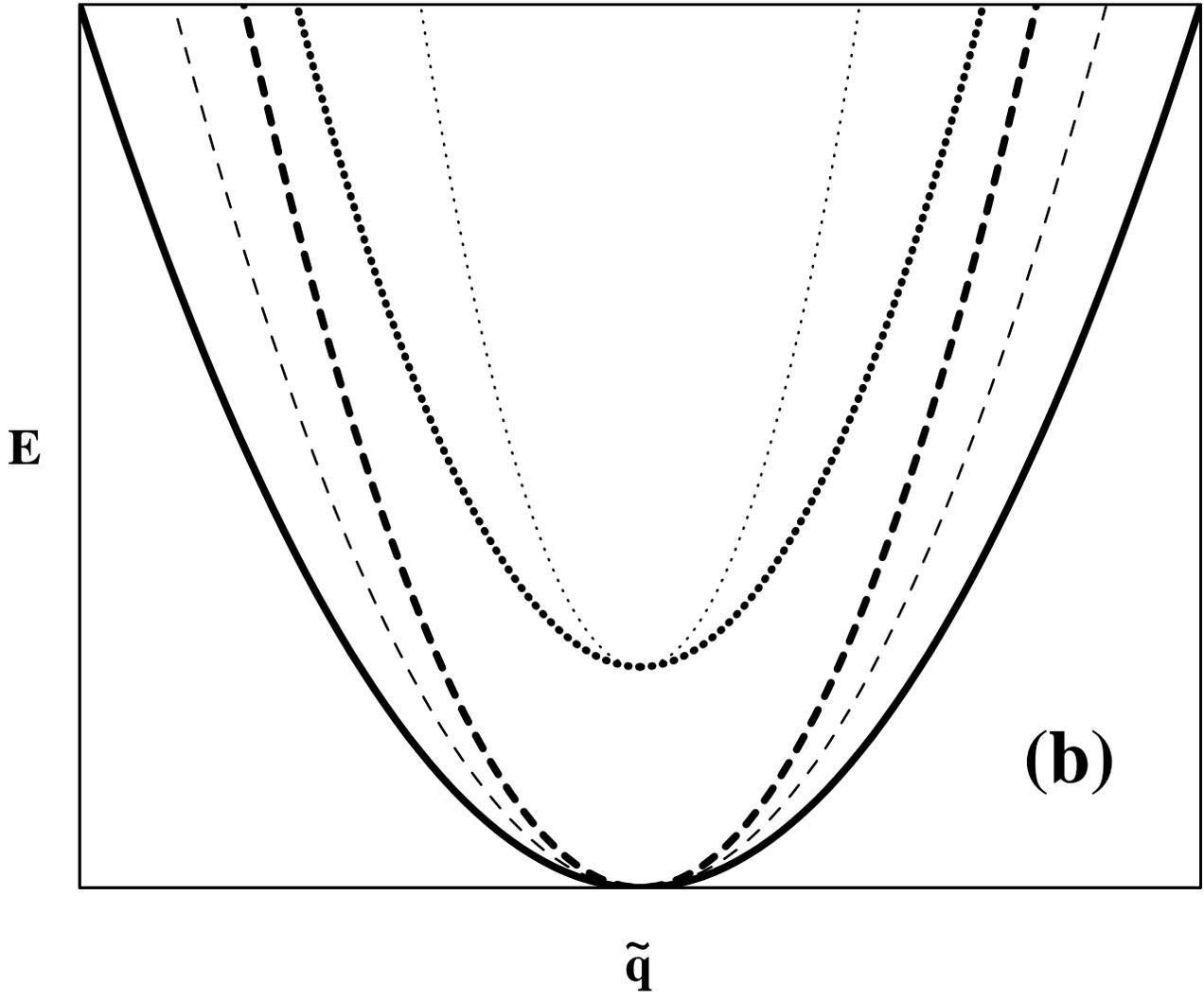
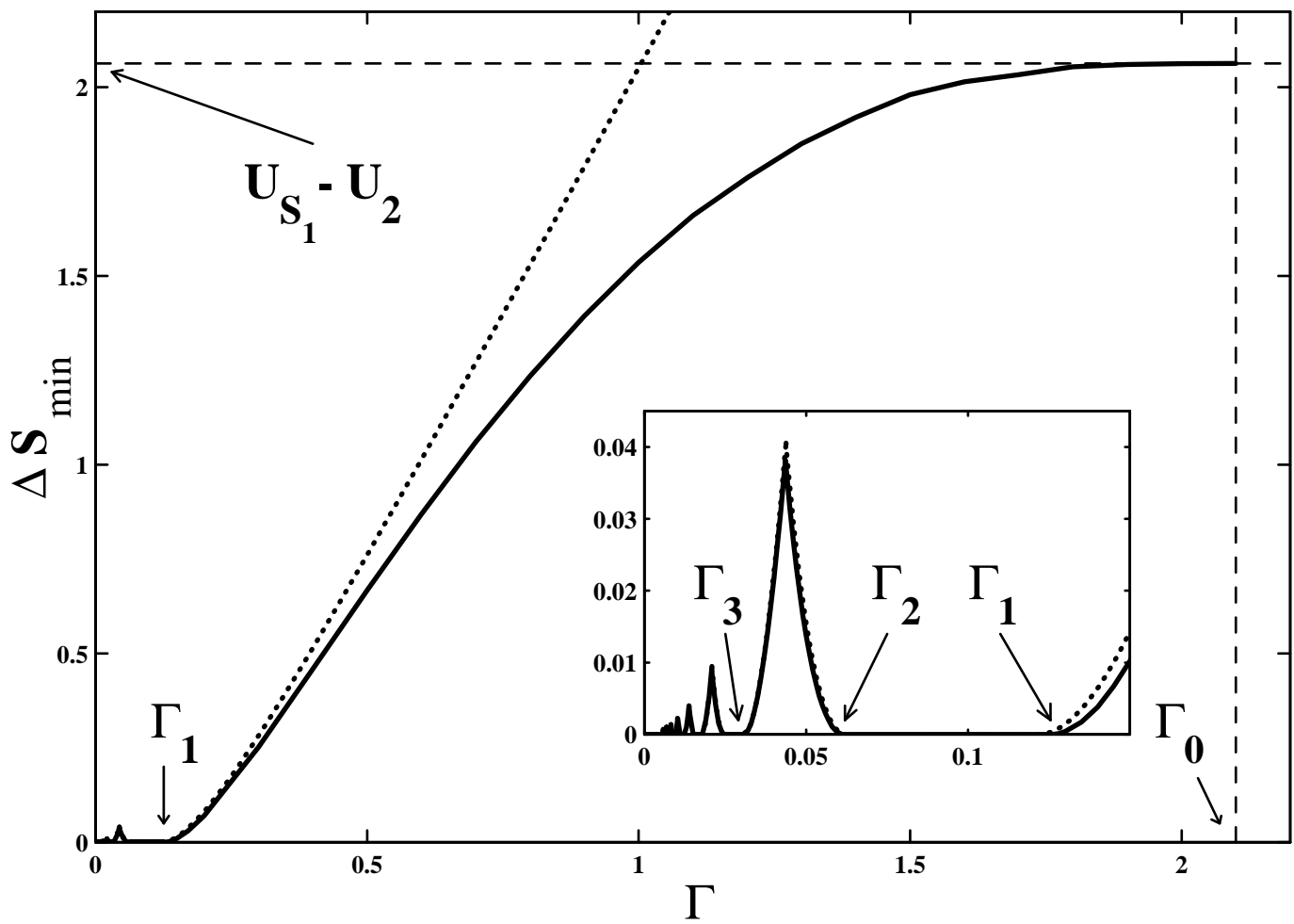
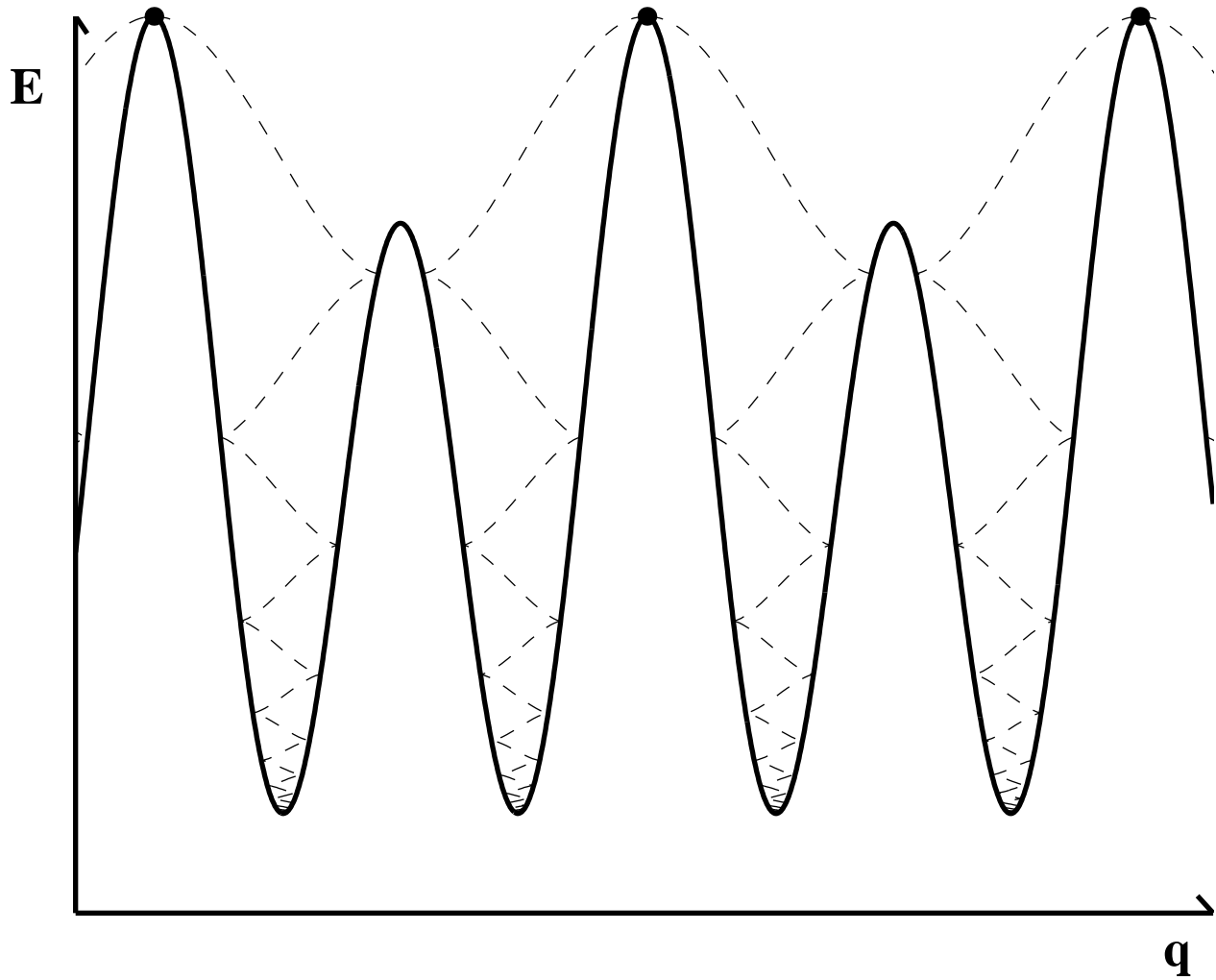
\end{document}